\documentclass[twoside,12pt]{article}
\usepackage{epsfig}
\usepackage{amsmath,amssymb}
\usepackage{url}

\usepackage{bm}
\newcommand{\Slash}[1]{\ooalign{\hfil/\hfil\crcr$#1$}}
\newcommand{\re}{\text{Re }}
\newcommand{\im}{\text{Im }}

\newcommand{\bra}[1]{\langle \, #1 \, |}
\newcommand{\ket}[1]{| \, #1 \, \rangle}

\newcommand{\kket}[1]{ \, #1 \, \rangle}
\usepackage{calligra}
\newcommand{\Jost}{\text{\textcalligra{f}\;}}

\newcommand{\figureBB}[2]{#2}

\newcommand{\be}{\begin{equation}}
\newcommand{\ee}{\end{equation}}
\newcommand{\bea}{\begin{eqnarray}}
\newcommand{\eea}{\end{eqnarray}}

\begin{document}

\title{ \vspace{1cm} QCD and the Strange Baryon Spectrum}

\author{Tetsuo\ Hyodo,$^{1}$ Masayuki Niiyama,$^2$\\
\\
$^1$Department of Physics, Tokyo Metropolitan University, Hachioji 192-0397, Japan\\
$^2$Department of Physics, Kyoto Sangyo University, Kyoto 603-8555, Japan}
\maketitle
\begin{abstract} 
The strange quark plays a unique role in QCD, reflecting its intermediate mass between the light and heavy quarks. In recent years, remarkable progress has been made in the spectroscopy of baryons with strangeness. Many new features of the strange baryon spectrum have been revealed by accurate experimental data with novel techniques, as well as systematic developments of theoretical framework to describe hadron resonances. The basic properties of strange baryons, namely, the pole positions, spin and parity, and decay branching ratios, are being determined accurately. As a consequence, the Particle Data Group have added new entries in the particle listings, such as the $\Lambda(1380)$ and the $\Omega(2012)$. The developments of the spectroscopy stimulate intensive discussion on the exotic internal structure of strange baryons beyond the ordinary three-quark configuration. In this review, we introduce the basics of QCD, the scattering theory, and the exotic internal structure of hadrons, emphasizing the importance of the pole positions of the scattering amplitude for the characterization of hadron resonances. We then summarize the current status of selected strange baryon resonances; $\Lambda(1405)$, $\Lambda(1670)$, $\Xi(1620)$, $\Xi(1690)$, and $\Omega(2012)$, from theoretical and experimental viewpoints.

\end{abstract}

\eject
\tableofcontents

\section{Introduction}

The spectrum of hadrons consists of wide variety of states, whose range is continuously increased by the reports of the newly observed hadrons~\cite{Zyla:2020zbs}. While the hadron spectrum should eventually be understood from the fundamental theory of the strong interaction, quantum chromodynamics (QCD), the nonperturbative nature of low-energy QCD still prevents us from the complete understanding of how the hadrons are constructed. Useful guiding principles to study the hadron spectrum are the symmetries in QCD. The dynamics of light quarks follows the constraints from chiral symmetry, which is exact in the massless limit of quarks. Quarks with a large mass are subject to heavy quark symmetry, which emerges in the limit of infinitely heavy mass of quarks. One can classify the up and down quarks in the light sector, and the charm and bottom quarks in the heavy sector, with suitably incorporating the symmetry breaking effects~\cite{Hosaka:2001ux,Scherer:2012xha,Manohar:2000dt}. From this viewpoint, the strange quark holds a unique position; $s$ is not as light as $u,d$ quarks but is too light to apply heavy quark symmetry. It is therefore a challenging task to study strange hadrons with keeping the connection with QCD symmetries.

Under the flavor SU(3) symmetry, $u$, $d$, $s$ quarks are treated identically in their interaction. 
However, since the strange quark does not have isospin and the number of strange quarks conserves under the strong interaction, it helps to simplify the internal structure of the strange hadrons compared with the hadrons made of only $ud$ quarks.
For example, the isospin structure of $\Lambda$ and $\Sigma$ hyperons
are determined by only $ud$ quarks in them, and a flavor exotic baryon that consists of
$uudd\bar{s}$ quarks can be identified from the existence of the $\bar{s}$ quark.
At the same time, the strangeness sector of hadrons contains several interesting states which are expected to have an exotic structure. 
For instance, the $\Lambda(1405)$ resonance~\cite{Hyodo:2011ur,Meissner:2020khl,Mai:2020ltx} has been studied for years and is still drawing attention. 
In the meson sector, the lowest scalar meson nonet has been intensively studied, due to the inverted ordering of the spectrum~\cite{Jaffe:2004ph}. 
Experimentally, the strange hadrons are accessible by the low-energy reactions with typical beam energy being a few GeV, 
as well as the high-energy collisions where a number of hadrons are produced in the final state. 
This is in contrast to the study of the hadrons in the heavy quark sector, which requires the energy to create at 
least one $\bar{c}c$ pair. Because the threshold energy of a charmed meson pair is $\sim 4$ GeV and that of a charmed baryon pair is $\sim 4.5$ GeV, to accumulate sufficient statistics, the charmed hadron spectroscopy is essentially limited in the high-energy experiments. On the other hand,
a variety of experiments were carried out to study the strange hadrons 
using meson, proton, photon, or electron beams additionally to high energy $e^+e^-$ or heavy ion collisions.
Among them, the experimental data of the two-body scattering in the strangeness baryon sector, 
such as $K^{-}p$ scattering cross sections, are beneficial to theoretical analysis, 
while direct scattering data are not available in the heavy quark sector.
Another experimental advantage of studying strange baryons stems from their rather narrow decay widths
in comparison with nucleon resonances.
In the case of the nucleon resonances, the resonance widths become wider even in the lower excited states, and
the partial wave analyses are necessary to separate the overlapping resonances. 
Strange baryons with relatively narrow decay widths are easier to identify experimentally in the invariant 
mass distributions.
For example, the first negative parity states of $N^*$ resonances, $N(1520)$ and $N(1535)$ have the Breit-Wigner widths of $110$ and $150$~MeV, respectively, while
corresponding states, the $\Lambda(1520)$ and the $\Lambda(1405)$, have $16$ and $\sim 50$~MeV respectively~\cite{Zyla:2020zbs}.
Thanks to the theoretical and experimental efforts, 
the properties of the strange baryons (the mass and width, or more precisely, 
the pole position) are being determined with accuracy. 
These are the basic information to study the properties of strangeness hadrons in nuclear medium~\cite{Tolos:2020aln}.

Once the basic properties of the hadrons, i.e., the pole positions, are settled, the next step is to clarify their internal structure of them, whether they have an ordinary structure of $q\bar{q}/qqq$, or they are constructed with some exotic configurations. There are many discussions on the possible exotic structures, such as the multiquark states, hadronic molecules, gluon hybrids, and so on~\cite{Klempt:2007cp,Klempt:2009pi,Hosaka:2016pey,Guo:2017jvc}. At first glance, the hadron structure seems to be identified by comparing the experimental data with a theoretical model with some specific configuration. But if one tackles this problem seriously, it turns out that this is not a straightforward task~\cite{Hyodo:2013nka}. There are several subtleties in the discussion of the structure of hadrons. For instance, the decomposition of the wave function in various components should be done with care, in order not to rely on specific models.

One of the most prominent issues in the discussion of the hadron structure is the unstable nature of the excited states. It should be emphasized that most of hadronic particles are unstable against the strong decay~\cite{Zyla:2020zbs}. In particular, the strange baryons in question have an appreciable decay width which should not be neglected in the discussion of the internal structure. From the theoretical viewpoint, it is now becoming a consensus that the excited hadrons should be treated as resonances in the hadron-hadron scattering~\cite{Meissner:2020khl}. In this sense, the Breit-Wigner mass and width of the peak structure are no longer suitable quantities to characterize the resonances, because the result can be reaction dependent due to the nonresonant background contributions. A theoretically unambiguous way to define the basic properties of hadron resonances is to determine the pole positions of the scattering amplitude in the complex energy plane. The pole position is in principle uniquely determined, and it represents the generalized eigenenergy of the Hamiltonian of the system. This is also reflected in the recent listings by the Particle Data Group (PDG)~\cite{Zyla:2020zbs}, where the pole positions are tabulated with higher priority than the traditional mass and width parameters. 

In this paper, we review the recent developments to understand the spectrum of baryons with strangeness, focusing on several specific states. We start from the introduction of a theoretical basis to extract information of the excited hadrons in Section~\ref{sec:theoreticalbasis}. In particular, we discuss in detail how one characterizes the excited hadrons as resonances in hadron scatterings. Next, we turn to the overview of selected strangeness baryons. Section~\ref{sec:S-1} deals with the strangeness $S=-1$ baryons, with special emphasis on the $\Lambda(1405)$, which is the most striking state in this field. The baryons with two and three strange quarks and with an anti-strange quark
are discussed in Sections~\ref{sec:S-2}, \ref{sec:S-3} and \ref{sec:S+1}, respectively. Summary of this review and possible future prospects are presented in the last section.

\section{Theoretical basis}\label{sec:theoreticalbasis}


\subsection{\it Quantum chromodynamics}

The strong interaction is governed by quantum chromodynamics (QCD), which is the color SU(3) quantum gauge theory. Here we summarize the properties of QCD, focusing on various symmetries. Lattice QCD approach for hadron spectroscopy is briefly introduced. Unless otherwise stated, we adopt the natural units where $c=\hbar=1$.

\subsubsection{\it Basics of QCD}
\label{subsubsec:basics}

The Lagrangian of QCD is given by
\begin{align}
    {\cal L}_{\mathrm{QCD}} 
    &= - \frac{1}{4} G_{\mu\nu}^{a} G^{a \mu\nu} 
    + \bar{q} 
    (i\Slash{D}-m_{q})q ,
    \label{eq:QCD}
\end{align}
with the field strength tensor $G_{\mu\nu}^{a}$ and the covariant derivative $D_{\mu}$
\begin{align}
    G_{\mu\nu}^{a}
    &= 
    \partial_{\mu}A_{\nu}^{a}
    -\partial_{\nu}A_{\mu}^{a}
    -gf^{abc}A_{\mu}^{b}A_{\nu}^{c}, \\
    D_{\mu}
    &= 
    \partial_{\mu}
    +igA_{\mu}^{a}T^{a} ,
\end{align}
where $g$ is the gauge coupling constant and $m_{q}$ is the quark mass. The gluons are represented by the gauge field $A_{\mu}^{a}$ $(a=1,\dots, 8)$ which belongs to the adjoint representation of color SU(3) symmetry. With the generator of the color SU(3) group $T^{a}$, the structure constant $f^{abc}$ is defined by the commutation relation as $[T^{a},T^{b}]=if^{abc}T^{c}$. The first term of Eq.~\eqref{eq:QCD} contains the kinetic terms of the gluons and the gluonic self interactions. The quarks are expressed by the Dirac fields $q$ and $\bar{q}$. The quark field $q$ ($\bar{q}$) belongs to the $\bm{3}$ ($\bar{\bm{3}}$) representation of color SU(3), and is given by the three-component column (row) vector in the color space. The quark-gluon coupling is given by the second term of the covariant derivative, in which the generator is expressed by the three-by-three Gell-Mann matrices as $T^{a}=\lambda^{a}/2$. The QCD Lagrangian is invariant under the local color SU(3) transformation. In Eq.~\eqref{eq:QCD}, we do not explicitly show the gauge fixing terms which is needed in the quantization procedure, and the $\theta$ term which breaks the CP symmetry in the strongly interacting sector.

In addition to colors, quarks have internal degrees of freedom of flavor. By explicitly writing the flavors, the quark field can be given by the six component column vector
\begin{align}
    q
    &=\begin{pmatrix}
    u \\ d \\ s \\ c \\ b \\ t
    \end{pmatrix} .
\end{align}
In the flavor space, the quark mass $m_{q}$ is given by the diagonal matrix $\text{diag}(m_{u},m_{d},m_{s},m_{c},m_{b},m_{t})$. The up $(u)$ and down $(d)$ quarks form the isospin $I=1/2$ doublet, with the third component assignment of $I_{3}=+1/2$ ($I_{3}=-1/2$) for the $u$ ($d$) quark. The strange $(s)$, charm $(c)$, bottom $(b)$, and top $(t)$ quarks have strangeness $S=-1$, charm $C=+1$, bottomness $B=-1$ and topness $T=+1$, respectively. These flavor quantum numbers are conserved under the strong interaction, because the QCD interactions are mediated by gluons which couple to quarks irrespective of their flavor. Because the top quark undergoes the Cabibbo favored weak decay before it forms a hadron via the strong interaction, the hadrons can be classified by the flavor quantum numbers of $I,I_{3},S,C,B$. The quark mass $m_{q}$ is not a direct observable due to the color confinement, and therefore the meaningful ``quark mass'' can only be given by specifying the details of the renormalization procedure. Under the $\overline{\rm MS}$ renormalization scheme at the renormalization scale $\mu=2$ GeV, the central values of the light quark masses are given by~\cite{Zyla:2020zbs}
\begin{align}
    m_{u}
    &=2.16\text{ MeV},\quad
    m_{d}=4.67\text{ MeV},\quad
    m_{s}=93 \text{ MeV} .
\end{align}
The running masses of the heavy quarks are evaluated at the renormalization scale $\mu=m_{c}$ for the charm quark and $\mu=m_{b}$ for the bottom quark in the $\overline{\rm MS}$ renormalization scheme, leading to~\cite{Zyla:2020zbs}
\begin{align}
    m_{c}
    &=1.27\text{ GeV},\quad
    m_{b}=4.18\text{ GeV}.\quad
\end{align}
The determination of the top quark mass involves additional subtleties, and it is estimated to be $m_{t}\sim 172$ GeV~\cite{Zyla:2020zbs}. It is remarkable that the quark masses are distributed in a wide energy range. This large variation of the energy scale is in fact the origin of the chiral and heavy quark symmetries in QCD as we discuss in the next section. 

An important property of QCD is the asymptotic freedom, namely, the decrease of the coupling constant at high energy~\cite{Gross:1973id,Politzer:1973fx}. Thanks to the asymptotic freedom, one can show that the perturbative calculation of QCD quantitatively  explains the violation of the Bjorken scaling observed in the deep inelastic scattering experiments. Renormalization group equation for the running coupling constant $\alpha_{s}=g^{2}/4\pi$ at the energy scale $\mu$ in QCD reads
\begin{align}
    \mu^{2}\frac{d\alpha_{s}}{d\mu^{2}}
    &=
    -\frac{(33-2N_{f})}{12\pi}\alpha_{s}^{2}+\dotsb ,
    \label{eq:RGE}
\end{align}
where $N_{f}$ is the number of active flavors at $\mu$, and the ellipsis stands for the higher order corrections. The number $33$ essentially stems from the gluons reflecting the non-Abelian nature of QCD, and the factor $-2N_{f}$ represents the contribution from quarks. For $N_{f}=6$, the coefficient of $\alpha_{s}^{2}$ is negative, showing the asymptotic free nature of QCD. The solution of the differential equation~\eqref{eq:RGE} is obtained as
\begin{align}
    \alpha_{s}(\mu^{2})
    &=\frac{12\pi}{(33-2N_{f})\ln\dfrac{\mu^{2}}{\Lambda_{\rm QCD}^{2}}}\left(1+\dotsb\right) ,
\end{align}
with a constant of integration $\Lambda_{\rm QCD}$, the energy scale at which the perturbative estimate of the coupling constant would diverge. This indicates that the perturbative calculation in QCD breaks down at some low-energy scale $\mu\sim \Lambda_{\rm QCD}$. Numerically, the scale is estimated to be $\Lambda_{\rm QCD}\sim 200$ MeV. This nonperturbative nature of the low-energy QCD is the origin of the rich and complicated dynamics of hadrons. 

\subsubsection{\it Symmetries in QCD}
\label{sec:symmetriesinQCD}

Symmetries have been playing central roles in modern physics. Let us recall an example in nonrelativistic quantum mechanics. When the potential is spherically symmetric, the angular momentum $\ell$ is a good quantum number, and the eigenstates of the Hamiltonian should appear with the $2\ell+1$ degeneracy. From the group theoretical viewpoint, the angular momentum operator is the generator of the infinitesimal transformation of the three-dimensional rotation. In this case, the Hamiltonian has an O(3) symmetry, and the $2\ell+1$ degeneracy is guaranteed for any form of the radial dependence of the potential. In other words, one can predict the $2\ell+1$ degeneracy of the eigenstates from O(3) symmetry without any calculations. If the symmetry is not exact, then the degeneracy becomes an approximate one. For instance, when the external magnetic field is applied in the $z$ direction, the rotational symmetry is broken, and the degeneracy is not manifest any more (Zeeman effect). Even in this case, the remnant of the symmetry can be seen as the approximate degeneracy of the states, and the energy splitting can be calculated from the strength of the magnetic field, which characterizes the degree of the symmetry breaking. In this way, the symmetry and its breaking can provide useful information of the observables in the system. 

The QCD Lagrangian~\eqref{eq:QCD} has several symmetries. It is invariant under the Lorentz transformation, CPT transformation, and color SU(3) gauge transformation. In addition to these exact symmetries, there are approximate symmetries; chiral symmetry, flavor symmetry, and heavy quark symmetry. In the following, we introduce these symmetries and their consequences.

Chiral symmetry is related to the right- and left-handed components of Dirac particles.~\cite{Hosaka:2001ux,Scherer:2012xha}. In general, the Dirac field $q$ can be decomposed into the right-handed field $q_{R}$ and the left-handed one $q_{L}$ as
\begin{align}
    q&=q_{R}+q_{L},
    \quad
    q_{R}
    =  P_{R}q,
    \quad 
    q_{L} 
    =  P_{L}q, 
\end{align}
where the projection operators $P_{R}$ and $P_{L}$ are defined by
\begin{align}
    P_{R}
    = \frac{1+\gamma_{5}}{2} , \quad
    P_{L}
    = \frac{1-\gamma_{5}}{2} ,
\end{align}
which satisfy the relations $P_{R}+P_{L}=1$, $P_{R}P_{L}=P_{L}P_{R}=0$, $P_{R}^{2}=P_{R}$, and $P_{L}^{2}=P_{L}$. Taking the Dirac conjugate, we obtain
\begin{align}
    \bar{q}&=\bar{q}_{R}+\bar{q}_{L},
    \quad
    \bar{q}_{R}
    =  \bar{q}P_{L} ,
    \quad
    \bar{q}_{L}
    =  \bar{q}P_{R}  .
\end{align}
With the right- and left-handed fields, the quark part of the QCD Lagrangian~\eqref{eq:QCD} is decomposed as
\begin{align}
    \bar{q} 
    (i\Slash{D}-m_{q})q 
    &= \bar{q}_{R} i\Slash{D}q_{R} 
    +\bar{q}_{L} i\Slash{D}q_{L}
    -(\bar{q}_{L}m_{q}q_{R}+\bar{q}_{R}m_{q}q_{L}) .
    \label{eq:decomposition}
\end{align}
We observe that the right- and left-handed components are separated in the kinetic term, while they are mixed in the mass term. Chiral transformation of $N_{f}$ flavor is defined as the independent unitary transformation (complex rotation) of the right- and left-handed components:
\begin{align}
    q_{R} 
    &\to Rq_{R},\quad
    R=e^{i\theta_{R}^{i}T^{i}} \in \textrm{U}(N_{f})_{R} ,
    \quad i=0,\dotsb, N_{f}^{2}-1 ,\label{eq:right} \\
    q_{L} 
    &\to Lq_{L},\quad
    L=e^{i\theta_{L}^{i}T^{i}} \in \textrm{U}(N_{f})_{L},\label{eq:left}
\end{align}
where $\theta_{R}^{i}$ and $\theta_{L}^{i}$ are the parameters of the transformation, $T^{i}(i\geq 1)$ are the generators of SU($N_{f}$), and $T^{0}=1/\sqrt{2N_{f}}$. Global U($N_{f})_{R}$$\otimes$U($N_{f})_{L}$ transformation is specified by a set of two $N_{f}\times N_{f}$ unitary matrices $(R,L)$. The vector and axial vector transformations are defined as 
\begin{align}
    q_{R} 
    &\to V q_{R},
    \quad
    q_{L} 
    \to V q_{L}, 
    \quad 
    V=e^{i\theta_{V}^{i}T^{i}}\in \textrm{U}(N_{f})_{V} ,
    \label{eq:vector}\\
    q_{R} 
    &\to A q_{R},\quad
    q_{L} 
    \to A^{\dag} q_{L} ,
    \quad
    A=e^{i\theta_{A}^{i}T^{i}} \in \textrm{U}(N_{f})_{A}
    \label{eq:axial}.
\end{align}
Namely, the vector transformation $(V,V)$ rotates the right- and left-handed fields in the same direction, while the axial transformation $(A,A^{\dag})$ does in the opposite direction. In general, the unitary group can be decomposed as ${\rm U}(N)={\rm U}(1)\otimes {\rm SU}(N)$, and the $N_{f}$ flavor chiral symmetry can be expressed as 
\begin{align}
    {\rm U}(1)_{V}
    \otimes {\rm U}(1)_{A}
    \otimes {\rm SU}(N_{f})_{R}
    \otimes {\rm SU}(N_{f})_{L} .
\end{align}
The kinetic term in Eq.~\eqref{eq:decomposition} is invariant under all these transformations, while the mass term is invariant only under ${\rm U}(1)_{V}$. This means that QCD has an exact phase symmetry ${\rm U}(1)_{V}$, which is manifested as the conservation of the quark number (the number of quarks minus the number of antiquarks). In the limit of vanishing quark mass (chiral limit) $m_{q}\to 0$, the QCD Lagrangian is invariant under all the chiral transformations, although the ${\rm U}(1)_{A}$ symmetry is broken by quantum anomaly~\cite{tHooft:1986ooh}. If there are $N_{f}$ quarks with a small mass, chiral symmetry
\begin{align}
    {\rm SU}(N_{f})_{R}
    \otimes {\rm SU}(N_{f})_{L} ,
\end{align}
is approximately realized in the QCD Lagrangian. 

Chiral symmetry is known to be broken spontaneously in vacuum~\cite{Nambu:1961tp,Nambu:1961fr,Goldstone:1961eq}. Symmetry is said to be spontaneously broken, if the symmetry of the Lagrangian (Hamiltonian) is not realized in the vacuum (eigenstate). We here recall again a nonrelativistic example of spontaneous symmetry breaking. Consider the quantum spin system with the Heisenberg model where the spins can point to any direction in the three-dimensional space. The Hamiltonian has three-dimensional rotation symmetry, which is realized in the eigenstate at sufficiently high temperature due to the thermal fluctuation. At low temperature, on the other hand, spins are aligned to a specific direction, breaking the rotational symmetry. The degree of the symmetry breaking can be measured by the order parameter, which is the expectation value of 
an operator that breaks the symmetry. In the Heisenberg ferromagnet, the order parameter is the magnetization, which becomes nonzero when the symmetry is spontaneously broken. In the case of massless QCD, one of the order parameters of the chiral symmetry breaking is the quark condensate, which is nonzero in the QCD vacuum $\ket{0}$:
\begin{align}
    \bra{0}\bar{q}q\ket{0}
    &\neq 0 .
\end{align}
An important feature of the quark condensate is its relation to the density of the states with vanishing eigenvalue of the QCD Dirac operator~\cite{Banks:1979yr}.
Note that the operator $\bar{q}q=\bar{q}_{L}q_{R}+\bar{q}_{R}q_{L}$ is not invariant under general ${\rm SU}(N_{f})_{R}\otimes {\rm SU}(N_{f})_{L}$ transformations. On the other hand, the flavor symmetry ${\rm SU}(N_{f})_{V}$ remains in the presence of the quark condensate. In fact, the vector symmetries are known to be unbroken spontaneously~\cite{Vafa:1983tf}. The spontaneous symmetry breaking pattern in QCD is thus given by 
\begin{align}
    \textrm{SU}(N_{f})_{R}\otimes \textrm{SU}(N_{f})_{L}
    &\to \textrm{SU}(N_{f})_{V} .
    \label{eq:SSB}
\end{align}
An important consequence of this spontaneous symmetry breaking is the emergence of the massless Nambu-Goldstone (NG) bosons. In the Lorentz invariant system, an NG boson appears for each generator of the broken symmetry. For $N_{f}$ flavor chiral symmetry, there are $N_{f}^{2}-1$ broken generators of the axial transformation, and the corresponding number of the NG bosons appear. For $N_{f}=2$, three pions ($\pi^{+},\pi^{-},\pi^{0}$) correspond to the NG bosons, and for $N_{f}=3$, the eight pseudoscalar mesons that form an octet ($\pi,K,\eta$) are identified as the NG bosons. Chiral symmetry also dictates the dynamics of the Nambu-Goldstone bosons. The celebrated low-energy theorems, such as Goldberger-Treiman relation~\cite{Goldberger:1958vp} and Weinberg-Tomozawa relation~\cite{Weinberg:1966kf,Tomozawa:1966jm}, are important constraints on the dynamics of hadrons from QCD. A systematic approach to incorporate the low-energy theorems has been developed as chiral perturbation theory~\cite{Weinberg:1978kz,Gasser:1983yg,Gasser:1984gg,Bernard:1995dp,Pich:1995bw,Bernard:2007zu,Scherer:2012xha}, which is the effective field theory of low-energy QCD based on chiral symmetry. On top of the spontaneous breaking, chiral symmetry is also broken explicitly by the quark masses. At this point, we need to discuss which flavor can be regarded as ``sufficiently light'' to apply constraints from chiral symmetry. In other words, we specify the typical energy scale of chiral symmetry $\Lambda_{\chi}$ and perform expansion in powers of $m_{q}/\Lambda_{\chi}$. Usually, $\Lambda_{\chi}$ is estimated by the scale of spontaneous chiral symmetry breaking, $\Lambda_{\chi}\sim 4\pi f\sim 1$ GeV. Here $f$ is the meson decay constant which determines the strength of the pion field in the axial vector current, and the factor $4\pi$ stems from the loop correction. The scale $\Lambda_{\chi}$ roughly corresponds to the mass of hadrons other than the NG bosons, such as the $\rho$ meson and the nucleon. From the quark masses shown above, it is clear that up and down quarks are sufficiently light, while chiral symmetry is not applicable for charm and bottom quarks. Strange quark can be classified as the light one, but one must keep in mind the substantial explicit symmetry breaking effect.

By writing Eq.~\eqref{eq:vector} for the quark field $q=q_{R}+q_{L}$, one can understand the vector SU$(N_{f})_{V}$ transformation as the rotation of quarks in the flavor space:
\begin{align}
    q 
    &\to V q, 
    \quad
    V=e^{i\theta_{V}^{i}T^{i}}\in \textrm{SU}(N_{f})_{V} .
    \label{eq:flavor}
\end{align}
Flavor symmetry is therefore exact, even in the presence of the quark masses, if $N_{f}$ quarks have an equal mass. The mass difference of quarks induces the explicit symmetry breaking. Flavor symmetry with $N_{f}=2$ is the isospin symmetry, which is known to be satisfied with a good accuracy. The symmetry breaking effect is typically a few MeV order ($M_{n}-M_{p}\sim 1.3$ MeV, $m_{\pi^{0}}-m_{\pi^{\pm}}\sim 4.6$ MeV, $m_{K^{0}}-m_{K^{\pm}}\sim 3.9$ MeV), which can be safely neglected in most cases. It must however be noted that the isospin symmetry breaking effect can be enlarged in some special cases. For instance, the typical energy scale of near-threshold states (binding or excitation energy) can be comparable with the threshold energy difference by the isospin symmetry breaking. A striking example is the $X(3872)$ state near the $D\bar{D}^{*}+c.c.$ threshold. The threshold energy of the $D^{0}\bar{D}^{*0}+c.c.$ state is 3871.68 MeV, and that of $D^{+}\bar{D}^{*-}+c.c.$ is 3879.91 MeV. Although the splitting of several MeV is in the expected magnitude of the isospin symmetry breaking, to discuss the $X(3872)$ at 3871.69 MeV~\cite{Zyla:2020zbs}, the energies of these channels should be treated separately. In the strangeness sector, flavor SU(3) symmetry is not an accurate symmetry in QCD, because of the substantial strange quark mass. The typical size of the symmetry breaking is of the order of few hundred MeV ($M_{\Lambda}-M_{N}\sim 177$ MeV, $M_{K^{*}}-M_{\rho}\sim 116.4$ MeV). Nevertheless, the classification of hadrons into SU(3) multiplets (singlet, octet, decuplet, $\dotsb$) has been an important guiding principle in hadron spectroscopy, because the symmetry breaking effect can be systematically incorporated. The mass splittings in a given SU(3) multiplet due to the leading order symmetry breaking by the strange quark mass can be calculated by the Gell-Mann--Okubo mass formula~\cite{GellMann:1962xb,Okubo:1961jc}
\begin{align}
  M(I,Y)
  &=
  a+bY+c\left[
  I(I+1)-\frac{1}{4}Y^{2}\right] ,
  \label{eq:GMO}
\end{align}
where $I$ is the isospin and $Y$ is the hypercharge, and $a,b,c$ are the parameters for each multiplet. This formula can be derived purely from the group theory, and therefore it should be valid irrespective of their internal structure. The first two terms can be understood as the common mass of the multiplet $(a)$ and the mass excess of the strange quark $(bY)$ (hypercharge $Y$ is linear in strangeness $S$), while the last term proportional to $c$ represents a nontrivial consequence of the SU(3) symmetry. The formula~\eqref{eq:GMO} indicates that there are only three degrees of freedom for any SU(3) multiplets. Thus, one can derive mass relations for a representation containing more than three isospin multiplet. A baryon octet contains four isospin states ($N$, $\Lambda$, $\Sigma$, $\Xi$) and there is one mass relation:
\begin{align}
  2(M_{N}+M_{\Xi})
  &
  =3M_{\Lambda}+M_{\Sigma} ,
  \label{eq:GMOoctet}
\end{align}
which is satisfied by the ground state baryon octet with high accuracy, $3M_{\Lambda}+M_{\Sigma}-2(M_{N}+M_{\Xi}))/(3M_{\Lambda}+M_{\Sigma})\sim 0.0059$. The equal spacing rule of the baryon decuplet
\begin{align}
  M_{\Sigma^{*}}-M_{\Delta}
  &=
  M_{\Xi^{*}}-M_{\Sigma^{*}}
  =
  M_{\Omega}-M_{\Xi^{*}} ,
  \label{eq:decupletsplitting}
\end{align}
was used to predict the $\Omega$ state from the information of $\Delta$, $\Sigma^{*}$ and $\Xi^{*}$.\footnote{Decuplet has four isospin states and the equal spacing rule~\eqref{eq:decupletsplitting} contains two relations. This is because a peculiar combination of $b$ and $c$ enters in the expressions of the mass, and there are only two degrees of freedom in the mass formula for decuplets. In general, symmetric representations having a triangle weight diagram ($\bm{6},\bm{10},\overline{\bm{10}},\dotsb$) exhibit the equal spacing rule.} In this way, even with the sizable symmetry breaking effect, approximate flavor SU(3) symmetry is an important clue to study hadron spectroscopy. At the same time, it should be emphasized that the symmetry breaking effect induces the mixing of the states in different multiplets, if they have the same quantum numbers. Namely, $\Sigma$ and $\Xi$ in octet can mix with the corresponding states in decuplet, which can disturb the simple mass relations in Eqs.~\eqref{eq:GMOoctet} and \eqref{eq:decupletsplitting}.

Charm and bottom quarks are heavy and beyond the applicability of the symmetries discussed above. For such heavy quarks, it is useful to consider the opposite  limit, infinitely heavy quark masses~\cite{Isgur:1989vq,Georgi:1990um,Neubert:1993mb,Manohar:2000dt}. In the $m_{q}\to \infty$ limit, heavy quark spin symmetry and heavy quark flavor symmetry are realized for hadrons with a single heavy quark.\footnote{We emphasize that the heavy quark symmetries are \textit{not} realized in hadrons containing multiple heavy quarks. For the system with a pair of heavy quark and antiquark, one should use the framework of nonrelativistic QCD (NRQCD)~\cite{Bodwin:1994jh,Brambilla:2004jw}. } In the $m_{q}\to \infty$ limit, the heavy quark is regarded as a static color source in a hadron with one heavy quark. The gluonic interaction with this static quark is given only by the color electric charge, and therefore the interaction is independent of the heavy quark spin. This means that the singly heavy hadron is invariant under the spin-flip of the heavy quark, known as heavy quark spin symmetry. This can be seen by decomposing the momentum of the heavy quark $p^{\mu}$ as 
\begin{align}
 p^{\mu} = m_{q} v^{\mu} + k^{\mu},
\end{align}
where $v^{\mu}$ is the four-velocity of the heavy quark and the residual momentum $k^{\mu}$ arises from the interaction with the light constituents. Because the typical momentum which is exchanged within a hadron can be estimated as the nonperturbative scale $\Lambda_{\rm QCD}$, the magnitude of each component of $k^{\mu}$ should be estimated by $\Lambda_{\rm QCD}$. Performing an expansion in powers of $\Lambda_{\rm QCD}/m_{q}$ for the quark part of the QCD Lagrangian~\eqref{eq:QCD}, we obtain the leading order term as 
\begin{align}
    \bar{Q}_{v} v \cdot iD Q_{v}
    + \dotsb ,
   \label{eq:HQET_Lagrangian}
\end{align}
with the velocity dependent heavy quark field
\begin{align}
   Q_{v}(x) = \frac{1+\Slash{v}}{2} e^{im_{q} v\cdot x} q(x),
\end{align}
where $(1+\Slash{v})/2$ is the projection to the positive energy components and the factor $e^{im_{Q} v\cdot x}$ subtracts the heavy quark mass from the definition of the energy. The covariant derivative $D$ generates the coupling to the gluons without spin flip, and the magnetic coupling which flips the spin of the heavy quark is in the $\mathcal{O}(\Lambda_{\rm QCD}/m_{q})$ contributions. Thus, the system shows heavy quark spin symmetry in the $m_{q}\to\infty$ limit. Equation~\eqref{eq:HQET_Lagrangian} is the leading order term in the heavy quark effective theory (HQET), and the symmetry breaking effect is systematically incorporated by the $\Lambda_{\rm QCD}/m_{q}$ expansion.
An important consequence of heavy quark spin symmetry is the spin doublet structure of singly heavy hadrons. The heavy hadron can be decomposed into the heavy quark and the rest called brown muck. If the brown muck has spin $j>0$, there are two possibilities of the spin of the total system, $j+1/2$ and $j-1/2$. These two states form the heavy quark spin doublet with a degenerated mass in the heavy quark limit. The approximate degeneracy can be seen in the pair of the ground state $J^{P}=0^{-}$ meson and $1^{+}$ meson ($D$, $D^{*}$), ($B$, $B^{*}$) in the $B=0$ sector, and the pair of the $1/2^{+}$ baryon and $3/2^{+}$ baryon ($\Sigma_{c}$, $\Sigma_{c}^{*}$), ($\Sigma_{b}$, $\Sigma_{b}^{*}$) in the $B=1$ sector.
If there are $N_{h}$ heavy quarks, the leading term in Eq.~\eqref{eq:HQET_Lagrangian} is invariant under the rotation in the flavor space. This U($N_{h}$) symmetry is called heavy quark flavor symmetry. The finite mass of heavy quarks explicitly breaks this flavor symmetry. Note that the breaking of heavy quark flavor symmetry is proportional to the inverse of the mass difference. For instance, the symmetry breaking in the physical charm and bottom quarks is estimated to be $\Lambda_{\rm QCD}(1/m_{c}-1/m_{b})\sim 0.1$, even though the absolute value of the mass difference is not small.

In closing, we comment on the examples of hadrons to which the symmetry argument can be applied. The hadrons with only light ($u,d,s$) quarks are dictated by chiral symmetry and low-energy theorems. Heavy-light hadrons, which contain one heavy quark and light degrees of freedom, follow the constraints of both chiral and heavy quark symmetries~\cite{Burdman:1992gh,Wise:1992hn}. For the strange baryons that are discussed in this review, an important guiding principle should be chiral symmetry.

\subsubsection{\it Lattice QCD}
\label{sec:lattice}

In recent years, lattice QCD becomes more and more important to study low-energy phenomena of the strong interaction. Lattice QCD is a nonperturbative formulation of QCD on the discretized space-time lattice with keeping the gauge invariance~\cite{Wilson:1974sk}. By truncating the space-time volume at finite extent, Euclidean path integrals can be numerically evaluated with the Monte Carlo method~\cite{Creutz:1980zw}. At this moment, lattice QCD is the only direct approach to the low-energy strong interaction physics from first principles. Lattice QCD has been used to study various phenomena of nonperturbative QCD~\cite{Rothe:1992nt,DeGrand:2006zz,Gattringer:2010zz,Aoki:2019cca}.

Hadron spectroscopy is one of the major subjects in lattice QCD. Thanks to the developments of the computational techniques, masses of the ground state hadrons can now be obtained by the full QCD calculation at physical point with high precision~\cite{Durr:2008zz,Aoki:2009ix,Aoki:2012st,Borsanyi:2014jba}. In essence, the masses of the ground states are obtained by evaluating the two-point correlation function
\begin{align}
   \Gamma(\tau)
   & =\langle
   O(\tau)O^{\dag}(0)
   \rangle ,
\end{align}
where $\tau$ is the Euclidean time and $O$ is an interpolating field for a hadron. Hence, the hadron is created at Euclidean time $0$ and annihilated at $\tau$. By inserting the complete set of the hadronic states, the correlation function can be decomposed as 
\begin{align}
   \Gamma(\tau)
   & = \sum_{n}C_{n}e^{-E_{n}\tau} ,
\end{align}
where $n$ is an index to label the intermediate states and $C_{n}$ and $E_{n}$ are the weight factor and the energy of the state $n$, respectively. As the Euclidean time $\tau$ is increased, the contribution from the higher energy states decays rapidly, and only the ground state component $C_{0}e^{-E_{0}\tau}$ remains at large $\tau$. To extract the ground state energy, it is common practice to plot the effective mass $M_{\rm eff}(\tau) =\ln[\Gamma(\tau-1)/\Gamma(\tau)]$ as a function of $\tau$. If the correlation function is dominated by the $C_{0}e^{-E_{0}\tau}$ term, the effective mass shows a plateau, from which we can read off the ground state energy $E_{0}$.

The effective mass method is however not an appropriate treatment for the hadron excited states which are unstable against the strong decay, because the lowest energy QCD state of such systems should in principle be a scattering state. In general, the scattering states form a continuum spectrum if the space-time volume is infinite, while the lattice calculation in the finite volume gives a discretized spectrum. Thus, one needs to extract the information of the infinite volume scattering from the finite volume discrete energy spectrum. This direction of the study was initiated by L\"uscher~\cite{Luscher:1986pf,Luscher:1990ux}, and further developed in Refs.~\cite{Doring:2011vk,Gockeler:2012yj}. Numerical calculations have been performed to determine the properties of physical hadron resonances~\cite{Aoki:2012tk,Briceno:2017max,Aoki:2020bew}. The most studied hadron resonance is the $\rho$ meson in the $p$-wave $I=1$ $\pi\pi$ scattering~\cite{Aoki:2007rd,Feng:2010es,Lang:2011mn,Aoki:2011yj,Pelissier:2012pi,Dudek:2012xn,Wilson:2015dqa,Bali:2015gji,Bulava:2016mks,Guo:2016zos,Alexandrou:2017mpi,Akahoshi:2020ojo}. Recent activities in the baryon number $B=0$ sector reach the calculation of the scalar mesons such as $\sigma$~\cite{Briceno:2017qmb} and $\kappa$~\cite{Dudek:2014qha,Wilson:2019wfr}. At the same time, the studies of the $B=2$ sector indicate several interesting (quasi-)bound states~\cite{Gongyo:2017fjb,Iritani:2018sra} (see also the critical discussion in Ref.~\cite{Haidenbauer:2019utu}). Unfortunately, baryon resonances in the $B=1$ sector have not yet been well explored by the scattering calculation on the lattice. This is partly a reason why we focus on the strange baryons in this review. We expect further developments in this direction in near future, which will shed new light on the study of the strange baryon resonances.

\subsection{\it Resonances in hadron scattering}
\label{sec:resonances}

As mentioned in the introduction, for hadron spectroscopy, it is inevitable to deal with resonances in hadron scatterings. In this section, after introducing the basics of the nonrelativistic scattering theory, we discuss how the signature of resonances appears in the scattering observables. Detailed account of these subjects can be found in Refs.~\cite{Taylor,Newton:1982qc} for scattering theory and in Refs.~\cite{Bohm,Kukulin,Moiseyev} for resonance physics.

\subsubsection{\it Scattering theory}

Here we introduce basic quantities of the scattering theory using the simplest system of scattering. Let us consider the nonrelativistic quantum scattering of distinguishable particles 1 and 2 with mass $m_{1}$ and $m_{2}$ in the three-dimensional space. The Hamiltonian of the system is given by 
\begin{align}
   H=H_{0}+V ,
\end{align}
where $H_{0}$ represents the kinetic energy operator and $V$ is the potential. We do not consider internal degrees of freedom such as spin, flavor, etc. We focus on the elastic single-channel scattering, and there are no coupled channels in the energy region under consideration. The potential $V$ is assumed to be local and spherical, and depends only on the relative distance of two particles. This means that the system has the rotational symmetry, the Hamiltonian commutes with the angular momentum operators, and the magnitude of the angular momentum $\ell$ and the magnetic quantum number $m$ are conserved quantum numbers. We consider short range potentials, whose strength vanishes at large distance sufficiently rapidly.

\begin{figure}[tbp]
\begin{center}
\figureBB{\includegraphics[width=0.5\linewidth,bb=54 201 520 426]{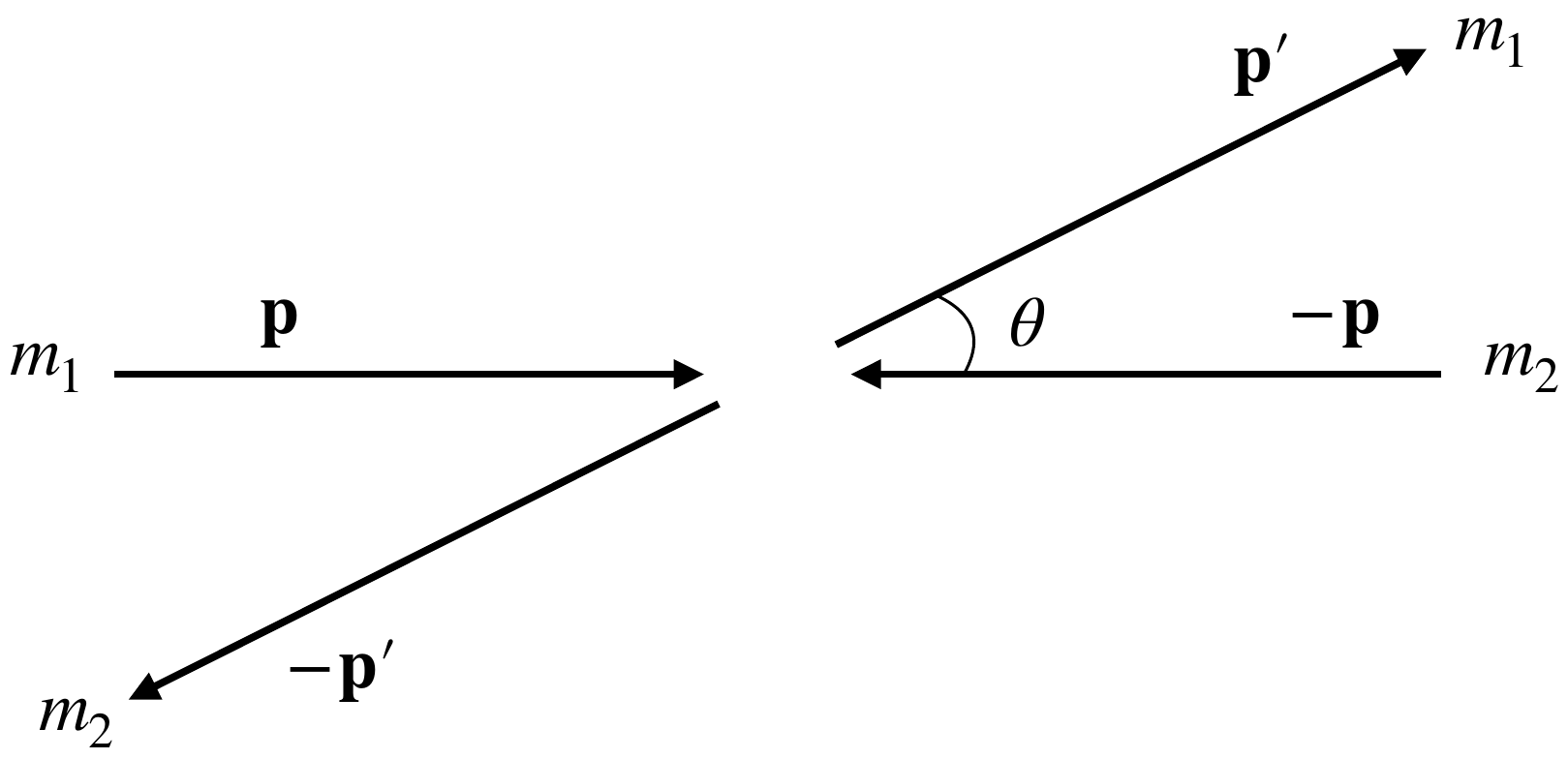}}
{\includegraphics[width=0.5\linewidth]{figs/kinematics.pdf}}
\end{center}
\caption{Schematic illustration of the kinematics of the scattering in the center-of-mass system.}
\label{fig:kinematics}
\end{figure}

The kinematics of the scattering is schematically shown in Fig.~\ref{fig:kinematics}. The initial state can be specified by the relative momentum $\bm{p}$, which means that the momentum of the particle 1 (2) is $\bm{p}$ ($-\bm{p}$) in the center-of-mass system. In the same way, the final state is specified by $\bm{p}^{\prime}$. In the elastic scattering, the magnitude of the momentum is unchanged, and we define $p=|\bm{p}|=|\bm{p}^{\prime}|$. The scattering angle $\theta$ is defined by the initial and final momenta as $\cos\theta=\bm{p}\cdot\bm{p}^{\prime}/p^{2}$. The scattering energy $E$ corresponding to the momentum $p$ is given by
\begin{align}
   E=\frac{p^{2}}{2\mu} ,
\end{align}
where the reduced mass is defined as $\mu=m_{1}m_{2}/(m_{1}+m_{2})$. The scattering wave function $\Psi$ is obtained by solving the time-independent Schr\"odinger equation $H\Psi=E\Psi$. The scattering process can be characterized by two parameters, the scattering energy $E$ (or the magnitude of the momentum $p$) and the scattering angle $\theta$. While physical scattering occurs only for $E> 0$ ($p> 0$), it is useful to perform an analytic continuation of $E$ ($p$) to the complex plane, as we discuss in the next section.\footnote{For physical scattering, we can use either $E$ or $p$, but for the analytic continuation to complex plane, the $S$ matrix and the scattering amplitude given below should be considered as meromorphic functions of $p$.} 

Next, we introduce the state vectors. In the momentum representation, the initial state is expressed by $\ket{\bm{p}}$, and the final state by $\bra{\bm{p}^{\prime}}$. These are the eigenstates of the noninteracting Hamiltonian $H_{0}\ket{\bm{p}}=p^{2}/(2\mu)\ket{\bm{p}}$. The normalization of the state vectors is given by
\begin{align}
   \bra{\bm{p}^{\prime}}\kket{\bm{p}}
   & =\delta^{3}(\bm{p}^{\prime}-\bm{p}) .
\end{align}
The initial and final states can also be expressed in the angular momentum representation, $\ket{E,\ell,m}$. The normalization of the state vectors in this representation reads
\begin{align}
   \bra{E^{\prime},\ell^{\prime},m^{\prime}}\kket{E,\ell,m}
   & =\delta(E^{\prime}-E)\delta_{\ell^{\prime}\ell}\delta_{m^{\prime}m} .
\end{align}
By writing the coordinate space wave functions explicitly and using the partial wave decomposition of the plane wave, one can show the relation between two representations
\begin{align}
   \bra{\bm{p}^{\prime}}\kket{E,\ell,m}
   & =\frac{1}{\sqrt{\mu p}}\delta(E^{\prime}-E)Y_{\ell}^{m}(\hat{\bm{p}}),\quad
   \hat{\bm{p}}=\frac{\bm{p}}{p} ,
\end{align}
with $Y_{\ell}^{m}(\hat{\bm{p}})$ being the spherical harmonics.

The transition from the initial state to the final state is represented by the scattering operator ${\sf S}$:
\begin{align}
   {\sf S}
   & =\Omega_{-}^{\dag}\Omega_{+}
   =\lim_{t\to+\infty}
   [e^{i\hat{H}_{0}t}e^{-i\hat{H}t}]
   \lim_{t\to -\infty}
   [e^{i\hat{H}t}e^{-i\hat{H}_{0}t}] ,
   \label{eq:Soperator}
\end{align}
where $\Omega_{\pm}$ are the M\o ller operators.
The $S$-matrix element $s_{\ell}(E)\in\mathbb{C}$ (also called ``$S$ matrix'') is defined through the matrix element of the ${\sf S}$ operator by the angular momentum representation as
\begin{align}
   \bra{E^{\prime},\ell^{\prime},m^{\prime}}
   {\sf S}
   \ket{E,\ell,m}
   & =\delta(E^{\prime}-E)\delta_{\ell^{\prime}\ell}\delta_{m^{\prime}m}
   s_{\ell}(E) .
\end{align}
Because of the rotational symmetry, the $S$ matrix is a function of the energy $E$ for each partial wave $\ell$. As long as we consider the hermitian Hamiltonian, the time evolution of the state is unitary, as seen in Eq.~\eqref{eq:Soperator}. In other words, the probability (square of the norm of the state) is conserved under the time evolution. The $\sf{S}$ operator therefore satisfies the unitarity condition:
\begin{align}
   \sf{S}^{\dag}\sf{S}
   & =1,
\end{align}
which gives a relation of the $S$ matrix as
\begin{align}
   s_{\ell}^{*}(E)s_{\ell}(E)=|s_{\ell}(E)|^{2}=1 .
\end{align}
This leads to the expression of the $S$ matrix by the phase shift $\delta_{\ell}(E)\in \mathbb{R}$:
\begin{align}
   s_{\ell}(E)
   & =\exp\{2i\delta_{\ell}(E)\} .
\end{align}
Using the intertwining relation for the M\o ller operators $H\Omega_{\pm}=\Omega_{\pm}H^{0}$, one can show the commutation relation $[H_{0},{\sf S}]=0$. This implies that the matrix element of the ${\sf S}$ operator by the state vectors with the momentum representation satisfies the energy conservation. It follows from the definition~\eqref{eq:Soperator} that the $\sf{S}$ operator reduces to the identity in the absence of the interaction, $V= 0$. Based on these facts, the on-shell T matrix $t(\bm{p}^{\prime}\gets\bm{p})\in\mathbb{C}$ is defined to express the net effect of the interaction as 
\begin{align}
   \bra{\bm{p}^{\prime}}({\sf S}-1)\ket{\bm{p}}
   & =-2\pi i\delta(E^{\prime}-E)t(\bm{p}^{\prime}\gets \bm{p}) ,
\end{align}
where the normalization factor of $-2\pi i$ is chosen such that the Born approximation of the T matrix is given by $\bra{\bm{p}^{\prime}}V\ket{\bm{p}}$. The scattering amplitude $f(E,\theta)$ is defined from the on-shell T matrix as  
\begin{align}
   f(E,\theta) 
   & =-(2\pi)^{2}\mu\; t(\bm{p}^{\prime}\gets \bm{p})  .
   \label{eq:scatteringamplitude}
\end{align}
This definition of the scattering amplitude $f(E,\theta) $ is equivalent to the one used in the boundary condition of the Schr\"odinger equation to obtain the scattering wave function $\psi_{\bm{p}}^{+}(\bm{r})$,
\begin{align}
   \psi_{\bm{p}}^{+}(\bm{r})
   & \propto e^{i\bm{p}\cdot\bm{r}}
   +f(E,\theta)\frac{e^{ipr}}{r}
   \quad 
   (r\to \infty) ,
   \label{eq:scatteringwave}
\end{align}
where $f(E,\theta)$ appears as the amplitude of the outgoing wave. Therefore, the differential cross section can be calculated as
\begin{align}
   \frac{d\sigma}{d\Omega}
   & =|f(E,\theta)|^{2} .
\end{align}
Performing the partial wave decomposition of the scattering amplitude
\begin{align}
   f(E,\theta)
   & =
   \sum_{\ell}(2\ell+1)f_{\ell}(E)P_{\ell}(\cos\theta),
\end{align}
with the Legendre polynomial $P_{\ell}(\cos\theta)$, we obtain the relation between the scattering amplitude and the $S$ matrix in $\ell$-th partial wave as
\begin{align}
   f_{\ell}(E)
   & =\frac{s_{\ell}(E)-1}{2ip} .
   \label{eq:fsrelation}
\end{align}
Thus, the scattering observables can be calculated from the scattering amplitude $f_{\ell}(E)$ or the $S$-matrix element $s_{\ell}(E)$. 

For a given potential $V$, the on-shell T matrix $t(\bm{p}^{\prime}\gets \bm{p})$ can be calculated by solving the Lippmann-Schwinger equation. Equivalently, the scattering amplitude can be obtained by solving the Schr\"odinger equation with an appropriate boundary condition. Let us describe this latter approach, because it clarifies the relation of the pole of the scattering amplitude and the generalized eigenstate of the Hamiltonian. To obtain the wave function of bound states (or in general, discrete eigenstates), one imposes two boundary conditions at $r\to 0$ and $r\to \infty$ on the general solution of the radial Schr\"odinger equation. The scattering wave function is determined by only the boundary condition at $r=0$, which provides continuous eigenstates. Because of the absence of the boundary condition at $r\to \infty$, the scattering wave functions are not square integrable, and the usual normalization condition cannot be applied. In other words, the normalization of the scattering wave function is in general not fixed. However, to extract the scattering amplitude, it is useful to define the scattering wave function with a fixed normalization, which is called the ``regular solution''. For the eigenmomentum $p=\sqrt{2\mu E}$ and the angular momentum $\ell$, the regular solution $\phi_{\ell,p}(r)$ is given by
\begin{align}
    \phi_{\ell,p}(r)
    &\to \hat{j}_{\ell}(pr)
    \quad (r\to 0)
    \label{eq:scat} ,
\end{align}
where $\hat{j}_{\ell}(z)$ is the Riccati-Bessel function.\footnote{The Riccati-Bessel (Riccati-Neumann) function $\hat{j}_{\ell}(z)$ [$\hat{n}_{\ell}(z)$] is related to the spherical Bessel (Neumann) function $j_{\ell}(z)$ [$n_{\ell}(z)$] as $\hat{j}_{\ell}(z)=zj_{\ell}(z)$ [$\hat{n}_{\ell}(z)=zn_{\ell}(z)$]. Note that the radial wave function $\chi_{\ell}(r)$ is related to the full wave function $\psi_{\ell,m}(\bm{r})$ as $\psi_{\ell,m}(\bm{r})=\frac{\chi_{\ell}(r)}{r}Y_{\ell}^{m}(\hat{\bm{r}})$ and if $V(r)=0$ the $r$ dependence of $\psi_{\ell,m}(\bm{r})$ is given by a linear combination of the spherical Bessel and Neumann functions. This means that the radial wave function $\chi_{\ell}(r)$ can be expressed by a linear combination of the Riccati-Bessel and Riccati-Neumann functions in the absence of the interaction.} Equation~\eqref{eq:scat} imposes two conditions: 1) $\phi_{\ell,p}(r)$ should vanish at $r\to 0$ and 2) the magnitude of $\phi_{\ell,p}(r)$ is normalized as $\phi_{\ell,p}(r)/\hat{j}_{\ell}(pr)\to 1$ at $r\to 0$. Once the boundary condition~\eqref{eq:scat} is imposed, one can solve (either analytically or numerically) the radial Schr\"odinger equation to obtain the regular solution $\phi_{\ell,p}(r)$ as a function of $r$. As seen in Eq.~\eqref{eq:scatteringwave}, the scattering information is included in the asymptotic behavior of the wave function at $r\to \infty$. Because the potential is assumed to vanish at $r\to \infty$, the asymptotic behavior of the regular solution can be given by the linear combination of the Riccati-Bessel and Riccati-Neumann functions, or equivalently, the Riccati-Hankel functions $\hat{h}^{\pm}_{\ell}(z)=\hat{n}_{\ell}(z)\pm i\hat{j}_{\ell}(z)$. The Jost function $\Jost_{\ell}(p)$ is defined as the coefficient of the Riccati-Hankel function as
\begin{align}
    \phi_{\ell,p}(r)
    &
    \to
    \frac{i}{2}[\ \Jost_{\ell}(p)\hat{h}_{\ell}^{-}(pr)
    -\Jost_{\ell}(-p)\hat{h}_{\ell}^{+}(pr)] .
    \quad (r\to \infty)
    \label{eq:regularlimit}
\end{align}
From the asymptotic behavior of the Riccati-Hankel functions
\begin{align}
    \hat{h}_{\ell}^{\pm}(z)
    &\to
    \exp[\pm i(z-\ell\pi/2)] \quad (z\to \infty),
\end{align}
we find that the term $\hat{h}_{\ell}^{-}(pr)\sim e^{-ipr}$ expresses the incoming wave, and $\hat{h}_{\ell}^{+}(pr)\sim e^{+ipr}$ the outgoing wave. Namely, the Jost function $\Jost_{\ell}(p)$ is the amplitude of the incoming wave. Now we are in a position to calculate the scattering observables. The $S$ matrix $s_{\ell}(p)$ is defined as the amplitude of the outgoing wave normalized by that of the incoming wave, so it can be expressed by the Jost function as
\begin{align}
    s_{\ell}(p)
    &= 
    \frac{\Jost_{\ell}(-p)}{\Jost_{\ell}(p)} .
    \label{eq:smatrixJost}
\end{align}
From Eq.~\eqref{eq:fsrelation}, we obtain the expression of the scattering amplitude by the Jost function as
\begin{align}
    f_{\ell}(p)
    = 
    \frac{\Jost_{\ell}(-p)-\Jost_{\ell}(p)}{2ip\Jost_{\ell}(p)}  .
    \label{eq:fJost}
\end{align}
In this way, the scattering observables can be calculated by the asymptotic behavior of the scattering wave function. Before closing this section, we note that if the Jost function vanishes at some momentum $p$,
\begin{align}
    \Jost_{\ell}(p)
    = 
    0 ,
    \label{eq:Jostzero}
\end{align}
then the scattering amplitude (and the $s$-matrix) diverges at $p$. The vanishing of $\Jost_{\ell}(p)$ in Eq.~\eqref{eq:regularlimit} means that the scattering wave function is purely given by the outgoing wave. This point will be important to relate the pole of the scattering amplitude and the eigenstate of the Hamiltonian in the next section. 

\subsubsection{\it Signals of resonances}

Traditionally, a resonance is identified by a peak in the cross section as a function of the scattering energy. The mass and width of the resonance correspond to the energy of the maximum of the peak and the half-width of the peak, respectively. This definition, however, does not uniquely characterize the resonance, because the peak of the spectrum is in general reaction dependent due to the nonresonant contributions. A theoretically well-defined characterization of a resonance is the pole of the scattering amplitude, which is in principle uniquely determined. In fact, the baryon part of PDG~\cite{Zyla:2020zbs} now tabulates the pole position of resonances, prior to the Breit-Wigner mass and width. In this section, we demonstrate that the pole of the scattering amplitude represents the generalized eigenstate of the Hamiltonian. In addition, we show that stable bound states and unstable resonances can be treated in a unified way, by utilizing the outgoing boundary condition.

Let us consider the same scattering problem as in the previous section, namely, nonrelativistic single-channel two-body scattering with reduced mass $\mu$ under the spherical and short-range potential. For simplicity, we deal with the $s$-wave scattering with the angular momentum $\ell=0$. In the energy region of the physical scattering $E>0$, the momentum $p=\sqrt{2\mu E}$ is real and positive. By solving the radial Schr\"odinger equation for $p>0$, we obtain the scattering solution of the radial wave function $\chi_{0,p}(r)$ which satisfies $\chi_{0,p}(r)\to 0$ at $r\to 0$. Because of the absence of the boundary condition at $r\to \infty$, we obtain the eigenstates for any $p>0$ and the scattering states form the continuous spectrum. At large distance where the potential vanishes, the wave function is given by the superposition of the plane waves:
\begin{align}
   \chi_{0,p}(r)
   &\to A^{-}(p)e^{-ipr}
   +A^{+}(p)e^{+ipr}\quad (r\to \infty),
\end{align}
where the coefficients $A^{-}(p)$ and $A^{+}(p)$ represent the amplitude of the incoming and outgoing waves, respectively. The explicit forms of $A^{\pm}(p)$ depend on the given potential. For instance, adopting the attractive square-well potential with depth $V_{0}$ and width $b$:
\begin{align}
   V(r)
   =\begin{cases}
   -V_{0} & 0\leq r \leq b \\
   0 & b<r
   \end{cases} ,
\end{align}
we obtain the coefficients 
\begin{align}
   A^{\pm}(p)= \frac{C}{2}
   \left[\sin(b\sqrt{p^{2}+2\mu V_{0}})\mp i\frac{\sqrt{p^{2}+2\mu V_{0}}}{p}\cos(b\sqrt{p^{2}+2\mu V_{0}})\right]e^{\mp ipb} .
   \label{eq:squarewell}
\end{align}
Because the scattering solution is not normalizable, the coefficient $C$ is arbitrary.

The eigenenergy of the bound state is negative, $E<0$. In this case, the eigenmomentum $p=\sqrt{2\mu E}$ is purely imaginary. Because of the branch cut, for a negative $E$, the momentum variable $\sqrt{-2\mu|E|}$ is indefinite, and one must specify the analytic continuation path from the positive $E$. For the bound state, we choose the path in the upper half energy plane, or equivalently, we define $p=\sqrt{-2\mu|E|+i0^{+}}=i\sqrt{2\mu|E|}$.
Defining $p= i\kappa$ with $\kappa>0$, the general solution of the radial wave function is
\begin{align}
   \chi_{0,i\kappa}(r)
   &\to A^{-}(i\kappa)e^{+\kappa r}
   +A^{+}(i\kappa)e^{-\kappa r} \quad (r\to \infty).
\end{align}
To obtain the bound state solution, we eliminate the increasing component $e^{+\kappa r}$, so that the wave function is square integrable. This is equivalent to demanding
\begin{align}
   A^{-}(i\kappa)=0
   \label{eq:boundcond} .
\end{align}
In fact, using the explicit form of $A^{-}(p)$ in Eq.~\eqref{eq:squarewell}, we obtain the bound state condition for the square-well potential
\begin{align}
   \tan(b\sqrt{-\kappa^{2}+2\mu V_{0}})
   =-\frac{\sqrt{-\kappa^{2}+2\mu V_{0}}}{\kappa} .
\end{align}
The bound state is a discrete eigenstate, because the solution is obtained only when $\kappa$ satisfies the condition~\eqref{eq:boundcond}. In this way, we have seen that the bound state condition is obtained from Eq.~\eqref{eq:boundcond}, which can be regarded as an analytic continuation of $A^{-}(p)=0$ with the momentum variable $p$ being pure imaginary $i\kappa$. 

The resonance solution can be obtained in the same way. In this case, we perform the analytic continuation of $p$ to general complex plane, and impose the boundary condition
\begin{align}
   A^{-}(p_{R})=0,\quad p_{R}\in \mathbb{C} .
   \label{eq:resonancecond}
\end{align}
If we find a solution $p_{R}$ away from the imaginary axis, the wave function $\chi_{0,p_{R}}(r)$ represents the resonance state. In fact, for the square well potential case of Eq.~\eqref{eq:squarewell}, the condition~\eqref{eq:resonancecond} provides infinitely many resonance solutions in the complex energy plane~\cite{AJP50.839}. When $p_{R}$ is complex, the corresponding eigenenergy is also complex:
\begin{align}
    E_{R}
    =\frac{p_{R}^{2}}{2\mu}
    &\equiv M_{R}-\frac{i}{2}\Gamma_{R} ,
    \label{eq:resonanceenergy}
\end{align}
where $M_{R}>0$ and $\Gamma_{R}>0$ are interpreted as the ``mass'' and ``width'' of the state (see the effect of the Breit-Wigner term discussed below).\footnote{From the analytic properties of the Jost function, one can show that the existence of a pole at $p=p_{R}$ indicates another pole at $p=-p_{R}^{*}$. This means that there should be a pair of poles at $E=E_{R}$ and $E=E_{R}^{*}$ in the complex energy plane. In other words, there is a pole with $\im [E_{R}]>0$ as well as the one with $\im [E_{R}]<0$ shown in Eq.~\eqref{eq:resonanceenergy}. Recalling the time dependence of the wave function $\Psi_{E}(t)\propto e^{-iEt}$, one finds that the pole with $\im [E_{R}]<0$ represents the state with decreasing probability $|\Psi_{E_{R}}(t)|^{2}\propto e^{-\Gamma_{R} t}$, while the other one denotes the state with increasing probability. These solutions are interpreted as the decaying resonance state and its time reversal, respectively.} Because $E_{R}$ is the eigenenergy of the Hamiltonian, one might wonder why the complex number is allowed as an eigenvalue. To show the reality of the eigenvalue of an hermitian (strictly speaking, self-adjoint) operator, one must consider the Hilbert space in a mathematically strict sense, i.e., the complete inner product space. Roughly speaking, the reality of the eigenenergy is guaranteed for the square integrable wave functions $\int dr|\chi_{0,p_{R}}(r)|<\infty$. For $p_{R}\in \mathbb{C}$ satisfying Eq.~\eqref{eq:resonancecond}, the corresponding wave function is 
\begin{align}
   \chi_{0,p_{R}}(r)
   &\to A^{+}(p_{R})e^{i\text{Re}[p_{R}]r}
   e^{-\text{Im}[p_{R}] r}
\end{align}
This function is square integrable for $\im [p_{R}]>0$, and therefore no complex energy state can appear in the upper half plane of $p$. In fact, only bound state solutions are allowed for $\im [p_{R}]>0$, which are $\text{Re}[p_{R}]=0$ and $\text{Im}[p_{R}] =\kappa$. On the other hand, in the lower half plane ($\im [p_{R}]<0$), the wave function is not square integrable, and therefore complex eigenenergy is allowed. Therefore, the resonance solutions found in this region can be understood as the eigenstates of the Hamiltonian, just as in the case of the bound state solutions. At the same time, we have to keep in mind that the resonance wave functions (whose amplitude increases at large distance) does not fit in the ordinary Hilbert space, so the resonances may be called ``generalized'' eigenstates.

\begin{figure}[tbp]
\begin{center}
\figureBB{\includegraphics[width=0.8\linewidth,bb=93 474 1727 985]{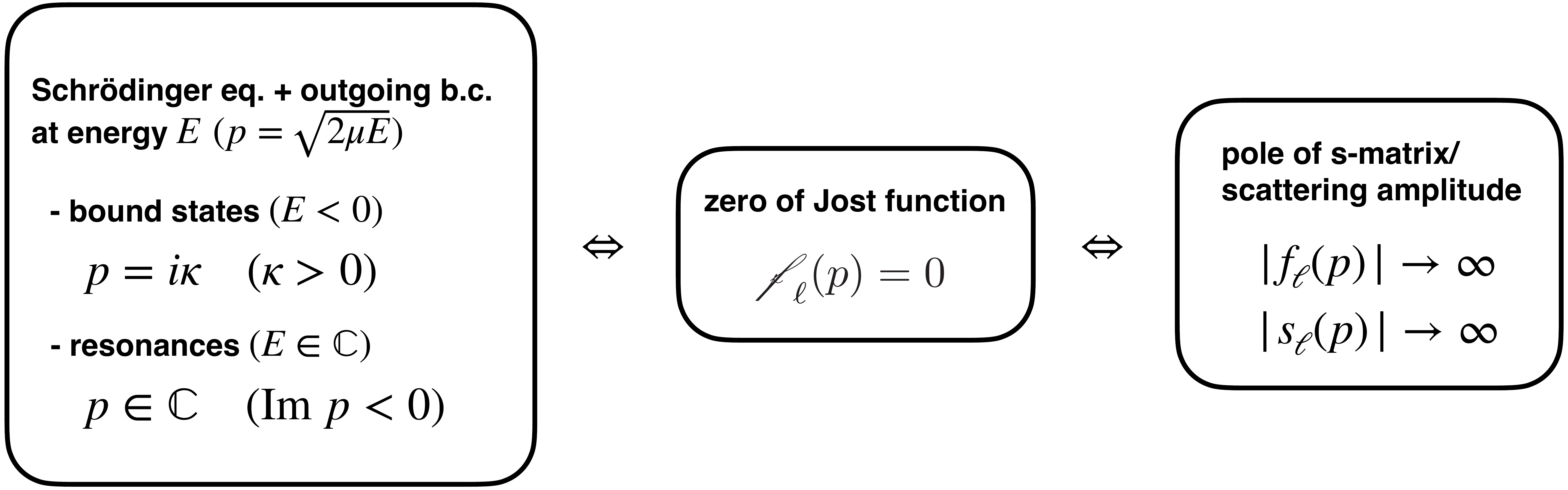}}
{\includegraphics[width=0.8\linewidth]{figs/resonance.pdf}}
\end{center}
\caption{Definition of resonances in the Schr\"odinger equation and in scattering theory. Resonances are identified as the generalized eigenstates with a complex eigenenergy by solving the Schr\"odinger equation with the outgoing boundary condition. This procedure is common to usual bound states. It follows from the asymptotic behavior of the scattering wave function~\eqref{eq:regularlimit} that the outgoing boundary condition is equivalent to the zero of the Jost function in scattering theory. From Eqs.~\eqref{eq:smatrixJost} and \eqref{eq:fJost}, this condition gives the pole of the $s$ matrix and the scattering amplitude.}
\label{fig:resonance}
\end{figure}

Because $A^{-}(p)$ is the amplitude of the incoming wave, Eq.~\eqref{eq:resonancecond} is referred to as the outgoing boundary condition. This reminds us of the zero of the Jost function~\eqref{eq:Jostzero} discussed in the previous section. In fact, by comparing the normalization of the wave functions, we find the relation of $A^{-}(p)$ and the Jost function $\Jost_{0}(p)$ as
\begin{align}
    \Jost_{0}(p)
    &=\bigl. A^{-}(p)\bigr|_{C=\frac{2p}{i\sqrt{p^{2}+2\mu V_{0}}}}
    =
   \left[\cos(b\sqrt{p^{2}+2\mu V_{0}})
   -i\frac{p}{\sqrt{p^{2}+2\mu V_{0}}}\sin(b\sqrt{p^{2}+2\mu V_{0}})\right]e^{ ipb}
\end{align}
Thus, Eq.~\eqref{eq:resonancecond} is equivalent to the vanishing of the Jost function~\eqref{eq:Jostzero}. As a consequence, the $S$ matrix and the scattering amplitude diverge at $p_{R}$. In other words, the resonance eigenstate is expressed by the pole of the $S$ matrix/scattering amplitude. These relations are schematically summarized in Fig.~\ref{fig:resonance}.

We have shown that the theoretically well-defined characterization of resonances is to determine the pole position of the scattering amplitude in the complex energy plane. On the other hand, physical scattering occurs only for real and positive energies, and therefore the pole at the complex energy is not directly accessible in experiments. Whereas the pole position is in principle uniquely determined, it is practically useful to show the characteristic behavior of observable quantities. Suppose that there is a resonance at $E=E_{R}=M_{R}-i\Gamma_{R}/2$ in the $\ell$-th partial wave. The scattering amplitude, having a pole at $E=E_{R}$, can be expressed by the Laurent series around $E_{R}$ as
\begin{align}
    f_{\ell}(E)
    &= 
    f_{\ell,{\rm BW}}(E)
    +f_{\ell,{\rm BG}}(E)  ,
    \label{eq:Laurentexpansion}
\end{align}
where $f_{\ell,{\rm BW}}(E)$ is the Breit-Wigner term containing the pole contribution
\begin{align}
    f_{\ell,{\rm BW}}(E)
    &= 
    \frac{Z_{R}}{E-E_{R}}
    = 
    \frac{Z_{R}(E-M_{R}-\frac{i}{2}\Gamma_{R})}{(E-M_{R})^{2}+\frac{1}{4}\Gamma_{R}^{2}}
    \label{eq:BreitWigner} ,
\end{align}
with $Z_{R}=-\Gamma_{R}/(2p_{R})$ is the complex residue of the pole, and $f_{\ell,{\rm BG}}(E)$ is called the nonresonant background contribution which is regular at $E=E_{R}$:
\begin{align}
    f_{\ell,{\rm BG}}(E) 
    &= \sum_{n=0}^{\infty}C_{n}(E-E_{R})^{n} .
\end{align}
From the right hand side of Eq.~\eqref{eq:BreitWigner}, we see that the pole term $f_{\ell,{\rm BW}}(E)$ varies rapidly with large amplitude near the resonance position $E\sim M_{R}$, in particular for the narrow width state. The background term is then regarded as a slowly varying function of $E$ with small magnitude, in comparison with the pole term. If one \textit{assume} that the background term is small and negligible, we can approximate the scattering amplitude by the Breit-Wigner term
\begin{align}
    f_{\ell}(E)
    &\approx f_{\ell,{\rm BW}}(E)
    \quad (f_{\ell,{\rm BG}}(E) \to 0) .
\end{align}
In this case, we find several traditional signatures of a resonance on the real energy axis:
\begin{itemize}
\item[(i)] the cross section peaks at $E=M_{R}$ with the half width $\Gamma_{R}$,
\item[(ii)] $\re[f_{\ell}(E)]=0$ and $\im[f_{\ell}(E)]$ becomes maximum at $E=M_{R}$, and
\item[(iii)] phase shift $\delta_{\ell}(E)$ increases rapidly and crosses $\pi/2$ at $E=M_{R}$.
\end{itemize}
Noting $Z=-\Gamma_{R}/(2p)<0$, (ii) directly follows from the right hand side of Eq.~\eqref{eq:BreitWigner}.\footnote{The momentum factor $p$ in the residue $Z$ stems from the one in the denominator of Eq.~\eqref{eq:fsrelation}. To derive the properties (i)-(iii), the background term is neglected in the $S$ matrix $s_{\ell}(E)$, and then translate it to the scattering amplitude $f_{\ell}(E)$ through Eq.~\eqref{eq:fsrelation}. In this case, the residue $Z$ is not a constant, because of $p$. For physical scattering, the momentum $p$ is real and positive, and therefore the residue $Z=-\Gamma_{R}/(2p)$ is real and negative. If one perform the Laurent expansion for the scattering amplitude $f_{\ell}(E)$ directly, the residue $Z_{R}=-\Gamma_{R}/(2p_{R})$ is a complex constant as in Eq.~\eqref{eq:BreitWigner}.} (i) is a consequence of the optical theorem $\sigma\propto \im f$ and the behavior of the imaginary part in (ii). From Eq.~\eqref{eq:fsrelation}, the condition $\re[f_{\ell}(M_{R})]=0$ requires that the $S$ matrix should be real, $s_{\ell}(M_{R})=\exp[2i\delta_{\ell}(M_{R})]\in \mathbb{R}$. To satisfy this except for the noninteraccting case, the phase shift should be $\pi/2$ (modulo $\pi$) at $E=M_{R}$. Because of the property (i), the real (imaginary) part of the pole position is regarded as mass (width$/2$). Here we emphasize that the features (i)-(iii) are realized only when the nonresonant background term is neglected.\footnote{In practice, the properties (i)-(iii) can be approximately realized when the magnitude of the background term is small. In addition, if the behavior of the background contribution is well understood, the resonance parameters can be extracted.} The 
contribution from the nonresonant background can modify these features. In fact, because the pole term and the background term are summed coherently in Eq.~\eqref{eq:Laurentexpansion}, taking the amplitude square, we obtain
\begin{align}
    |f_{\ell}(E)|^{2}
    &= 
    |f_{\ell,{\rm BW}}(E)|^{2}
    +|f_{\ell,{\rm BG}}(E)|^{2} 
    +2\re [f_{\ell,{\rm BW}}(E)f_{\ell,{\rm BG}}^{*}(E)] ,
\end{align}
where the last term represents the interference of the pole and the background. The experimentally observed spectrum can also be influenced by such interference term. In addition, if there exists a threshold opening near the resonance, then we must treat a much complicated coupled-channel scattering amplitude. In this case, a resonance pole below the threshold does not directly affect the scattering amplitude above the threshold, and vice versa. As a consequence, the validity of the Breit-Wigner term is limited at the threshold energy, and it cannot be extended over the threshold. The kinematical effects induced by a threshold, such as cusp structures and triangle singularities~\cite{Guo:2019twa}, can produce some peak like structure in the spectrum even in the absence of the resonance pole. In this way, one should be cautious about the use of the Breit-Wigner function to fit a peak in the spectrum, because it is valid {\it only} in the idealized situation; the width of the resonance is sufficiently narrow, the background contribution is properly understood, and no threshold exists in the energy region of the peak structure. It is therefore important to determine 
the pole position from the careful analysis of the 
experimental data, rather than the simple Breit-Wigner fit. Although the determination of the pole position is a challenging task experimentally, it is a necessary step
to pin down the basic properties of hadron resonances. 

\subsection{\it Internal structure of hadrons}\label{sec:internal}

Hadrons are made from quarks and gluons, but they are constructed in a highly complicated way, reflecting the nonperturbative dynamics of QCD. It is therefore natural to ask what kind of internal structure they have. Traditionally, the success of constituent quark models suggests that the mesons are composed of $\bar{q}q$ and the baryons are composed of $qqq$~\cite{Eichten:1974af,Isgur:1978xj,Godfrey:1985xj,Capstick:1986bm,Capstick:2000qj}. It turns out that there are some exceptions which do not fit well in the quark model description. Because these hadrons are expected to have an unconventional structure beyond $\bar{q}q$ and $qqq$, they are called exotic hadrons. Recently, investigations along this direction are further accelerated by the findings of the $XYZ$ states in heavy quark sector~\cite{Brambilla:2010cs,Hosaka:2016pey,Guo:2017jvc,Olsen:2017bmm}. The study of exotic hadrons thus becomes a major subject in hadron physics. On the other hand, there is no unique definition for the word ``exotic hadrons'', and the ambiguity of the definition sometimes causes confusions in the discussion. Let us therefore first consider some suitable classification scheme of exotic hadrons.

First of all, any hadronic states that are realized in nature should obey the rule of the strong interaction. In this sense, there is nothing ``exotic'' from the viewpoint of QCD. To define the exotic hadrons, one should find regularity of some property of hadrons, which is satisfied by most of the observed hadrons. One can then classify the exceptions of this regularity as exotics. At this point, we emphasize that the classification should be done in a theoretically well defined manner. A proper classification must be given without referring to any specific models, such as constituent quark models. Rather, we should rely on the {\it conserved quantum numbers} which are well defined in QCD. For this purpose, we can utilize the spin-parity $J^{PC}$, the flavor quantum numbers (isospin, strangeness, etc.), and the baryon number $B$, which are based on symmetries of QCD.\footnote{While the isospin SU(2) symmetry is an approximate one in QCD, the total $I_{3}$ quantum number is conserved due to the independent conservations of $u$ quark number and $d$ quark number. One can rephrase it by the conservation of the electric charge.} Using these quantum numbers, exotic 
hadron candidates with $B=0$ and $B=1$ can then be classified into three categories:
\begin{itemize}

\item[(i)] quantum number exotics : hadrons whose quantum numbers cannot be reached by $\bar{q}q$/$qqq$

\item[(ii)] quarkonium associated exotics : hadrons whose quantum numbers can (in principle) be reached by $\bar{q}q$/$qqq$, but it is plausible that they contain $\bar{c}c$ or $\bar{b}b$

\item[(iii)] other exotics

\end{itemize}
In the following, we discuss these classes in detail, giving possible candidates in each class.

(i) : the clearest examples of exotic structure are the quantum number exotics. This can be further classified into the flavor exotics and the $J^{PC}$ exotics. The flavor exotics are the hadrons whose flavor quantum number requires more than three valence quarks. They are also called ``manifestly/genuine exotic hadrons'', for their exotic nature is manifested in the valence quark configuration. Theoretically, the flavor exotics can be specified by the well-defined quantum number exoticness~\cite{Hyodo:2006yk,Hyodo:2006kg}, which counts the number of quark-antiquark pairs in addition to $\bar{q}q$/$qqq$ in the minimal valence configuration. Experimental identification of flavor exotics is also straightforward due to the flavor conservation in the strong interaction. Possible candidates of the flavor exotics, whose experimental evidence has been once given, are $\Theta^{+}\sim uudd\bar{s}$~\cite{Nakano:2003qx}, $\Xi^{--}(1860)\sim ddss\bar{u}$~\cite{Alt:2003vb}, $\Theta_{c}\sim uudd\bar{c}$~\cite{Aktas:2004qf}, and $X(5568)\sim bu\bar{d}\bar{s}$~\cite{D0:2016mwd}. Unfortunately, these states were not confirmed by the follow-up experiments and their existence is not established so far. The $J^{PC}$ exotics are the mesons whose $J^{PC}$ quantum number cannot be constructed from the $\bar{q}q$ configuration. It follows from the symmetry under the exchange of quark and antiquark that $J^{PC}=0^{+-},1^{-+},2^{+-},\dotsb$ are not obtained by the $\bar{q}q$ configuration. In PDG, $\pi_{1}(1400)$ and $\pi_{1}(1600)$ have $J^{PC}=1^{-+}$~\cite{Zyla:2020zbs} and hence classified as the $J^{PC}$ exotics. Minimal valence configurations of these states should be $\bar{q}q\bar{q}q$ or $\bar{q}qg$. We emphasize that the absence of the flavor quantum number exotics is a highly nontrivial fact. There is no rule to forbid such configuration in QCD, just as in the case of color confinement. It is therefore important to look for possible quantum number exotics experimentally. At the same time, theoretical effort is required to clarify the mechanism of non-appearance of quantum number exotics in the hadron spectrum. 

(ii) : several quarkonium associated exotics have been observed recently. Representative examples are the tetraquarks $Z_{b}^{\pm}\sim \bar{b}b\bar{u}d/\bar{b}b\bar{d}u$~\cite{Belle:2011aa} and the pentaquarks $P_{c}\sim \bar{c}cuud$~\cite{Aaij:2015tga,Aaij:2019vzc}. Compared with the states in (i), the $\bar{c}c$ or $\bar{b}b$ pair can in principle be annihilated, and the $Z_{b}^{\pm}$ ($P_{c}$) state have the same quantum number with $\bar{u}d/\bar{d}u$ ($uud$). Of course the existence of the $\bar{c}c$ or $\bar{b}b$ pair in these states is almost certain from their mass and decay products, but one cannot distinguish $P_{c}$ from the highly excited proton by the conserved quantum number in QCD. Once we accept the existence of the $\bar{c}c$ ($\bar{b}b$) pair, these states cannot be the ordinary the $\bar{q}{q}$ meson or $qqq$ baryon. Although the number of observed quarkonium associated exotics is increasing, they occupy only a small fraction in the hadron spectrum. It is not clear why they are rare, but at the same time, the existence of the quarkonium associated exotics indicates that there may be a difference from the quantum number exotics. Thus, the quarkonium associated exotics will bring us an important clue to understand the construction mechanism of hadrons from quarks and gluons.

(iii) : there are hadrons whose quantum numbers are describable by $\bar{q}q$ or $qqq$, but considered to have an exotic structure. Most of the so-called exotic hadron candidates fall into this category. Famous examples are the lowest lying scalar mesons and the $\Lambda(1405)$ resonance in the light quark sector, and $X(3872)$~\cite{Choi:2003ue} and $D_{s}(2317)$~\cite{Aubert:2003fg} in the heavy sector. Motivated by the failure of the prediction by quark models, many configurations, such as multiquarks and hadronic molecules, have been proposed to explain their properties. It should however be noted that there are no conserved quantum numbers that distinguish these hadrons from the ordinary $\bar{q}q$ or $qqq$ states. This means that a hadron in this class is a mixture of the exotic structure and the ordinary configuration and one needs to introduce a measure to characterize the internal structure beyond the conserved quantum numbers. One promising quantity is the compositeness of hadrons~\cite{Baru:2003qq,Hanhart:2011jz,Hyodo:2011qc,Aceti:2012dd,Xiao:2012vv,Hyodo:2013iga,Hyodo:2013nka,Sekihara:2014kya,Guo:2015daa,Kamiya:2015aea,Kamiya:2016oao}, which is based on the field renormalization constant to distinguish composite and elementary particles~\cite{Weinberg:1965zz}. From the experimental viewpoint, the first step to study these exotics is the accurate determination of the basic properties, the resonance pole positions. Of course, the pole position does not give the information on the internal structure by itself, but a meaningful conclusion should only be achieved with the reliable basic properties. The second step will be to measure an observable that reflects the internal structure. In this regard, the theoretical task is to define a sensible measure of the internal structure, and to relate it with the experimentally observable quantities. Thus, collaborative efforts of theory and experiment are desired to find out a way to understand these exotic hadrons.

\section{$S=-1$ baryons}\label{sec:S-1}

\begin{figure}[tbp]
\begin{center}
  \figureBB{\includegraphics[width=14cm, bb=0 0 502 500]{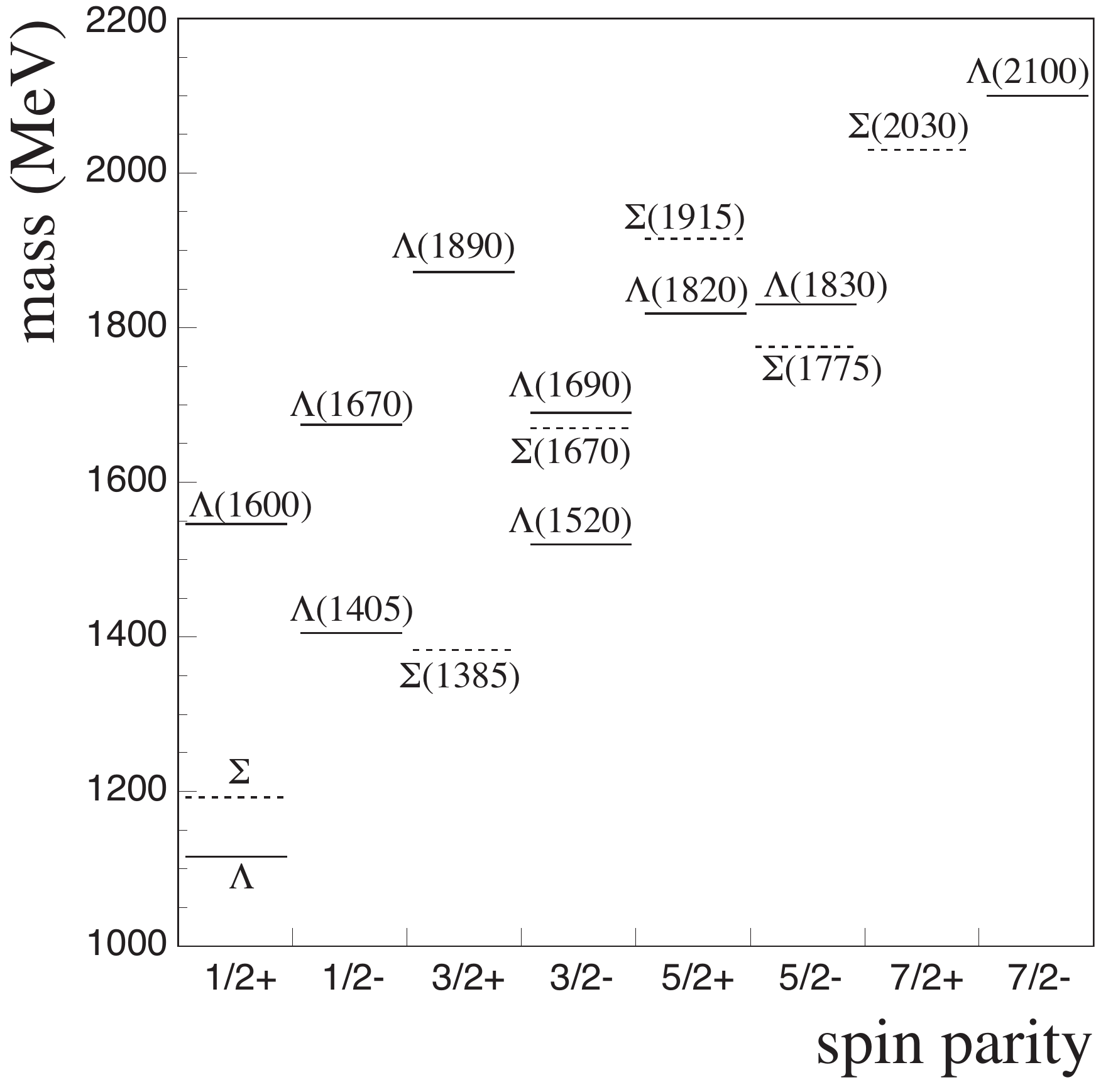}}
  {\includegraphics[width=14cm]{figs/mass_hyp.pdf}}
 \caption{\label{fig:mass_hype.eps} The mass spectrum of 
$S=-1$ hyperons that are listed as four-star resonances by the PDG~\cite{Zyla:2020zbs}. The solid and dashed lines present $\Lambda$ and $\Sigma$ families, respectively. }
\end{center}
\end{figure}

The members of $S=-1$ baryons are composed of an $s$ quark and
two of $u$ or $d$ quarks, and are classified into isospin $I=0$ and $I=1$
families, $\Lambda$ and $\Sigma$ hyperons, respectively.
Figure~\ref{fig:mass_hype.eps} shows the mass spectrum of
four-star $S=-1$ hyperons in the PDG~\cite{Zyla:2020zbs}.

\subsection{\it Recent progress in the $\Lambda(1405)$ studies}
\label{sec:L1405}

The $\Lambda(1405)$ is the lowest lying resonance with $J^{P}=1/2^{-}$, which has been continuously studied for more than 60 years since its theoretical prediction by Dalitz and Tuan~\cite{Dalitz:1959dn,Dalitz:1960du}. Detailed description of the investigations before 2011, including historical developments, can be found in a review article~\cite{Hyodo:2011ur}. Since then, there have been several important theoretical and experimental developments. In the following, we summarize recent achievements in the study of the $\Lambda(1405)$. Experimental results are summarized in Sections~\ref{sec:KbarNscattering}, \ref{sec:L1405_lineshape} and \ref{sec:spinparity}, and theoretical studies are reviewed in the subsequent sections. See also the recent reviews~\cite{Meissner:2020khl,Mai:2020ltx}.

\subsubsection{\it $\bar{K}N$ scattering data and the $\Lambda(1405)$}
\label{sec:KbarNscattering}

The $\Lambda(1405)$ baryon is the $S=-1$ isospin $I=0$ resonance, which 
can couple to 
$\pi\Sigma$ and $\bar{K}N$ channels. The 
mass of the $\Lambda(1405)$ is just below the $\bar{K}N$ threshold,
and it decays to $\pi\Sigma$ with 100\% branching fraction.
Experimentally, the $\Lambda(1405)$ can be identified as a resonance peak in 
the $\pi\Sigma$ invariant mass spectrum. 
The lineshapes and the spin-parity of the $\Lambda(1405)$ are 
studied by reconstructing the $\Lambda(1405)$
in $\pi\Sigma$ final states.
On the other hand, the $\Lambda(1405)$ plays an important role
in the $\bar{K}N$ scattering near the threshold due to the strong coupling
to this channel.
The properties of the $\Lambda(1405)$ can be investigated both from 
the $\pi\Sigma$ and $\bar{K}N$ channels.

\begin{figure}[tbp]
\begin{center}
  \figureBB{\includegraphics[width=14cm, bb=42 289 435 626]{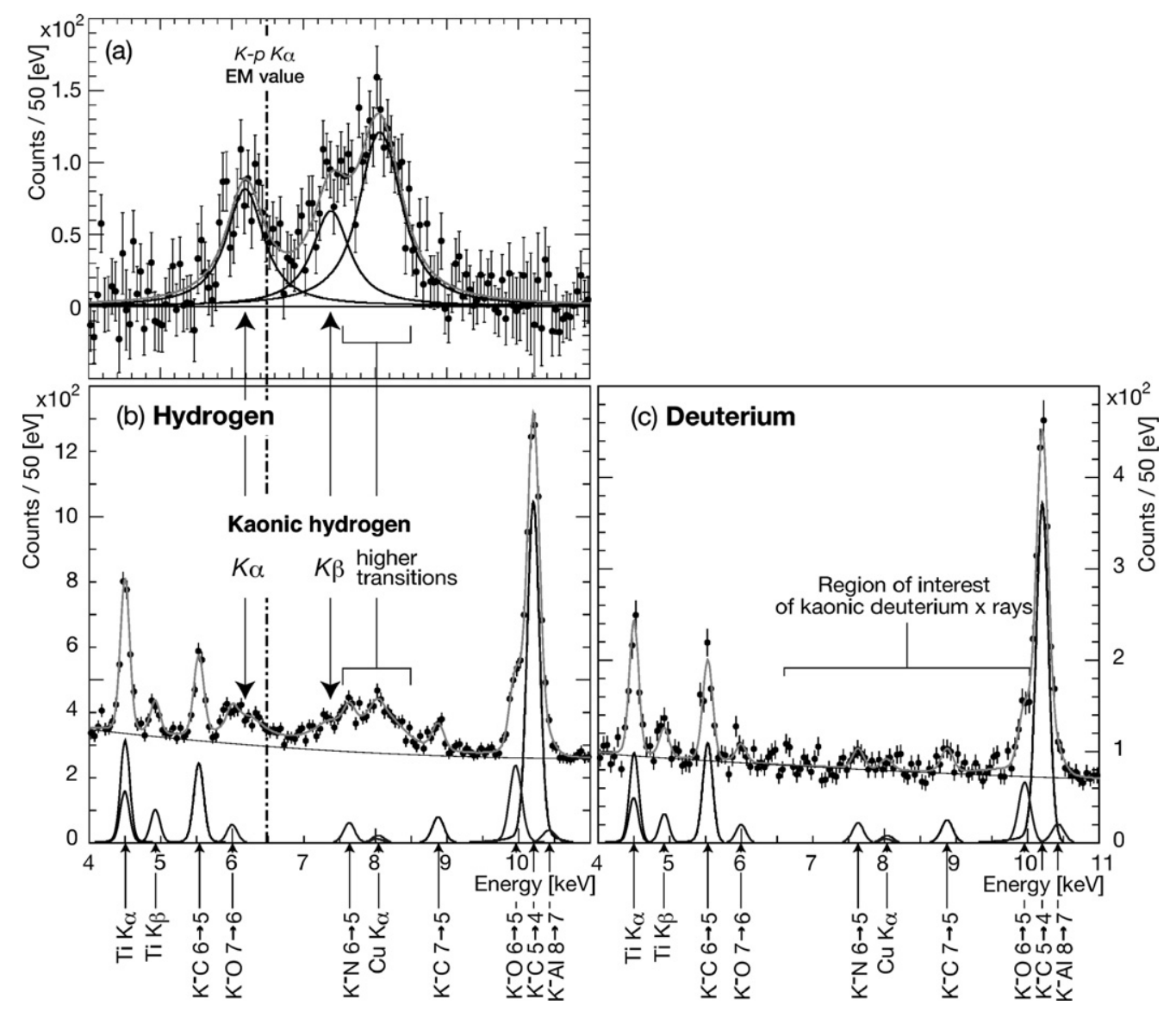}}
  {\includegraphics[width=14cm]{figs/SIDDHARTA_result.pdf}}
\begin{minipage}[t]{16.5 cm}
\caption{A global simultaneous fit result of the X-ray energy spectra of hydrogen and deuterium data.
Adapted from Ref.~\cite{Bazzi:2012eq}.
\label{siddharta_result}}
\end{minipage}
\end{center}
\end{figure}

The $K^-p$ scattering data near the threshold were obtained
using low energy kaon beams. Due to the finite life time of
$K^-$, the intensity of low energy kaon beam is low,
and precision of the scattering cross section is quite limited.
Besides, 
the $K^-p $ scattering length, which is the combination of the 
$I=0$ and $I=1$ scattering amplitudes at threshold, can be obtained from the level
shift ($\Delta E$) by strong interaction 
and the width ($\Gamma$) of the 1s level of the kaonic hydrogen.
The details of low energy $K^-p$ scattering data 
and old measurements of the kaonic hydrogen X rays
are summarized in Ref~\cite{Hyodo:2011ur}.
The SIDDHARTA collaboration has performed the newest measurement of 
the kaonic hydrogen X rays at DA$\Phi$NE (Fig.~\ref{siddharta_result}).
They found the repulsive shift 
\[
 \Delta E = -283\pm 36 (\mathrm{stat.}) \pm 6(\mathrm{syst.})~\rm{eV~~and~~} \Gamma = 541\pm 89(\mathrm{stat.})\pm 22(\mathrm{syst.})~\rm{eV},
\]
which
is consistent with the existence of the quasi-bound state of 
$\bar{K}N$~\cite{Bazzi:2011zj,Bazzi:2012eq}.
In order to access the antikaon-neutron interaction, 
X-ray spectroscopy of kaonic deuterium atoms is planned
by the E57 collaboration at J-PARC~\cite{Hashimoto:2019qfy, Marton:2019lni, J-PARC_P57}
and the SIDDHARTA 2 collaboration at DA$\Phi$NE~\cite{Scordo:2018ufs,Marton:2019lni},
and isospin dependent scattering lengths will be measured in near future.

\begin{figure}[tbp]
\begin{center}
  \figureBB{\includegraphics[width=15cm, bb=72 550 536 747]{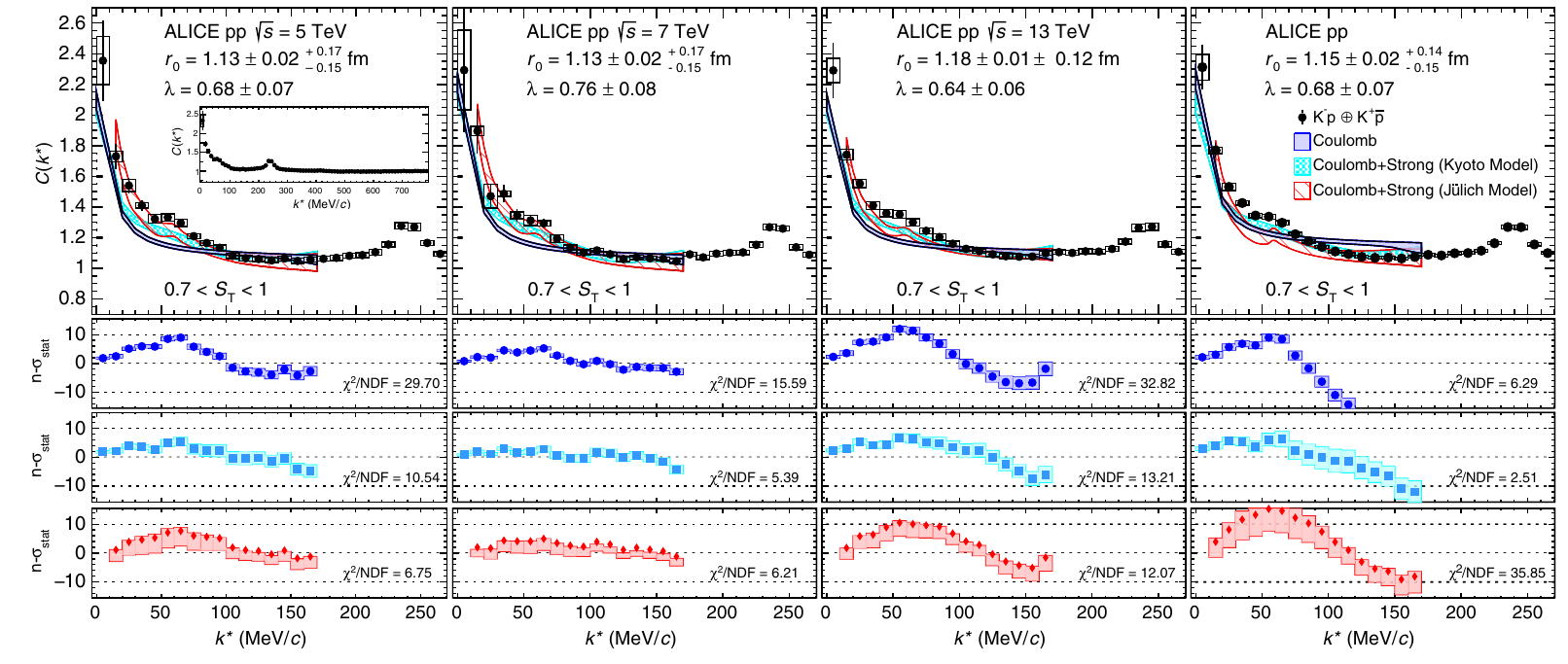}}
  {\includegraphics[width=15cm]{figs/Alice_KP_results.pdf}}
\begin{minipage}[t]{16.5 cm}
\caption{
$(K^{-} p \oplus K^{+} \bar{p})$ correlation functions obtained
 (from left to right) from $pp$ collisions at $\sqrt{s} = 5, 7, 13$~TeV.
The fourth panel shows the combined results at the three colliding energies; 
the number of pairs in each data sample has been used as weight. Adapted from Ref.~\cite{Acharya:2019bsa}.
\label{Alice_KP_results}}
\end{minipage}
\end{center}
\end{figure}

Recently, the ALICE collaboration demonstrated a new method to measure 
the $\bar{K}N$ interaction using the $pp$ collision data at 
$\sqrt{s} = 5, ~7$ and 13~TeV~\cite{Acharya:2019bsa} (Fig.~\ref{Alice_KP_results}).
They performed femtoscopic measurements of the correlation 
function at low relative momentum of $K^+ p$ ($K^-\bar{p}$) and
$K^-p$ ($K^+\bar{p}$) pairs, and they observed a cusp structure
around a relative momentum of 58~MeV in the measured correlation function of
 $K^-p$ ($K^+\bar{p}$) pairs, which corresponds
to the threshold of the isospin partner channel 
$\bar{K^0}n$ ($K^0\bar{n}$) due to the mass difference among isospin multiplets.
The measured correlation functions were compared to several models.
Although their results are sensitive to the {\it source size}, $r_0$,
the $\bar{K}N$ interaction was investigated.
Theoretical calculation of the $K^{-}p$ correlation function was performed in Ref.~\cite{Kamiya:2019uiw}. By using the meson-baryon coupled-channel potential developed in Ref.~\cite{Miyahara:2018onh}, the measured correlation function is well reproduced. Because the potential in Ref.~\cite{Miyahara:2018onh} was constructed to reproduce the $K^{-}p$ scattering data including the above mentioned kaonic hydrogen measurement by SIDDHARTA, one can say that the ALICE result is consistent with the SIDDHARTA data, within the framework of Ref.~\cite{Kamiya:2019uiw}. It should, however, be noted that the calculation of the correlation function requires the construction of the meson-baryon potential, as well as the determination of the parameters such as the source size. Nevertheless, the ALICE data, with its excellent quality, will be important for the future studies of the $\bar{K}N$ interaction. 

\begin{figure}[tbp]
\begin{center}
  \figureBB{\includegraphics[width=7cm,bb=0 0 223 136]{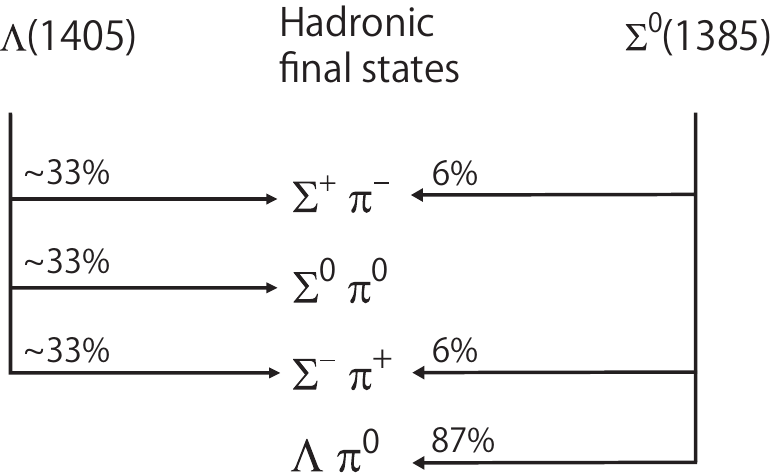}}
  {\includegraphics[width=7cm]{figs/L1405_S1385_br.pdf}}
\begin{minipage}[t]{16.5 cm}
\caption{Branching fractions of the $\Lambda(1405)$ and the $\Sigma^{0}(1385)$
into hadronic final states.\label{L1405_S1385_br}}
\end{minipage}
\end{center}
\end{figure}

\subsubsection{\it Lineshape of $\pi\Sigma$ invariant mass spectra}\label{sec:L1405_lineshape}

In this section, we review recent results of the $\Lambda(1405)$ obtained
from the $\pi\Sigma$ final states.
An experimental difficulty to study the $\Lambda(1405)$ is to 
separate the isospin $I=0$ component from the $I=1$ component which couples to the $\Sigma(1385)$.
The mass difference of these two baryons are smaller than their widths of
them, and thus, they overlap with each other in $\pi^\pm\Sigma^\mp$ invariant mass spectra.
Figure~\ref{L1405_S1385_br} shows the hadronic branching fractions of 
these baryons.
The $\Lambda(1405)$ baryon decays into a $\pi\Sigma$ pair with 100\% 
branching fraction while the $\Sigma(1385)$ mainly decays to $\pi\Lambda$.
The $\Sigma(1385)$ baryon is an $I=1$ resonance and cannot decay into 
a $\pi^0\Sigma^0$ pair, whereas, it can decay into a $\pi^\pm\Sigma^\mp$ pair.
The isospin of $\pi^0\Sigma^0$ pairs can be $I=0$ and $I=2$. However,
the $I=2$ amplitude is nonresonant and assumed to be negligible. 
Thus, the invariant mass spectra of $\pi^0\Sigma^0$ pairs can be regarded as a pure $I=0$ 
amplitude.
In order to reconstruct $\pi \Sigma$ pairs, we need to identify 
a neutral particle such as a photon or a neutron;
the main decay modes of $\Sigma$ baryons are 
$\Sigma^+\rightarrow p \pi^0$ ($\sim 52$\%),
$\Sigma^+\rightarrow n \pi^+$ ($\sim 48$\%), 
$\Sigma^-\rightarrow n \pi^-$ ($\sim 99.8$\%), 
and $\Sigma^0\rightarrow \Lambda \gamma$ ($\sim 100$\%). 
Thus, the reconstruction of $\Sigma$ baryons is rather difficult
compared with that of $\Lambda \rightarrow p \pi^-$ where only 
charged particles exist in the final state.

For a long time, the experimental data of the $\Lambda(1405)$ were limited to
low statistics data obtained using bubble chambers.
The results of bubble chamber experiments are summarized in 
Ref.~\cite{Hyodo:2011ur}.
Since 2003, the $\Lambda(1405)$ has been studied using modern detectors
and high intensity beams.
In these decades, experimental information of the $\Lambda(1405)$ 
have increased rapidly owing to intensive studies with high statistics data.
In order to understand the nature of the $\Lambda(1405)$, 
experimental studies have been performed to observe
the lineshape of the invariant mass of $\pi\Sigma$ pairs, 
the spin-parity quantum number, 
and the production cross sections.

The lineshape of $\pi\Sigma$ invariant mass contains the information of
the pole position and the decay width of the $\Lambda(1405)$.
Under the assumption of negligible $I=2$ component, the $\pi\Sigma$ invariant mass spectra for isospin 0 and 1 components can be described as 
\begin{equation}
\frac{d \sigma\left(\pi^{+} \Sigma^{-}\right)}{d M_{I}} \propto
\frac{1}{3}\left|T^{(0)}\right|^{2}+\frac{1}{2}\left|T^{(1)}\right|^{2}+\frac{2}{\sqrt{6}}
\mathrm{Re}\left(T^{(0)} T^{(1) *}\right),
\end{equation}

\begin{equation}
 \frac{d \sigma\left(\pi^{-} \Sigma^{+}\right)}{d M_{I}} \propto
 \frac{1}{3}\left|T^{(0)}\right|^{2}+\frac{1}{2}\left|T^{(1)}\right|^{2}-\frac{2}{\sqrt{6}}
 \mathrm{Re}\left(T^{(0)} T^{(1) *}\right),
\end{equation}
\begin{equation}
 \frac{d \sigma\left(\pi^{0} \Sigma^{0}\right)}{d M_{I}} \propto \frac{1}{3}\left|T^{(0)}\right|^{2},
\end{equation}
where $T^{(I)}$ and $M_I$ represent the $\pi\Sigma$ amplitude and the
invariant mass with isospin $I$, respectively.
The isospin interference term $\mathrm{Re}\left(T^{(0)} T^{(1)
*}\right)$ makes the difference of the charged $\pi^\pm\Sigma^\mp$
spectra.
Based on this observation, Ref.~\cite{Nacher:1998mi} theoretically calculated the production of the $\Lambda(1405)$ in the $\gamma p \rightarrow K^+ \pi\Sigma$ reaction, predicting
the different lineshapes in 
$\pi^-\Sigma^+$, $\pi^0\Sigma^0$, and $\pi^+\Sigma^-$ channels.

\begin{figure}[tbp]
\begin{center}
  \figureBB{\includegraphics[width=15cm,bb=0 0 502 156]{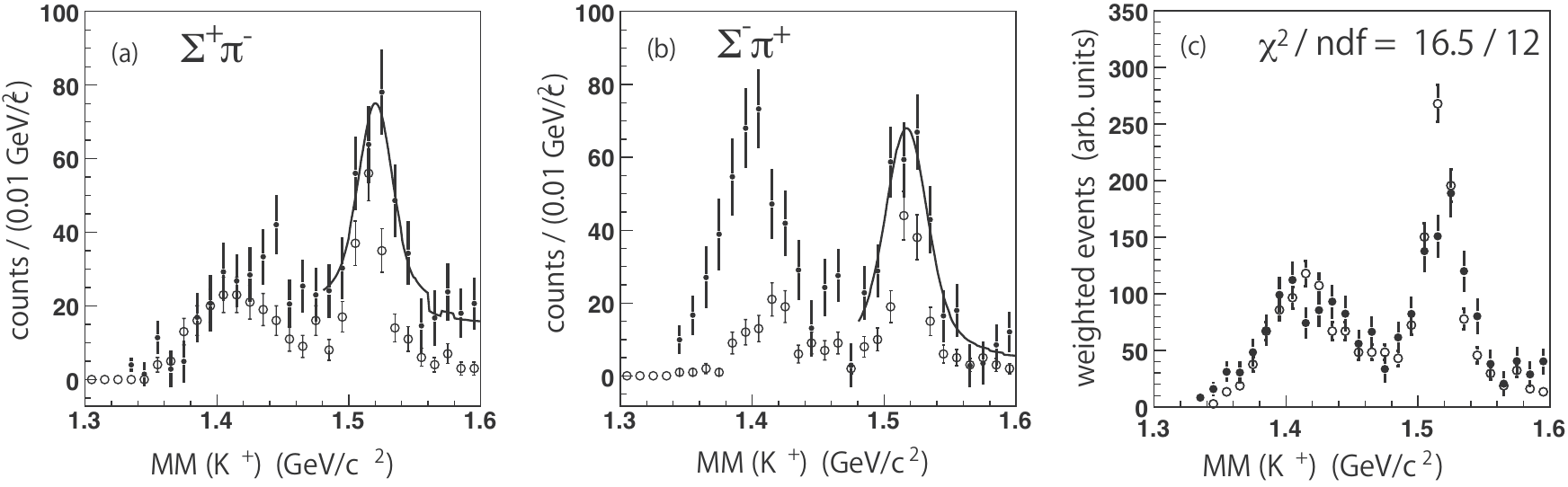}}
  {\includegraphics[width=15cm]{figs/niiyama-PhysRevC78.pdf}}
\begin{minipage}[t]{16.5 cm}
\caption{
Missing mass for the $\gamma p \rightarrow K^+ X$ reaction~\cite{Niiyama:2008rt}.
(a) $K^+\Sigma^+\pi^-$ final state. (b) $K^+\Sigma^-\pi^+$
final state. Solid lines in (a) and (b) show fit results of
$K^+\Lambda(1520)$ plus nonresonant
($K^+\Sigma\pi$)  production. (c) The combined spectra
 of the $\Sigma^+\pi^-$ and
$\Sigma^-\pi^+$ and decay modes. Closed and open circles show
 spectra obtained by Ref.~\cite{Niiyama:2008rt} and by Ref.~\cite{Ahn:2003mv},
 respectively.\label{niiyama-PhysRevC78}
}
\end{minipage}
\end{center}
\end{figure}

After the theoretical prediction, the LEPS Collaboration measured the 
lineshapes of $\pi^-\Sigma^+$ and
$\pi^+\Sigma^-$ invariant mass spectra using $\gamma p \rightarrow K^+ \pi\Sigma$
reaction~\cite{Ahn:2003mv} with the photon energy 
1.5-2.4~GeV. The contribution of the $K^*(890)$
production was excluded in the invariant mass of 
$K^+\pi^-$, and they observed a peak structure around 1.4~GeV.
The observed lineshapes were consistent with the theoretical predictions
for the $\Lambda(1405)$ photoproduction by Ref.~\cite{Nacher:1998mi}.
However, 
the experimental spectra contain the $\Sigma\pi$ pairs from the decay of
the $\Sigma(1385)$, and the amount of the $\Sigma(1385)$ decay contribution
was not separated.
In the subsequent study, the LEPS collaboration measured the production
cross section of the $\Lambda(1405)$ and the $\Sigma(1385)$ 
in the $\gamma p \rightarrow K^+ \Lambda(1405)$ and 
$\gamma p \rightarrow K^+ \Sigma^{0}(1385)$ reactions 
by detecting the $\Lambda(1405) \rightarrow \pi^\pm\Sigma^\mp $ decay and
$\Sigma(1385)^0 \rightarrow \pi^0\Lambda $ decay, respectively~\cite{Niiyama:2008rt}.
In the $\pi^0\Lambda$ final state, only $I=1$ amplitude contributes
and we can identify the $\Sigma^{0}(1385)$. The amount of the $\Sigma^0(1385)$
in the $\pi^\pm\Sigma^\mp$ final state was estimated 
using the known branching fractions of the $\Sigma^0(1385)$ decay.
The absolute value of the differential cross section
$d\sigma/d(\cos\theta)$
was obtained as 0.43~$\mu$b (0.072~$\mu$b) for the photon energy $1.5<E_\gamma <2.0$~GeV
($2.0<E_\gamma <2.4$~GeV).
They observed the difference in the charged $\pi\Sigma$ spectra again 
(Fig.~\ref{niiyama-PhysRevC78}),
however, the shape of the peak was not consistent with the previous measurement,
likely because of the different kinematical region of the final state pion. 
Since the LEPS first observation is consistent with the theoretical
prediction, the second one contradicts the prediction of Ref.~\cite{Nacher:1998mi}.

\begin{figure}[tbp]
\begin{center}
  \figureBB{\includegraphics[width=15cm,bb=0 0 360 239]{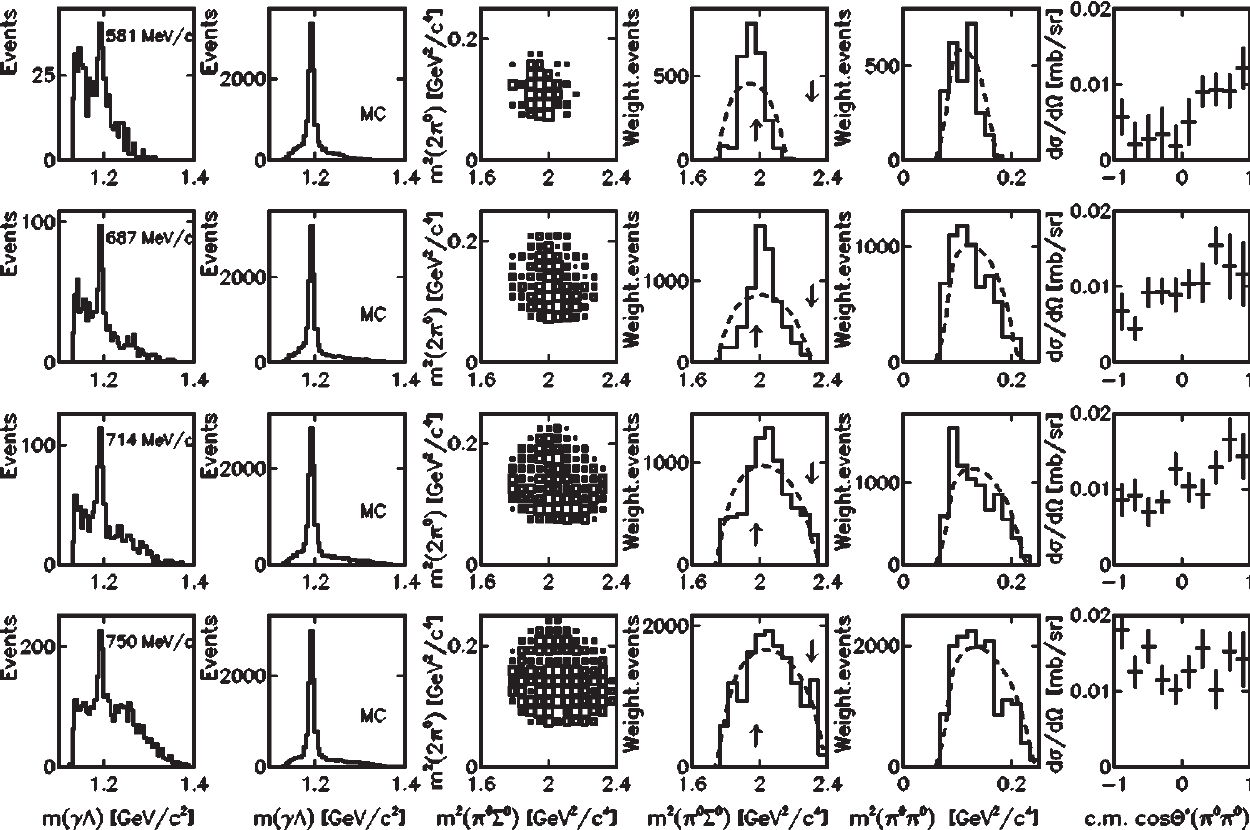}}
  {\includegraphics[width=15cm]{figs/Crystal_ball_results.pdf}}
\begin{minipage}[t]{16.5 cm}
\caption{
Dalitz plot projections to $\pi^{0}\Sigma^{0}$ (column 4) in the $K^- p \rightarrow \pi^0\pi^0\Sigma^0$ reaction by the Crystal Ball collaboration. Adapted from Ref.~\cite{Prakhov:2004an}.\label{Crystal_ball_results}
}
\end{minipage}
\end{center}
\end{figure}

The Crystal Ball collaboration observed the neutral $\pi^0\Sigma^0$ spectrum 
in the $K^- p \rightarrow \pi^0\pi^0\Sigma^0$ reaction in the $K^-$ momentum
range of $514-750$~MeV~\cite{Prakhov:2004an} (Fig.~\ref{Crystal_ball_results}). 
The $\pi^0\Sigma^0$ channel is ideal to investigate the $\Lambda(1405)$ spectrum, 
since the $\pi^0\Sigma^0$ spectrum does not contain
the $I=1$ amplitude with the $\Sigma(1385)$.
The authors of Ref.~\cite{Magas:2005vu} pointed out that the peak position
of the spectrum locates at 1.42~GeV, and they discussed the two pole
structure of the $\Lambda(1405)$.

\begin{figure}[tbp]
\begin{center}
  \figureBB{\includegraphics[width=7cm,bb=0 0 247 173]{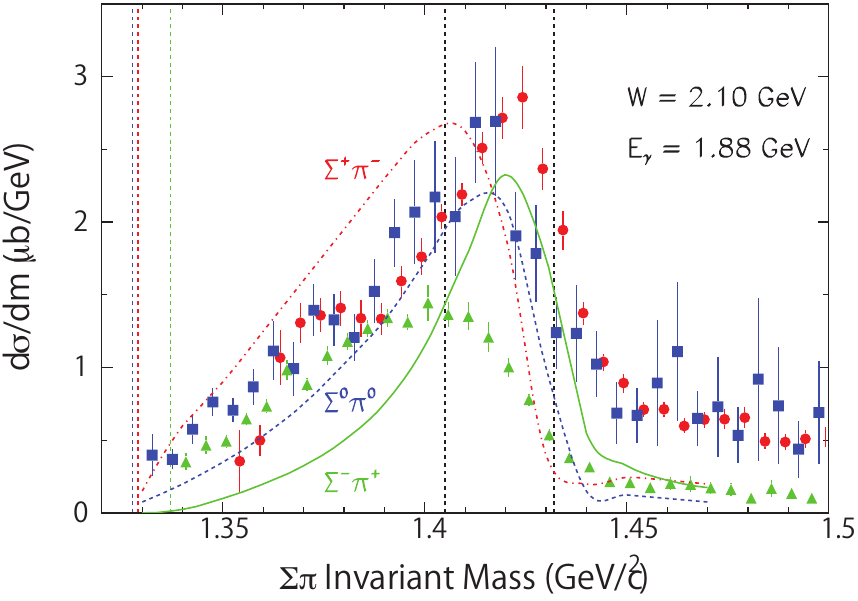}}
  {\includegraphics[width=7cm]{figs/CLAS_lineshape.pdf}}
\begin{minipage}[t]{16.5 cm}
\caption{
$\pi\Sigma$ mass distributions in the $\gamma p \rightarrow K^+ \pi\Sigma$ reaction by the CLAS collaboration. Adapted from Ref.~\cite{Moriya:2013eb}.
\label{CLAS_lineshape}}
\end{minipage}
\end{center}
\end{figure}

The CLAS collaboration measured the lineshapes of all charge combinations of
$\pi\Sigma$ invariant mass
using very high statistics data~\cite{Moriya:2013eb} (Fig.~\ref{CLAS_lineshape}).
The difference of the lineshapes of the charged $\pi\Sigma$
spectra were confirmed, and the observed differences contradict the theoretical predictions 
of Ref.~\cite{Nacher:1998mi}.
They separated isospin amplitudes using a Breit-Wigner model, 
and obtained two $I=1$ amplitudes with a centroid at 
$1394\pm 20$~MeV and $1413\pm10$~MeV, here, the fit quality
was fairly good and the reduced $\chi^2$ was 2.15 at the best.
The centroid of the $I = 0$ $\Lambda(1405)$ strength was found at
the $\pi\Sigma$ threshold, and they suggest that 
the observed shape is determined by channel coupling.
The authors of Ref.~\cite{Roca:2013av} implemented five parameters 
to the chiral unitary model and fitted 
the $\pi^0\Sigma^0$ spectra obtained by CLAS.
The model reproduce the CLAS results successfully with $\chi^2/ndf =
0.6\sim 1.76$, showing the two-pole structure discussed in later sections.
Using the same high statistics data, the CLAS collaboration measured the differential photoproduction 
cross sections of the $\Sigma^{0}(1385)$,
the $\Lambda(1405)$, and the $\Lambda(1520)$ in the $\gamma p \rightarrow K^+ Y^*$
reactions in the photon beam energy from near the production threshold to 
the center-of-mass energy $W$ of 2.85~GeV with very high precision~\cite{Moriya:2013hwg}.
The CLAS data cover large $K^+$ angular regions, while the previous LEPS measurements
cover the very forward scattering angle of $K^+$ for these hyperon production and
the very backward $K^+$ angle for $\Lambda(1520)$.
The production cross sections of the $\Sigma^{0}(1385)$ and the $\Lambda(1520)$ seem consistent
between CLAS and LEPS results in the close angular regions. However, for the 
$\Lambda(1405)$, these two results are consistent in the low photon energy region, but
CLAS do not observe the reduction of the production rate in the high photon energy region.

\begin{figure}[tbp]
\begin{center}
  \figureBB{\includegraphics[width=7cm,bb=55 449 261 730]{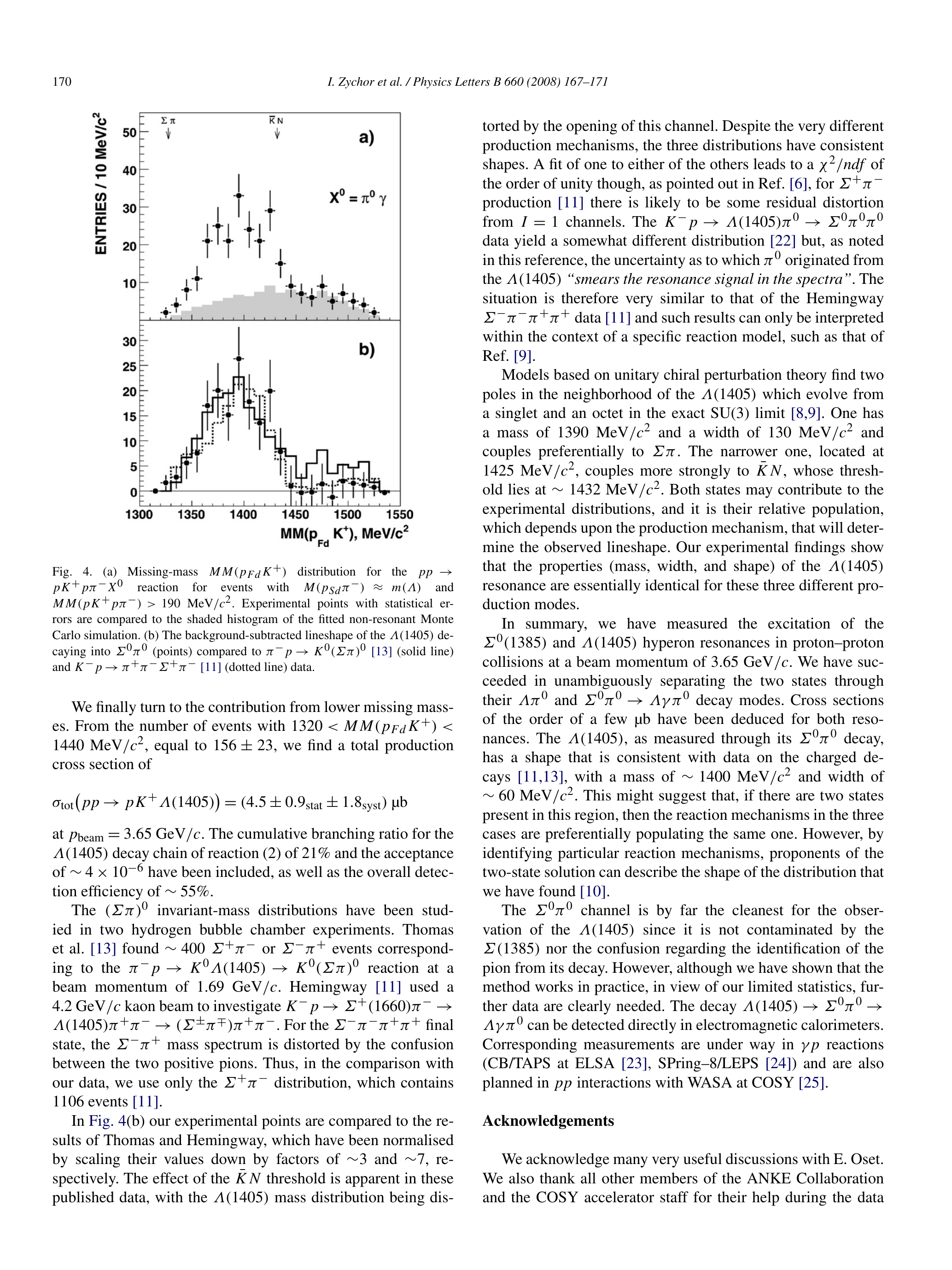}}
  {\includegraphics[width=7cm]{figs/COSY_Zychor_ANKE_PLB660.pdf}}
\begin{minipage}[t]{16.5 cm}
\caption{ 
Missing-mass $MM(p_{Fd}K^+)$ distribution for the 
$p p \rightarrow p K^{+} p \pi^{-} X^{0}$ reaction
with the ANKE spectrometer at COSY- J\"ulich. Adapted from Ref.~\cite{Zychor:2007gf}.
\label{COSY_lineshape}}
\end{minipage}
\end{center}
\end{figure}

\begin{figure}[tbp]
\begin{center}
  \figureBB{\includegraphics[width=7cm,bb=322 342 528 729]{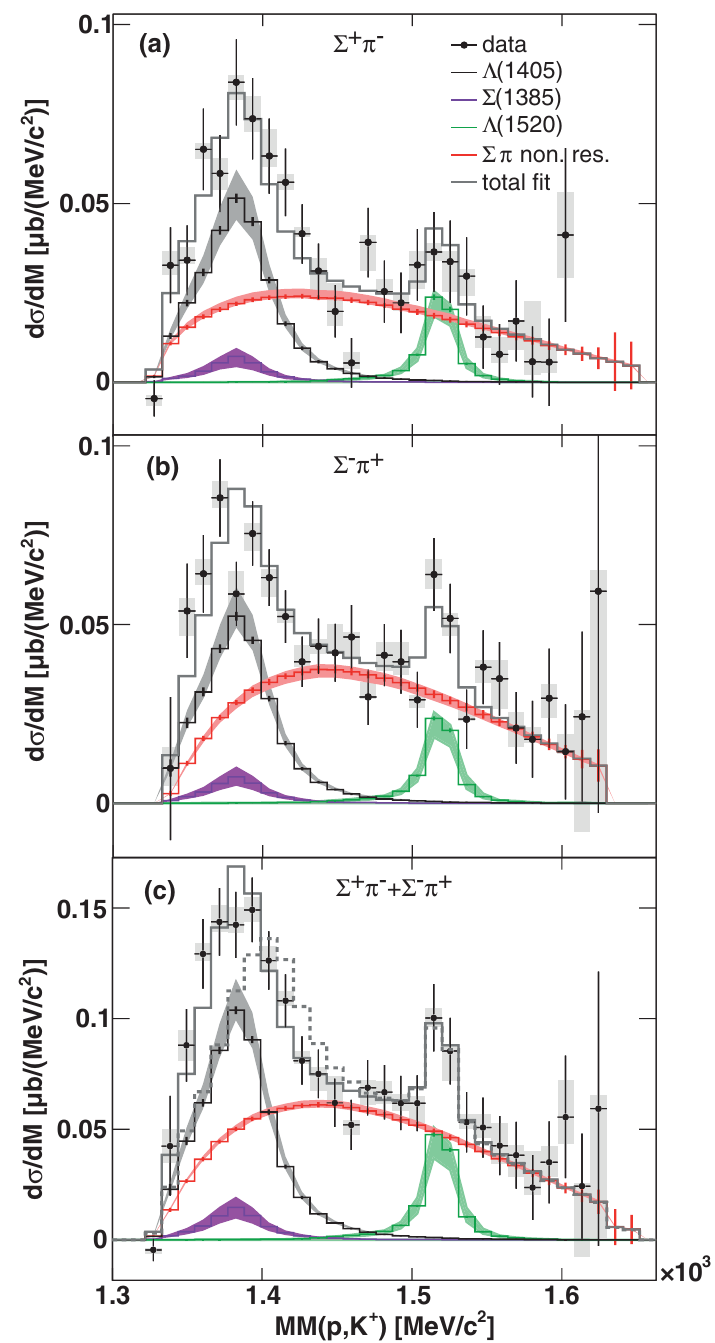}}
  {\includegraphics[width=7cm]{figs/HADES_L1405_pp_collisions_PhysRevC.pdf}}
\begin{minipage}[t]{16.5 cm}
\caption{
Missing mass $MM(p, K^+)$ distributions for events attributed 
to the $\Sigma^+\pi^-$ decay channel
(a) and the $\Sigma^-\pi^+$ decay channel (b) by the HADES collaboration.
Adapted from Ref.~\cite{Agakishiev:2012xk}.
\label{HADES_lineshape}}
\end{minipage}
\end{center}
\end{figure}

The $\Lambda(1405)$ production from the $pp$ collision was studied 
in the reactions 
$p p \rightarrow \pi^0\Sigma^0 p K^+ $ 
and $p p \rightarrow \pi^{\mp} \Sigma^{\pm} p  K^{+}$
at COSY-J\"{u}lich~\cite{Zychor:2007gf} and at HADES-GSI~\cite{Agakishiev:2012xk},
respectively.
The neutral $\pi^0\Sigma^0$ spectrum was measured by the COSY collaboration
from the missing mass of $p p \rightarrow p K^+ X$ reaction (Fig.~\ref{COSY_lineshape}). In order to detect
$\pi^0 \Sigma^0 \rightarrow \pi^0 \Lambda \gamma$ in the final state, 
they selected the events with a $\Lambda$ and with the constraint 
on the missing mass of $p p \rightarrow p K^+ \Lambda X$ larger than 190~MeV
for the $\pi \gamma$ in the final state.
The peak position of the $\Lambda(1405)$ was found at 1.405~GeV. 
The total cross section of the $pp \rightarrow pK^+ \Lambda(1405)$ was obtained as
4.5~$\mu b$ at the proton beam momentum of 3.5~GeV.
The HADES collaboration measured the lineshape of $\pi^\pm\Sigma^\mp$ in
the $p p \rightarrow p K^+ \pi^\pm\Sigma^\mp \rightarrow p K^+ \pi^\pm n \pi^\mp$ 
reactions at the 3.5~GeV kinetic
proton beam energy (4.3~GeV proton beam momentum)~\cite{Agakishiev:2012xk} (Fig.~\ref{HADES_lineshape}). 
The neutron in the final state and $\Sigma^\pm$ in the intermediate state
were reconstructed from the missing mass of the
$p p \rightarrow p K^+ \pi^\pm \pi^\mp X$ reaction and 
$p p \rightarrow p K^+ \pi^\mp X$ reaction, respectively.
The invariant mass spectra of $\pi^\pm\Sigma^\mp$ were obtained
from the missing mass of the
$p p \rightarrow p K^+ X$ reaction for the events with a neutron and $\Sigma$
were identified.
The contribution of the $\Sigma^0(1385) \rightarrow \pi^\pm\Sigma^\mp$ decay was
estimated from the $\Sigma^0(1385) \rightarrow \pi^0\Lambda$ decay where the only $I=1$
amplitude contributes and the branching fractions of the $\Sigma^0(1385)$ are known.
From the peak corresponding to $\pi^0$ in the missing mass spectrum of 
$p p \rightarrow p K^+ \Lambda X$ reaction, 
the yield of the $\Sigma^0(1385)$ was obtained, and the contribution into the
$\pi^\pm\Sigma^\mp$ spectra were turned out to be small.
In the same spectrum, 
the contribution of 
the $\Lambda(1405) \rightarrow \Sigma^0 \pi^0 \rightarrow \Lambda \gamma \pi^0$ 
decay was seen at the higher mass side of $\pi^0$. However, 
due to the limited statistics, 
the  analysis of the lineshape of the $\Lambda(1405)$ in the neutral $\pi^{0}\Sigma^0$
decay channel was not possible.
The spectra of $\pi^\pm\Sigma^\mp$ after 
the efficiency and acceptance-correction 
showed a peak position below 1.4~GeV. 
The total production cross section of the 
$\Lambda(1405) $ was obtained at this energy as 
$9.2\pm 0.9\pm 0.7^{+3.3}_{-1.0}$~$\mu$b, and the 
 polar angle distribution of the cross section was isotropic in the $p-p$ center-of-mass system.
The reason for the relatively low mass of the peak position ($\sim 1380$ MeV) was theoretically studied in Ref.~\cite{Bayar:2017svj}, where a possible mechanism was proposed in relation with the triangle singularity.

\begin{figure}[tbp]
\begin{center}
  \figureBB{\includegraphics[width=8cm,bb=57 287 269 728]{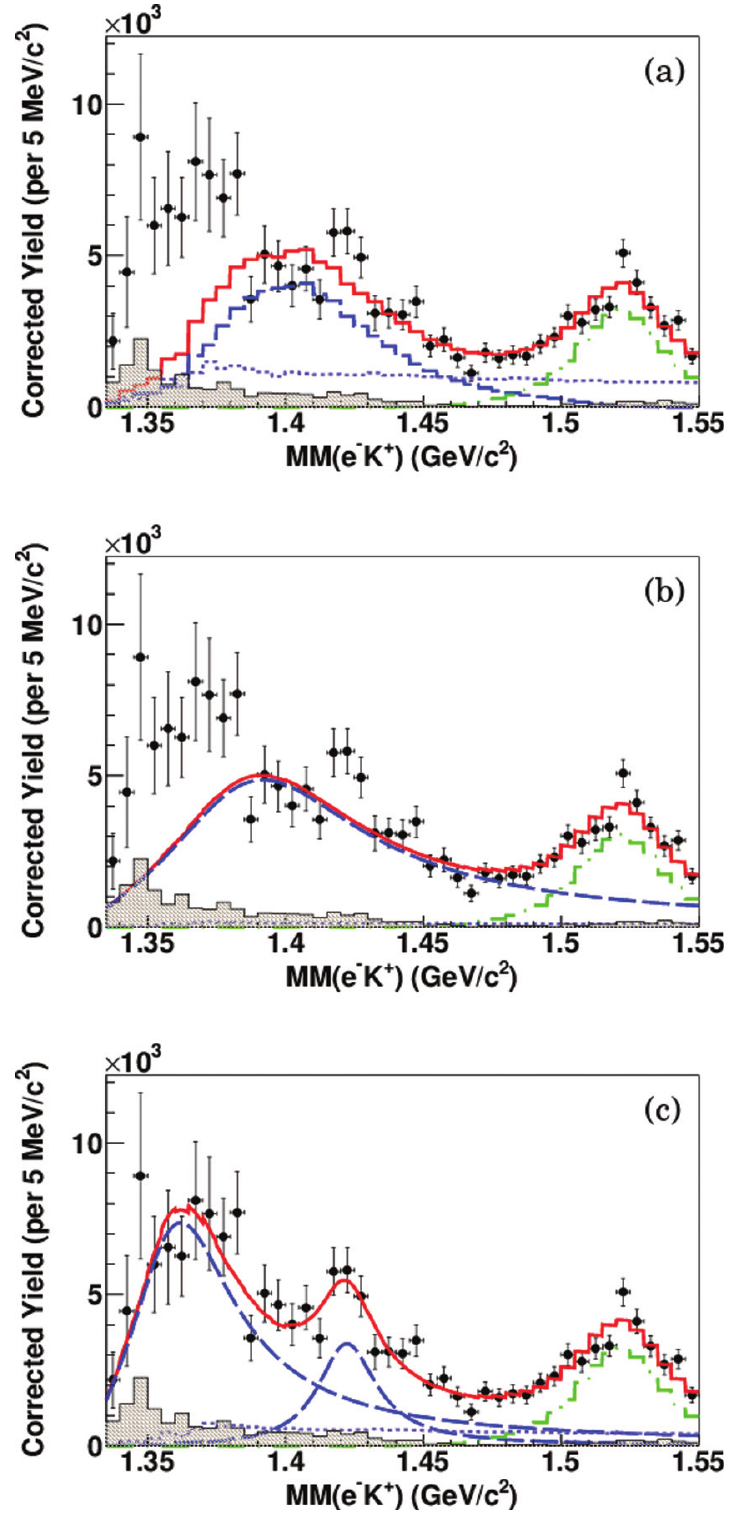}}
  {\includegraphics[width=8cm]{figs/CLAS_ee_1405.pdf}}
\begin{minipage}[t]{16.5 cm}
\caption{
Fits of the missing mass of $e^-K^+$
for $1.5 < Q^2 < 3.0$~GeV$^{2}$ by the CLAS collaboration.
Adapted from Ref.~\cite{Lu:2013nza}.
\label{CLAS_ee_lambda1405}
}
\end{minipage}
\end{center}
\end{figure}

The lineshape of $\pi^+\Sigma^-$ invariant mass near the $\Lambda(1405)$ region
in the $e^- p \rightarrow e^- K^+ \pi^+\Sigma^-$ reaction
was measured for the first time at CLAS
in the range of $1.0 < Q^2 < 3.0$~GeV$^{2}$~\cite{Lu:2013nza} (Fig.~\ref{CLAS_ee_lambda1405}).
The contamination from the $\Sigma(1385)$ was estimated from the $\pi^0\Lambda$ channel,
and was turned out to be negligible.
Two peak structures were observed at 1.368~GeV and 1.423~GeV, and
with increasing photon virtuality the mass distribution shifts toward 
the higher mass pole, suggesting two-pole structure of the $\Lambda(1405)$.

Very recently, the J-PARC E31 collaboration has reported the measurement
of the $K^- d \rightarrow n\Sigma^0\pi^0$ reaction~\cite{Kawasaki:2019wwj}.
Since the $\Lambda(1405)$ cannot be formed directly from $K^-p$ scattering
in free space, they used the reaction of $d(K^-,n)$ with an
incident momentum of 1~GeV. They measured the momenta of
neutrons scattered at forward angles, and that of protons and
negative pions in the large angular region.
They reconstructed $\Lambda$'s from proton-$\pi^-$ pairs, and
aimed to identify the $ K^- d\rightarrow n \Sigma^0 \pi^0$ reaction
by selecting $\gamma\pi^0$ produced events in the missing mass spectrum of
the $ K^- d\rightarrow n \Lambda X$ reaction.
They observed a significant number of events below $\bar{K}N$
threshold, and are finalizing the analysis to extract the
contribution of the $\Lambda(1405)$.

The AMADEUS collaboration at DA$\Phi$NE aims to investigate the 
$K^-N$ interaction at low energy, they analyzed data taken
with KLOE detector and obtained $\pi^0\Sigma^0$ invariant
mass spectrum from $K^-$ captures in ${}^{12}$C nuclei~\cite{Piscicchia:2018umq}.
They also measured the $K^- n \rightarrow \Lambda \pi^-$ amplitude.
These results can be used to further increase the understanding of $K^-N$ 
interaction at low energy.
 
\subsubsection{\it Spin and parity}
\label{sec:spinparity}

\begin{figure}[tbp]
\begin{center}
  \figureBB{\includegraphics[width=7cm,bb=0 0 792 612]{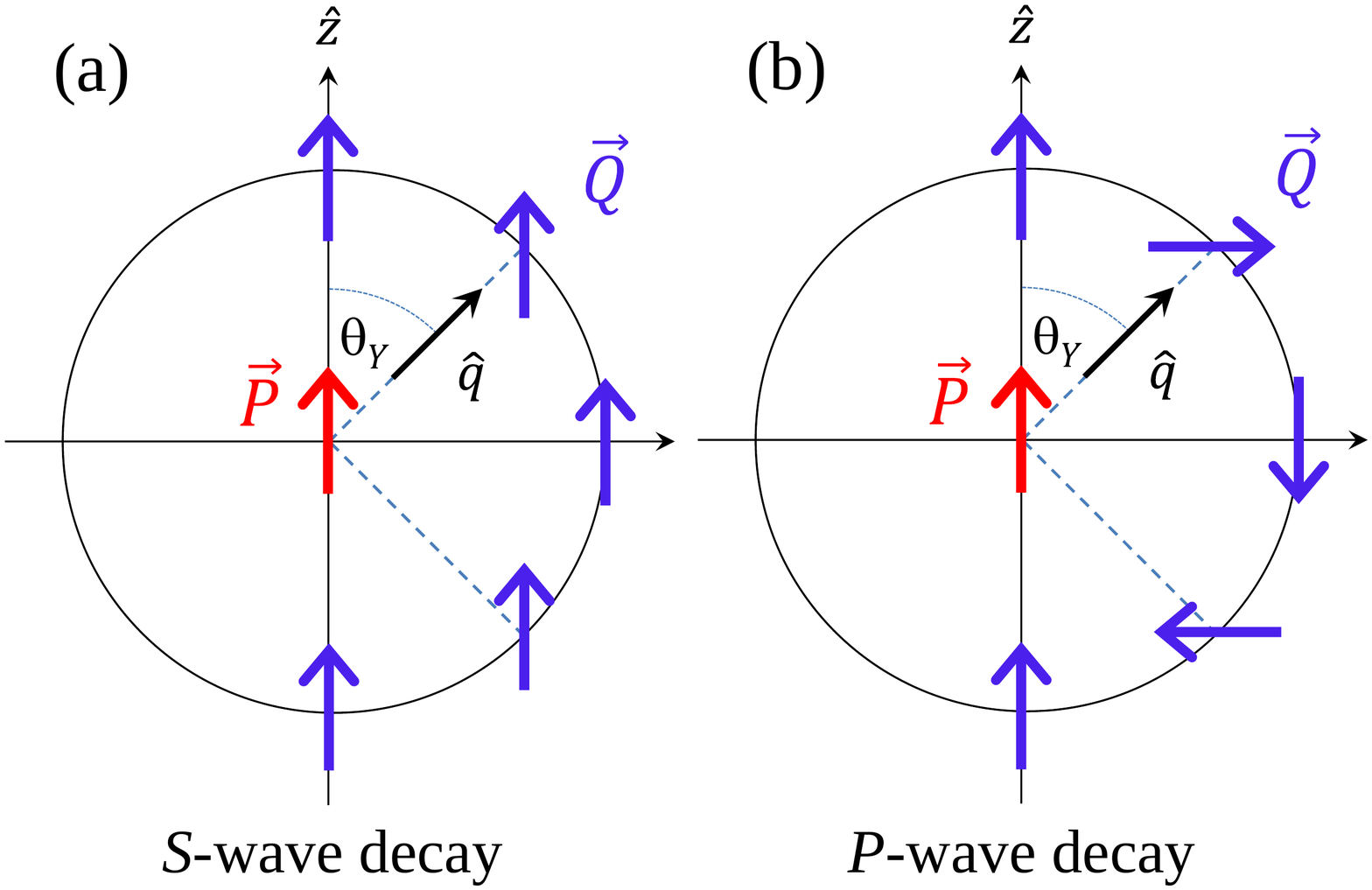}
  \includegraphics[width=7cm,bb=322 127 554 245]{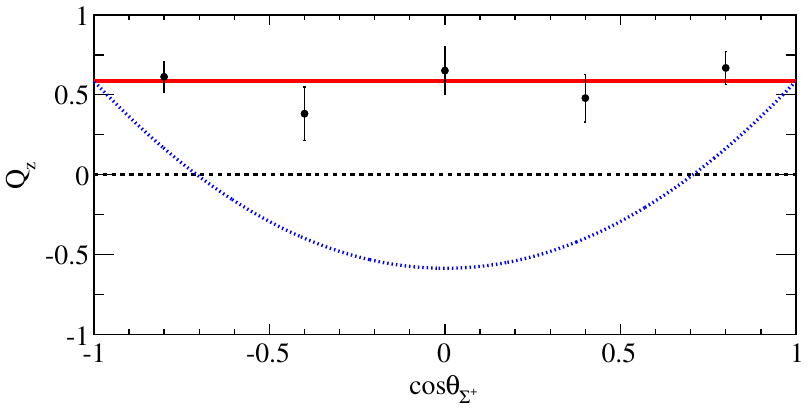}}
  {\includegraphics[width=7cm]{figs/clas_spin_parity.pdf}
  \includegraphics[width=7cm]{figs/CLAS_JP_result.pdf}}
\begin{minipage}[t]{16.5 cm}
\caption{
Left: polarization transfer from $Y^*$ to $Y$ in the decay
$Y^* \rightarrow Y \pi$, where $Y^*$ has spin $1/2$. Right: polarization $Q_z$ of $\Sigma^+$ versus $\cos \theta_{\Sigma^{+}}$ for
$2.65<W<2.75$~GeV and $0.70<\cos \theta_{K^{+}}^{\mathrm{cm} .}<0.80$.
The average is shown as the solid line. The dotted curve is the expectation for $p$-wave decay, and dashed line shows the no-polarization case.
Adapted from Ref.~\cite{Moriya:2014kpv}.
\label{fig:clas_spin_parity}}
\end{minipage}
\end{center}
\end{figure}

The spin of the $\Lambda(1405)$ was assigned as 1/2 from past
experiments~\cite{Engler:1965zz,Thomas:1973uh,Hemingway:1984pz}.
However, the parity of the $\Lambda(1405)$ 
had not been determined directly but assumed as negative
since the observed invariant mass spectra of the $\Lambda(1405)$ 
drop 
rapidly near the $\bar{K}N$ threshold, 
which indicate $s$-wave coupling to $\bar{K}N$,
and thus, $J^{P}=1/2^{-}$ is preferred.
Recently, the parity of the $\Lambda(1405)$ was
determined directly for the first time 
using high statistics data taken by the CLAS
collaboration~\cite{Moriya:2014kpv}.
The decay angular distribution of the $\Lambda(1405)\rightarrow \pi^-\Sigma^+$ and
the variation of the $\Sigma^+$ polarization ($\vec{Q}$)
with respect to the $\Lambda(1405)$ polarization direction ($\vec{P}$) 
determines the parity.
Figure~\ref{fig:clas_spin_parity} (a) shows the $s$-wave decay ($J^{P}=1/2^{-}$) of 
$Y^* \rightarrow Y\pi$. In this case, the direction of $\vec{Q}$ is independent of the decay angle
$\theta_Y$. On the other hand, in the case of the $p$-wave decay ($J^{P}=1/2^{+}$), 
the direction of $\vec{Q}$ rotates around the $\vec{P}$ vector. Thus, the
parity of the $\Lambda(1405)$ can be determined from the polarization of $\Sigma^+$
around the polarization vector of the $\Lambda(1405)$.
The quantization axis of the $\Lambda(1405)$ spin was 
selected as the direction out of the production plane
which was determined as 
$\hat{z} = \vec{p}_\gamma \times \vec{p}_{K^+}/|\vec{p}_\gamma \times \vec{p}_{K^+}|$.
The angular distributions of the $\Lambda(1405) \rightarrow \pi^-\Sigma^+$ decay,
$\Sigma^+ \rightarrow \pi^+ n$ decay and $\Sigma^+ \rightarrow \pi^0 p$ decay were measured.
The contamination of the background events was approximately 16\% and was mainly 
from the $\Sigma^{0}(1385)$.
In the right panel of Fig.~\ref{fig:clas_spin_parity}, 
the Poralization $Q_{z}$ for one bin of the total energy is shown. 
For an $s$-wave decay, $Q_{z}$ is independent of $\theta_{Y}$, while 
$Q_{z}$ changes its sign for a $p$-wave decay as indicated by the dotted curve. 
The observed $Q_{z}$ shows that
the spin-parity of the $\Lambda(1405)$
was consistent with $J^{P}=1/2^{-}$, while
the $1/2^{+}$ combination was strongly disfavored.

\subsubsection{\it Chiral SU(3) dynamics}
\label{subsec:chiralSU3}

Now we turn to the theoretical studies of the $\Lambda(1405)$.
The $\Lambda(1405)$ is a resonance in the $\pi\Sigma$ scattering, and locates slightly below the $\bar{K}N$ threshold. For the description of the $\Lambda(1405)$, therefore, it is necessary to deal with the coupled-channels meson-baryon scattering with strangeness $S=-1$. An elaborate approach, called chiral SU(3) dynamics, has been formulated in a series of works~\cite{Kaiser:1995eg,Oset:1997it,Oller:2000fj,Lutz:2001yb}, by combining the unitarity in coupled-channels scattering and chiral perturbation theory for low-energy meson-baryon interaction. This approach respects chiral symmetry of QCD, and the accuracy of the result can be sharpened by systematically introducing terms with higher chiral orders. 
The scattering amplitude $T_{ij}$, which is a matrix in the channel space (such as $K^{-}p, \pi^{0}\Sigma^{0},$ etc.), is obtained by solving the coupled-channels scattering equation
\begin{align}
   T_{ij}
   =V_{ij}+V_{ik}G_{k}T_{kj},
   \label{eq:scattering}
\end{align}
with the interaction kernel $V_{ij}$ and the loop function $G_{i}$. By constructing $V_{ij}$ from chiral perturbation theory, the low-energy constraints from chiral symmetry are encoded. In addition, the iterative substitution of $T_{ij}$ in the right hand side gives the resummation of infinite series of multiple scattering, which guarantees the coupled-channel unitarity.

In chiral perturbation theory, the meson-baryon interaction $V$ can be sorted out by chiral order $\mathcal{O}(p^{n})$, starting from $n=1$~\cite{Ecker:1994gg,Bernard:1995dp,Pich:1995bw,Bernard:2007zu,Scherer:2012xha}. The terms with small $n$ are dominant at low energy, and the terms up to the next-to-leading order $\mathcal{O}(p^{2})$ can be schematically written as
\begin{align}
   V
   =V_{\rm WT}
   +V_{\rm Born}
   +V_{\rm NLO}
   +\dotsb ,
\end{align}
where the ellipsis stands for the higher order terms of $\mathcal{O}(p^{3})$. As in the case of the chiral effective field theory for the nuclear force, the accuracy of the theory increases when the higher order terms are included, but we need sufficient amount of experimental data to fix the low-energy constants (LECs) which cannot be determined by the symmetry principle. In the leading order (LO) terms of $\mathcal{O}(p^{1})$, the dominant contribution for the $s$-wave scattering comes from the Weinberg-Tomozawa term $V_{\rm WT}$, which is the meson-baryon four-point contact interaction. It should be noted that the chiral low-energy theorem completely determines the properties of the Weinberg-Tomozawa term~\cite{Weinberg:1966kf,Tomozawa:1966jm}, such as the sign (whether the interaction is attractive or repulsive) and the strength of the coupling. 
Besides the meson decay constants which are determined by the spontaneous breaking of chiral symmetry, $V_{\rm WT}$ depends only on the flavor structures of the target hadron and the two-body system, thanks to the conservation of the vector current.
This means that, for instance, the WT term for the $\pi D$ scattering is the same with that for the $\pi N$ scattering, and it is possible to make a prediction even in the absence of experimental data.
The Born terms $V_{\rm Born}$ are given by the s- and u-channel exchange of ground state baryons. 
Chiral symmetry constrains the three-point meson-baryon (Yukawa) vertex in $V_{\rm Born}$ to be the axial vector coupling. The value of the axial charge depends on the target hadron. The Born terms
are formally counted also as $\mathcal{O}(p^{1})$ in the chiral counting, but they mainly contribute to the $p$-wave scattering, and their $s$-wave component is in a higher order than $V_{\rm WT}$ in the nonrelativistic expansion~\cite{Weinberg:1996kr}. This means that the leading meson-baryon interaction in the low energy limit is model-independently given by $V_{\rm WT}$ according to chiral symmetry. The phenomenological success of the model with only $V_{\rm WT}$~\cite{Oset:1997it} indicates that the chiral symmetry constraint indeed works in reality. In order to deal with the precise experimental measurements, such as those from SIDDHARTA~\cite{Bazzi:2011zj,Bazzi:2012eq}, we need to increase the precision of the theoretical framework as well~\cite{Ikeda:2011pi,Ikeda:2012au,Guo:2012vv,Mai:2014xna}. This can be achieved by the inclusion of the next-to-leading order (NLO) terms $V_{\rm NLO}$ which are the contact interactions of $\mathcal{O}(p^{2})$. 

The scattering equation~\eqref{eq:scattering} is an integral equation reflecting the off-shell nature of the interaction kernel $V_{ij}$. It is, however, practically useful to adopt the on-shell factorized form which still satisfies the unitarity condition (see Refs.~\cite{Oset:1997it,Oller:2000fj,Hyodo:2011ur,Mai:2020ltx} for more details).
Because the leading Weinberg-Tomozawa term is a four-point contact interaction, the momentum integration in the scattering equation~\eqref{eq:scattering} diverges at ultraviolet. The ultraviolet divergence of the loop function $G$ is usually tamed by the dimensional regularization scheme. 
In this scheme, the finite part of the loop function $G_{i}$ is determined by the subtraction constant, which is related to the ultraviolet cutoff parameter~\cite{Oller:2000fj}. Because the meson-baryon loop function is counted as $\mathcal{O}(p^3)$, the renormalization procedure of the meson-baryon scattering in chiral perturbation theory is achieved at $\mathcal{O}(p^3)$. In the unitarized framework with the $\mathcal{O}(p^2)$ interaction ($V=V_{\rm WT}+V_{\rm Born}+V_{\rm NLO}$), therefore, the subtraction constants should be fixed by the experimental data.
As mentioned above, because the Weinberg-Tomozawa term $V_{\rm WT}$ is uniquely determined by chiral symmetry, there is no free parameter in $V_{\rm WT}$. The Born terms $V_{\rm Born}$ contain the axial vector coupling constants, usually denoted as $F$ and $D$. These are empirically determined by the axial charge of the nucleon and the hyperon nonleptonic decay. Thus, basically the 
subtraction constants are the free parameters in the leading order models of $\mathcal{O}(p^{1})$, where the terms $V_{\rm WT}$ and $V_{\rm Born}$ are used as $V$ in Eq.~\eqref{eq:scattering}. The NLO terms $V_{\rm NLO}$ contain seven contact terms with different momentum structure in the on-shell scheme, each of which has one LEC. The $\mathcal{O}(p^{2})$ models therefore contain seven additional free parameters on top of the 
subtraction constants. At present, the available experimental data of the $K^{-}p$ system can determine the LECs at $\mathcal{O}(p^{2})$, but is not sufficient to work in $\mathcal{O}(p^{3})$.

\subsubsection{\it Two resonance poles}

The pole of the scattering amplitude can be obtained by analytically continuing the scattering amplitude $T_{ij}$ to the complex energy plane. As we show in Section~\ref{sec:resonances}, the pole of the scattering amplitude corresponds to the eigenstate of the Hamiltonian of the system. In general, there is one pole in the energy region of one resonance, and the resonance mass $M_{R}$ and the width $\Gamma_{R}$ can be read off from the complex pole position $z$ as 
\begin{align}
   M_{R}
   =\re z,\quad
   \Gamma_{R}
   =-2\ \im z .
\end{align}
In the case of the $\Lambda(1405)$, it is reported in the PDG that there are two poles in the scattering amplitude with $I=0$ and $S=-1$ between the $\bar{K}N$ and $\pi\Sigma$ thresholds~\cite{Zyla:2020zbs}. One pole (high-mass pole) lies near the $\bar{K}N$ threshold with relatively small imaginary part. The other pole (low-mass pole) appears near the $\pi\Sigma$ threshold, and its imaginary part is large. This indicates that the ``$\Lambda(1405)$'' resonance is not a single state but is expressed by a superposition of two eigenstates. In fact, in the latest version of the PDG particle listings, the low-mass pole has been included as a two-star resonance the $\Lambda(1380)$, and the $\Lambda(1405)$ is used to mean the high-mass pole around 1420 MeV, although the traditional Breit-Wigner mass and width are still shown in the summary table section. In relation to the two-pole structure, one should note that there is only one resonance signature in the scattering amplitude (zero crossing of the real part and peak of the imaginary part), and this is \textit{not} realized as a two-peak structure. In other words, one peak structure of the $\Lambda(1405)$ spectrum is produced by the cooperative effect from the two eigenstate poles.

In chiral SU(3) dynamics, the existence of two poles was first reported in Ref.~\cite{Oller:2000fj}, and confirmed by many subsequent works (see the recent reviews~\cite{Meissner:2020khl,Mai:2020ltx}). The quantitative determination of the pole positions will be discussed in Section~\ref{subsec:determination}. The appearance of the two poles in this energy region has also been found in other approaches. For instance, an old study using the cloudy bag model found two poles in the $\Lambda(1405)$ energy region~\cite{Fink:1989uk}. Two poles were also found in the J\"ulich meson exchange model~\cite{Haidenbauer:2010ch}, in the dynamical coupled-channels model~\cite{Kamano:2014zba,Kamano:2015hxa}, and in Hamiltonian effective field theory~\cite{Liu:2016wxq}. These works use the scattering equation to obtain the dynamical scattering amplitude, but the interactions are constructed with different strategies. Thus, the two-pole nature of the $\Lambda(1405)$ is not a result specific to chiral dynamics.

The origin of the two poles has been studied in Ref.~\cite{Jido:2003cb}. The channel basis $(i,j)$ of the Weinberg-Tomozawa term $V_{\rm WT}$ can be transformed into those with SU(3) representations by the SU(3) Clebsch-Gordan coefficients. There are four channels with the $\Lambda(1405)$ quantum numbers, $\bm{1}$, $\bm{8}$, $\bm{8}^{\prime}$, and $\bm{27}$. Because $V_{\rm WT}$ is SU(3) symmetric, it becomes diagonal in the SU(3) basis. The interaction is attractive in the $\bm{1}$, $\bm{8}$ and $\bm{8}^{\prime}$ channels, and each channel forms a bound state, when the SU(3) symmetric hadron masses are used. One of these three bound states (octet) evolves into the $\Lambda(1670)$ resonance, and the other two (singlet and octet) evolve into the two resonance poles of the $\Lambda(1405)$, along with the gradual breaking of SU(3) symmetry towards the physical point. Therefore, the origin of the two poles in the $\Lambda(1405)$ region is attributed to the two attractive components in the Weinberg-Tomozawa term. The same argument holds in more physical isospin basis~\cite{Hyodo:2007jq}, where the relevant channels are $\pi\Sigma$, $\bar{K}N$, $\eta\Lambda$ and $K\Xi$. In this basis, $V_{WT}$ has off-diagonal couplings which represent the channel transition, and the diagonal components in $\pi\Sigma$, $\bar{K}N$ and $K\Xi$ are attractive. In the absence of the off-diagonal channel coupling, the $\bar{K}N$ attraction provides a bound state below the threshold, and the $\pi\Sigma$ attraction generates a resonance above the threshold. In this way, two eigenstates are produced between the $\pi\Sigma$ and $\bar{K}N$ thresholds, and they evolve into the low-mass pole and the high-mass pole through the channel coupling. Another attraction in the $K\Xi$ channel is the origin of the $\Lambda(1670)$. As mentioned above, the property of the Weinberg-Tomozawa term is determined by chiral symmetry, and therefore the appearance of two attractive interactions in the $\Lambda(1405)$ region is a model-independent consequence of chiral symmetry. 

The two-pole structure has its significance in hadron spectroscopy, because it is related to the number of eigenstates in this sector. For instance, the classification of baryon resonances into the SU(3) multiplets is highly affected by the number of states in a given energy region (unless the state purely belongs to a singlet). Namely, the existence of an excited $\Lambda$ state in the octet representation indicates that there should also be its flavor partners having the same $J^{P}$ in the same energy region. Although the determination of the pole positions does not immediately give information on the internal structure of the resonance (three-quark state, meson-baryon molecule, or something else), it is an important starting point of the discussion of the internal structure. At the same time, the two-pole structure also has some observable implications. In general, a resonance state can have different coupling strengths for various channels. It was shown in Ref.~\cite{Jido:2003cb} that the high-mass pole of the $\Lambda(1405)$ strongly couples to $\bar{K}N$ and the low-mass pole has a large coupling to $\pi\Sigma$, as expected from their origin in the isospin basis. This means that the scattering amplitudes $T(\bar{K}N\to\pi\Sigma)$ and $T(\pi\Sigma\to\pi\Sigma)$ are influenced by the two poles with different weights, and show different behaviors~\cite{Jido:2003cb}. As a consequence, the lineshape of the $\Lambda(1405)$ can be reaction-dependent, because the relative weights of $T(\bar{K}N\to\pi\Sigma)$ and $T(\pi\Sigma\to\pi\Sigma)$ depend on the reaction mechanism. In fact, as shown in Section~\ref{sec:L1405_lineshape}, there are sizable deviations of the lineshape of the $\Lambda(1405)$ with different reaction processes. At the same time, we should keep in mind that the nonresonant background term and the isospin interference can also modify the shape of the spectrum. Comparison with experimental data will be discussed in the next section.

\subsubsection{\it Determination of pole positions}
\label{subsec:determination}

Here we review the recent studies to pin down the pole positions of the $\Lambda(1405)$, focusing on the works adopted in the PDG~\cite{Ikeda:2011pi,Ikeda:2012au,Guo:2012vv,Mai:2014xna} which performed the uncertainty analysis using the NLO chiral SU(3) dynamics with the SIDDHARTA constraint. To determine the pole positions quantitatively, sufficient accuracy of the experimental data is required. Currently available data can be classified as follows: 
\begin{itemize}
\item[(i)] the total cross sections of the $K^{-}p$ scattering (elastic and inelastic channels)~\cite{Abrams:1965zz,Sakitt:1965kh,Kim:1965zzd,Csejthey-Barth:1965izu,Mast:1975pv,Bangerter:1980px,Ciborowski:1982et,Evans:1983hz}, 
\item[(ii)] the threshold branching ratios~\cite{Tovee:1971ga,Nowak:1978au}, 
\item[(iii)] level shift and width of the kaonic hydrogen~\cite{Bazzi:2011zj,Bazzi:2012eq} (Section~\ref{sec:KbarNscattering}), and
\item[(iv)] the $\pi\Sigma$ invariant mass distribution in various reaction (Section~\ref{sec:L1405_lineshape}).
\end{itemize}
The digitalized data can be found in the GitHub repository by Mai~\cite{Maidata}.
The total cross section $\sigma_{ij}$ (from channel $j$ to channel $i$) at the total energy $\sqrt{s}$ can be calculated from the scattering amplitude $T_{ij}$ as
\begin{align}
   \sigma_{ij}(\sqrt{s})
   \propto |T_{ij}(\sqrt{s})|^{2} ,
   \label{eq:crosssection}
\end{align}
where the proportionality constant is given by the phase space factor. The branching ratios (ii) are basically the ratios of the cross sections to various final states at the $K^-p$ threshold, $ \sigma_{ij}(\sqrt{s}=m_{K^{-}}+M_{p})$. Thus, the data (i) and (ii) can be directly related to the theoretical meson-baryon scattering amplitude square $|T_{ij}|^{2}$. The level shift $\Delta E$ and width $\Gamma$ of the kaonic hydrogen measurement (iii) are related to the complex $K^{-}p$ scattering length $a_{K^{-}p}$ by the improved Deser formula~\cite{Meissner:2004jr,Meissner:2006gx}
\begin{align}
   \Delta E-\frac{i\Gamma}{2}
   =-2\mu^{2}\alpha^{3}a_{K^{-}p}[1-2\mu\alpha(\ln\alpha-1)a_{K^{-}p}]
   +\dotsb ,
   \label{eq:ImprovedDT}
\end{align}
where $\mu$ is the reduced mass of the $K^{-}p$ system and $\alpha$ is the electromagnetic fine structure constant. The scattering length is theoretically given by the diagonal $K^{-}p$ scattering amplitude $T_{ij}$ at threshold. In this way, the data (i), (ii), and (iii) are related to the two-body scattering amplitude $T_{ij}$, and therefore can be used as direct experimental constraints on $T_{ij}$. However, the $\pi\Sigma$ spectra (iv) cannot be calculated solely from $T_{ij}$. Because only the $\pi\Sigma$ channels are kinematically open at the energy of the $\Lambda(1405)$, a possible way to determine the $\pi\Sigma$ amplitude is the low-energy $\pi\Sigma$ elastic scattering experiment, which is not accessible by the 
current experimental technique. Instead, the $\pi\Sigma$ pair in the $\Lambda(1405)$ energy region is experimentally obtained by some reaction processes, such as $\gamma p\to K^{+}(\pi\Sigma)$ and $pp\to K^{+}p(\pi\Sigma)$, as presented by Section~\ref{sec:L1405_lineshape}. In this case, the invariant mass distribution is schematically given by
\begin{align}
   \sigma_{j}(M_{I})
   \propto \left|\sum_{i}C_{i}G_{i}(M_{I})T_{ij}(M_{I})\right|^{2} ,
   \label{eq:Minv}
\end{align}
where $M_{I}$ is the invariant mass of the $\pi\Sigma$ pair in the final state (channel $j$). The coefficient $C_{i}$ determines the relative weight of the initial channel $i$, which can depend on various kinematics (initial energy, scattering angle, $M_{I}$, etc.) as well as the reaction itself. This procedure introduces additional uncertainty in the analysis. Hence, the $\pi\Sigma$ spectra (iv) are not the direct constraints on $T_{ij}$. One can either construct reaction models to determine $C_{i}$ explicitly~\cite{Nacher:1998mi,Nacher:1999ni,Hyodo:2003jw,Hyodo:2004vt,Geng:2007vm} or parametrize $C_{i}$~\cite{Oller:2000fj,Roca:2013av,Roca:2013cca} to determine $T_{ij}$. Very recently, the $K^{-}p$ correlation function from the high-energy collisions has been measured by the ALICE collaboration~\cite{Acharya:2019bsa}. The correlation function is obtained at very low energies, with much better precision than the old cross section measurements. With the developments of the theoretical calculation of the correlation function~\cite{Kamiya:2019uiw}, the correlation function data will bring new constraints on the $\Lambda(1405)$ in future.

\begin{figure}[tbp]
\begin{center}
  \figureBB{\includegraphics[width=9cm,bb=0 0 537 501]{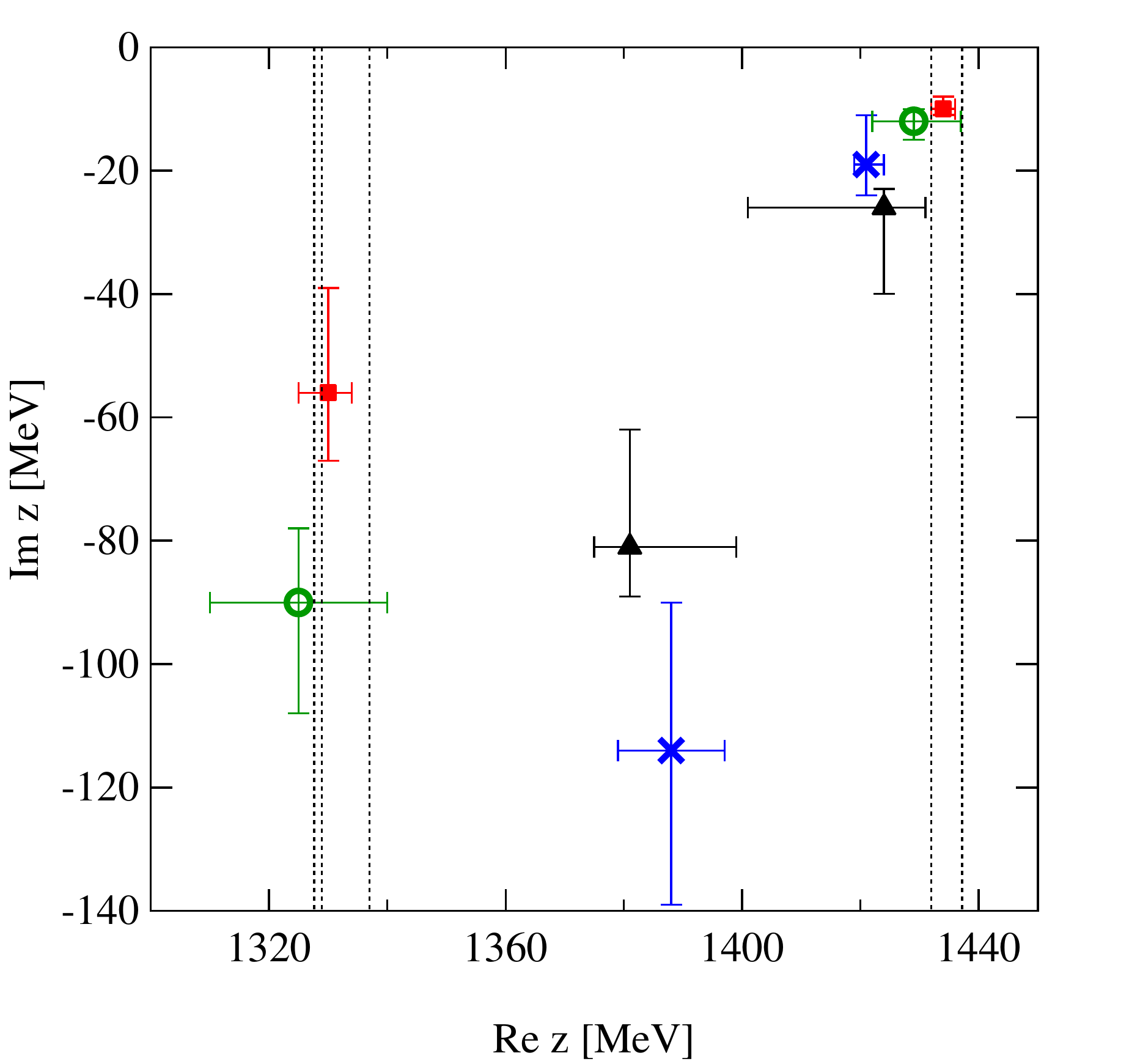}}
  {\includegraphics[width=9cm]{figs/poles.pdf}}
\begin{minipage}[t]{16.5 cm}
\caption{
Pole structure of the $\Lambda(1405)$ region~\cite{Zyla:2020zbs}. Filled triangles, 
crosses, filled squares, and open circles represent the results of Refs.~\cite{Ikeda:2011pi,Ikeda:2012au} (NLO model), Ref.~\cite{Guo:2012vv} (Fit II), Ref.~\cite{Mai:2014xna} (Solution \#2), and Ref.~\cite{Mai:2014xna} (Solution \#4), respectively. Dotted lines stand for the threshold energies of meson-baryon channels, $\pi^{0}\Sigma^{0}$,  $\pi^{-}\Sigma^{+}$,  $\pi^{+}\Sigma^{-}$, $K^{-}p$, $\bar{K}^{0}n$ (from left to right).
\label{fig:poles}}
\end{minipage}
\end{center}
\end{figure}

Currently, the PDG tabulates four sets of the pole positions of the $\Lambda(1405)$~\cite{Zyla:2020zbs}, obtained from the analyses in Refs.~\cite{Ikeda:2011pi,Ikeda:2012au,Guo:2012vv,Mai:2014xna}. In Refs.~\cite{Ikeda:2011pi,Ikeda:2012au}, systematic $\chi^{2}$ fitting with the uncertainty analysis was performed in the next-to-leading order (NLO) chiral SU(3) dynamics. The direct experimental constraints (i), (ii), and (iii) were used to fix the free parameters. It is shown that the SIDDHARTA measurement of the kaonic hydrogen is consistent with the cross section data (i) and (ii), in contrast to the previous DEAR measurement~\cite{Beer:2005qi} which causes some controversy~\cite{Borasoy:2004kk,Borasoy:2005ie,Oller:2005ig,Borasoy:2005fq,Oller:2006ss}. In addition, the SIDDHARTA measurement turns out to give a stringent constraint on the $\Lambda(1405)$. By comparing with the almost same analysis but without the kaonic hydrogen data (iii)~\cite{Borasoy:2006sr}, the constraint from the SIDDHARTA measurement significantly reduces the uncertainty in the extrapolation of $T_{ij}$ below the $\bar{K}N$ threshold where the $\Lambda(1405)$ exists. It is also shown that the model with only the Weinberg-Tomozawa term $V_{\rm WT}$ works reasonably well, when the cutoff parameters are properly adjusted. This supports the phenomenological success of such models in earlier studies. The work in Ref.~\cite{Guo:2012vv} used the $K^{-}p\to \eta\Lambda$ cross sections~\cite{Starostin:2001zz} and the $\pi\Lambda$ phase shift at the $\Xi^{-}$ mass~\cite{Huang:2004jp,Chakravorty:2003je} in addition to (i), (ii), and (iii). The $\pi\Sigma$ invariant mass distributions from the $\Sigma^{+}(1660)\to \pi^{+}\pi^{-}\Sigma^{+}$ decay~\cite{Hemingway:1984pz} and the $K^{-}p\to \pi^{0}\pi^{0}\Sigma^{0}$ reaction~\cite{Prakhov:2004an} were also used to constrain the model. The treatment of the meson decay constants was studied in detail. Reference~\cite{Mai:2014xna} included the $\pi\Sigma$ spectra by the CLAS photoproduction data via Eq.~\eqref{eq:Minv} in the analysis. It is shown that eight solutions can be found with the constraints (i), (ii), and (iii), and the inclusion of the $\pi\Sigma$ spectra rules out six of them~\cite{Mai:2014xna}. The resulting pole positions in the complex energy plane from these analyses~\cite{Ikeda:2011pi,Ikeda:2012au,Guo:2012vv,Mai:2014xna} are shown in Table~\ref{tab:poles} and plotted in Fig.~\ref{fig:poles}. In all cases, two poles are found in this energy region. The position of the high-mass pole (which locates near the $\bar{K}N$ threshold) is converging in a small region, thanks to the strong constraint from the SIDDHARTA data at the $\bar{K}N$ threshold. In contrast, there exists sizable uncertainty in the position of the low-mass pole.

\begin{table}
\begin{center}
\begin{minipage}[t]{16.5 cm}
\caption{Pole structure of the $\Lambda(1405)$ region~\cite{Zyla:2020zbs}. 
\label{tab:poles}}
\end{minipage}
\begin{tabular}{lll}
\hline
approach & high-mass pole [MeV] & low-mass pole [MeV] \\ \hline
Refs.~\cite{Ikeda:2011pi,Ikeda:2012au} NLO 
  & $1424^{+7}_{-23}- i 26^{+3}_{-14}$
  &  $1381^{+18}_{-6}- i 81^{+19}_{-8}$ \\
Ref.~\cite{Guo:2012vv} Fit II 
  & $1421^{+3}_{-2}- i 19^{+8}_{-5}$ 
  & $1388^{+9}_{-9}- i 114^{+24}_{-25}$ \\
Ref.~\cite{Mai:2014xna} solution \#2
  & $1434^{+2}_{-2} - i \, 10^{+2}_{-1}$ 
  & $1330^{+4~}_{-5~} - i \, 56^{+17}_{-11}$\\
Ref.~\cite{Mai:2014xna} solution \#4
  & $1429^{+8}_{-7} - i \, 12^{+2}_{-3}$ 
  & $1325^{+15}_{-15} - i \, 90^{+12}_{-18}$\\
\hline
\end{tabular}
\end{center}
\end{table}   

Let us briefly introduce some related works. A nonrelativistic model with separable potentials were constructed in Ref.~\cite{Cieply:2011nq} including the SIDDHARTA constraint. Nuclear medium effects on the $\bar{K}N$ amplitude were also discussed. A separable potential model was further studied in Ref.~\cite{Shevchenko:2012np}. Under the SIDDHARTA constraint, two potentials $V^{1,\text{SIDD}}$ and $V^{2,\text{SIDD}}$ are constructed, where the former generates one pole for the $\Lambda(1405)$ and the latter describes the two-pole $\Lambda(1405)$. The pole positions of the $\Lambda(1405)$ are found to be $1426-i 48$ MeV in the one-pole model and $1414-i58$ MeV and $1386-i104$ MeV in the two-pole model. The partial wave analysis (PWA) was performed for the low-energy meson-baryon scattering in Refs.~\cite{Anisovich:2019exw,Anisovich:2020lec}, where two-pole and one-pole solutions were obtained. In the one-pole solution, the central value of the pole position is at $1421-23i$ MeV, while in the two-pole solution, the results are at $1423-20i$ MeV and at $1380-90i$ MeV~\cite{Anisovich:2020lec}. Note however that the resonance parameters are determined by fitting the Breit-Wigner parametrization to the photoproduction data, rather than searching for the poles of the scattering amplitude in the complex energy plane. The results in Refs.~\cite{Shevchenko:2012np,Anisovich:2020lec} indicate that the position of the pole near the $\bar{K}N$ threshold is no longer at 1405 MeV, even in the model without the low-mass pole. The CLAS photoproduction data was analyzed in Refs.~\cite{Roca:2013av,Roca:2013cca} using the Weinberg-Tomozawa interaction with fine tuning of the coupling strengths. The NLO chiral dynamics was also used to study the higher energy region, including the $K^{-}p\to K\Xi$ reaction~\cite{Feijoo:2015yja}. All these studies~\cite{Roca:2013av,Roca:2013cca,Feijoo:2015yja} found two poles in the $\Lambda(1405)$ region, along the same line with Fig~\ref{fig:poles}. See also the comparison of several models in Ref.~\cite{Cieply:2016jby} which also studied the fate of poles in the zero-coupling limit. The dynamical coupled-channels model in Refs.~\cite{Kamano:2014zba,Kamano:2015hxa} is the most systematic analysis for the hyperon resonances above the $\bar{K}N$ threshold. This analysis, however, did not use the SIDDHARTA constraint in the analysis. Nevertheless, there are two poles in the $\Lambda(1405)$ region.

Several recent studies have examined the approximations commonly adopted in the calculation of chiral SU(3) dynamics. Let us discuss three issues, namely,
\begin{itemize}

  \item $p$-wave contribution,
  \item on-shell factorization, and
  \item improved Deser formula~\eqref{eq:ImprovedDT}.
\end{itemize}
First, most of the previous analyses used the $s$-wave meson-baryon scattering amplitude for the fitting to the cross section data. At low energy, the dominant contribution should be given by the $s$-wave amplitude, but as the energy increases, the higher partial waves can contribute to the cross sections. In addition, the angular dependence of the differential cross sections cannot be reproduced by the $s$-wave amplitude. The NLO analysis was performed with an explicit $p$-wave $J^{P}=1/2^{+}$ meson-baryon amplitude in Ref.~\cite{Sadasivan:2018jig} (see also Ref.~\cite{Jido:2002zk} for the inclusion of the $p$-wave contributions). When the $\Sigma(1385)$ resonance in the $J^{P}=3/2^{+}$ component was explicitly included, the central values of the pole positions of the $\Lambda(1405)$ were found to be $1430-i15$ MeV and $1364-i43$ MeV~\cite{Sadasivan:2018jig}. Because the result is in fair agreement with those in Table~\ref{tab:poles}, the $p$-wave contribution in the cross section does not affect the pole determination of the $\Lambda(1405)$ very much. 
Second, in most cases, an algebraic form of the scattering equation~\eqref{eq:scattering} was used instead of the integral equation. Such amplitude is obtained by applying the on-shell factorization to the integral equation~\cite{Oset:1997it}, or by using the formulation of the N/D method with the unitarity constraint~\cite{Oller:2000fj} (see also Ref.~\cite{Hyodo:2011ur}). In Ref.~\cite{Revai:2017isg} it is argued that the two-pole structure of the $\Lambda(1405)$ is caused by the on-shell factorization. 
It should however be noted that the counter examples to this statement have already been published before Ref.~\cite{Revai:2017isg}. 
Namely, Refs.~\cite{Nieves:2001wt,GarciaRecio:2002td,Mai:2012dt} and a recent work~\cite{Morimatsu:2019wvk} found two poles in the relevant energy region without the on-shell factorization.
In addition, the specific model used in Ref.~\cite{Revai:2017isg} was shown to violate the chiral low-energy theorem~\cite{Bruns:2019bwg}. It was also shown in Ref.~\cite{Bruns:2019bwg} that, by improving the model to satisfy the chiral symmetry constraint, two poles were generated for the $\Lambda(1405)$. Thus, the two-pole structure of the $\Lambda(1405)$ is not directly related to the algebraic form of the scattering amplitude (on-shell factorization).
Third, to relate the kaonic hydrogen measurement with the $K^{-}p$ scattering length, the improved Deser formula~\cite{Meissner:2004jr} shown in Eq.~\eqref{eq:ImprovedDT} was commonly adopted. This formula is useful, because the shift and width are directly related to the scattering length in a model-independent manner. Although the formula is systematically derived using the effective field theory,  some higher order corrections are neglected. A part of such corrections can be incorporated by the resummation of a class of diagrams to all orders, leading to the resummed formula~\cite{Baru:2009tx}
\begin{align}
   \Delta E-\frac{i\Gamma}{2}
   =-\frac{2\mu^{2}\alpha^{3}a_{K^{-}p}}{1+2\mu\alpha(\ln\alpha-1)a_{K^{-}p}}
   +\dotsb .
   \label{eq:resummedDT}
\end{align}
To check the validity of these formulas, it is necessary to introduce some $\bar{K}N$ potential equivalent to chiral SU(3) dynamics, and solve the Schr\"odinger equation with both Coulomb and strong interactions. Such analysis was performed in Ref.~\cite{Hoshino:2017mty}, using an equivalent local potential in coordinate space~\cite{Miyahara:2015bya} based on the scattering amplitude in Refs.~\cite{Ikeda:2011pi,Ikeda:2012au}. It is shown that the deviation of the result in Eq.~\eqref{eq:ImprovedDT} from the exact solution is 10-11 eV, and that in the resumed formula~\eqref{eq:resummedDT} is of the order of eV in the kaonic hydrogen.\footnote{Note however that the deviation is as large as a few hundreds of eV in the kaonic deuterium.} This indicates that the use of the improved Deser formula~\eqref{eq:ImprovedDT} to extract the $K^{-}p$ scattering length, in particular its resummed version~\eqref{eq:resummedDT}, is quantitatively verified, given the systematic uncertainty of 36 eV in the SIDDHARTA measurement.
 
\subsubsection{\it Predictions of the $\Lambda(1405)$ spectra}
\label{sec:prediction}

By combining the meson-baryon scattering amplitude with some reaction models, it is possible to predict the $\pi\Sigma$ invariant mass spectrum which can be compared with experiments directly. In fact, this strategy was adopted to analyse the experimental data of photoproductions and $pp$ collisions, as discussed in Section~\ref{sec:L1405_lineshape}. Here we present examples of the predictions of $\pi\Sigma$ spectra, which will be tested in forthcoming experiments. 

Let us first consider the $K^{-}$ induced reaction with a deuteron  target, $K^{-}d\to n(\pi\Sigma)$ where the $\Lambda(1405)$ can be seen in the $\pi\Sigma$ spectrum. This reaction was experimentally studied in 1970s by the $K^{-}d$ capture at rest in Ref.~\cite{Tan:1973at}, and by the in-flight experiment with beam momenta of 686-844 MeV$/c$ in Ref.~\cite{Braun:1977wd}. The in-flight reaction with 1 GeV/c $K^{-}$ beam is currently performed by the J-PARC E31 collaboration. 
Because the deuteron is the bound state of a $pn$ pair, theoretical investigation of the $K^{-}d\to n(\pi\Sigma)$ reaction requires the construction of three-body to three-body ($K^{-}pn\to \pi \Sigma n$) amplitude. In order to take into account the complete three-body dynamics, it is necessary to employ the Faddeev type three-body equations. In practice, truncated two-step approaches are commonly adopted, where only the single scattering process (impulse approximation) plus meson exchange contributions are included. Because the kinematics of the reaction varies with the initial beam momenta, theoretical studies have been performed for each experimental condition. The stopped $K^{-}$ reaction of Ref.~\cite{Tan:1973at} was studied in Ref.~\cite{Revai:2012fx} using Faddeev equations. Neutron spectra with the incident kaon energy of $E_{K^{-}}^{\rm cm}=1$-50 MeV were presented. With the two-step approach, Ref.~\cite{Jido:2010rx} studied the $K^{-}d$ reaction with the kaons from the $\phi$ decay (incident $K^{-}$ momentum of 80-130 MeV/c), having possible experiment at DAFNE in mind. The in-flight kaon reaction of Ref.~\cite{Braun:1977wd} was studied in several works with two-step approach~\cite{Jido:2009jf,Miyagawa:2012xz,Jido:2012cy,YamagataSekihara:2012yv}. In particular, the treatment of the intermediate kaon propagator in the two-step process has been discussed in detail~\cite{Miyagawa:2012xz,Jido:2012cy}. 
Contribution from the $p$-wave $\Sigma(1385)$ and the pion exchange diagrams were examined in Ref.~\cite{YamagataSekihara:2012yv}. 
In-flight reaction with J-PARC E31 kinematics was studied in a full three-body calculation using the coupled-channel Alt-Grassberger-Sandhas (AGS) equations~\cite{Ohnishi:2015iaq}. The $\pi\Sigma$ spectra are shown to be sensitive to the $\bar{K}N$ interaction employed. A pronounced maximum of the spectrum was found at around 1450 MeV. Within the two-step approach, Ref.~\cite{Kamano:2016djv} studied this reaction with the $\bar{K}N$ interaction from the dynamical coupled-channels model~\cite{Kamano:2015hxa}. It was pointed out that the high momentum of the initial kaon requires the $\bar{K}N$ amplitude above the threshold. Full Faddeev type calculation was performed in Ref.~\cite{Miyagawa:2018xge}, also in comparison with the preliminary data of E31 experiment.

Recently, three-body weak decays of heavy hadrons are found to be useful to study hadron resonances through the final state interactions (see the review~\cite{Oset:2016lyh}). The reactions involving the $\Lambda(1405)$ have also been studied in several works. The reaction $\Lambda_b \rightarrow J/\psi (\pi \Sigma)$ was studied in Ref.~\cite{Roca:2015tea} (subsequent studies can be found in Refs.~\cite{Feijoo:2015cca,Feijoo:2018den}). In Ref.~\cite{Miyahara:2015cja}, the $\Lambda_{c}\to \pi^{+}(\pi \Sigma)$ decay was discussed with the same mechanism. In both cases, the heavy $b$ or $c$ quark in the initial baryon decays weakly, and a $q\bar{q}$ pair is created to form the three-body states $J/\psi MB$ or $\pi^{+}MB$. The $MB\to \pi\Sigma$ amplitude is then included to generate the $\Lambda(1405)$ in the final state interaction. An interesting observation is that the intermediate $MB$ pair is produced purely in the isospin $I=0$ combination, due to the favored mechanisms in the weak decay process. This implies that a clean signal of the $\Lambda(1405)$ is expected in these processes, where the $I=1$ contamination would be neglected. This isospin filtering effect is more evident in the semileptonic decay $\Lambda_c\to \nu l^+\pi\Sigma$~\cite{Ikeno:2015xea} where the leptons do not carry isospin, although the identification of the semileptonic decay is experimentally more challenging. Similar investigations have been performed for the $\chi_{c0}(1P)\to \bar{\Lambda}(\pi\Sigma)$ decay~\cite{liu:2017efp}, the $\Lambda^0_b \to \eta_c  (\pi \Sigma)$ decay~\cite{Xie:2017gwc}, and the $\Xi_{b}\to D^{0}(\pi\Sigma)$ decay~\cite{Miyahara:2018lud}.

\subsubsection{\it Lattice QCD}

The first principle lattice QCD calculation is a powerful tool to study the hadron spectrum. The masses of the ground states are well reproduced, and the excited states are now described as resonances in the hadron-hadron scattering, thanks to the recent developments (see Section~\ref{sec:lattice}). Concerning the $\Lambda(1405)$, unfortunately, such scattering calculation has not been achieved yet, and the properties of the $\Lambda(1405)$ have been studied by the traditional effective mass techniques with the two-point function. Here we introduce recent lattice QCD investigations on the $\Lambda(1405)$. Detailed account of the previous works~\cite{Melnitchouk:2002eg,Nemoto:2003ft,Lee:2005mr,Burch:2006cc,Ishii:2007ym,Takahashi:2009bu} can be found in Ref.~\cite{Hyodo:2011ur}.

Lattice QCD study for the $\Lambda(1405)$ with near physical point was performed in Ref.~\cite{Menadue:2011pd} in which the lowest quark mass corresponds to the pion mass of 156 MeV. By diagonalizing the cross correlation function with different three-quark interpolating fields, the lowest negative parity $\Lambda$ state was found around 1.5 GeV which can be identified with the $\Lambda(1405)$. Such low-mass the $\Lambda(1405)$ on the lattice was confirmed by the BGR collaboration~\cite{Engel:2012qp} where the $J^{P}=1/2^{\pm}$ and $3/2^{\pm}$ $\Lambda$ states were systematically studied. Flavor SU(3) decomposition of the states was also performed. Further study on the $\Lambda$ spectrum was performed in Ref.~\cite{Gubler:2016viv}, by varying the mass of the heavy valence quark from strangeness to charm, and by analyzing the flavor SU(3) components. The internal structure of the $\Lambda(1405)$ was also studied on the lattice. In Ref.~\cite{Hall:2014uca}, the magnetic form factor of the $\Lambda(1405)$ was measured at $Q^{2}\simeq 0.16$ GeV$^{2}$ using the setup of Ref.~\cite{Menadue:2011pd}. It is found that the strange quark contribution to the magnetic form factor is almost vanishing near the physical point, while it is comparable with those from the up and down quarks at the heavier quark mass region. In nonrelativistic quantum mechanics, the magnetic moment of a composite system stems from the spin of the constituent particle and/or the angular momentum from the orbital motion of the constituents. In the $\bar{K}N$ molecule component, the strange quark is contained in the spinless antikaon, and there is no orbital motion of $\bar{K}$ because the $\bar{K}N$ system is combined in $s$ wave. This means that the vanishing of the strange quark contribution to the magnetic form factor near the physical point is consistent with the $\bar{K}N$ molecular picture for the $\Lambda(1405)$.

As mentioned above, the scattering calculation on the lattice will be the key to ultimately understand the $\Lambda(1405)$ in QCD. Because the lattice calculation is performed in a finite space-time volume, the energy eigenvalues are discretized due to the boundary conditions. It is theoretically possible to predict the spectrum of the finite-volume energy levels by employing some scattering models. The finite volume spectrum of the $\Lambda(1405)$ sector has been studied with the nonrelativistic EFT~\cite{Lage:2009zv},
with the J\"ulich hadron-exchange model~\cite{Doring:2011ip}, with the Hamiltonian effective field theory~\cite{Liu:2016wxq}, and with the chiral SU(3) dynamics~\cite{MartinezTorres:2012yi,Molina:2015uqp,Tsuchida:2017gpb}. In Ref.~\cite{Molina:2015uqp}, the finite volume spectrum of the lattice data in Ref.~\cite{Menadue:2011pd} was analyzed.
In general, a resonance signature is indicated by the appearance of an additional energy level on top of the shifted scattering states in finite volume. It should, however, be noted that the number of the additional level does not reflect the poles in the complex energy plane in the infinite volume, but rather determined by the behavior of the scattering amplitude on the real energy axis~\cite{Tsuchida:2017gpb}. In the case of the $\Lambda(1405)$, therefore, there appears only one additional energy level in finite volume, even though there are two poles in the complex energy plane in infinite volume~\cite{Tsuchida:2017gpb}. These studies, combined with the accurate lattice QCD calculation, will provide further information on the nature of the $\Lambda(1405)$ in future.

\subsection{\it The excited $\Lambda$ states above the $\bar{K}N$ threshold}

In the energy region from 1600~MeV to 1700~MeV, the PDG lists three $\Lambda$ resonances:
the $\Lambda(1600)$ with $J^P=1/2^+$, 
the $\Lambda(1670)$ with $J^P=1/2^-$, and
the $\Lambda(1690)$ with $J^P=3/2^-$. In this energy region where the $\bar{K}N$ channel is open, the $K^{-}p$ scattering data can be used to study these resonances.
The pole positions and spin-parity were identified using partial-wave-analysis of the $\overline{K}N$ 
scattering including multichannels in the final states~\cite{Kamano:2015hxa,Sarantsev:2019xxm}.
These states have the same spin-parity quantum numbers with the ground state $\Lambda$, 
the $\Lambda(1405)$ and the $\Lambda(1520)$, respectively, 
and can be radial excitation of these states. However, the internal structures of these are not understood well.
Among them, the $\Lambda(1670)$ appears with a narrow width ($\sim 30$~MeV) compared with
the others, $\sim 200$~MeV for the $\Lambda(1600)$ and $\sim 70$~MeV for the $\Lambda(1690)$.

A theoretical calculation based on a quark model~\cite{PhysRevD.21.1868} assigns the $\Lambda(1670)$ as
an SU(3) octet partner of $N(1535)$. Another
calculation using the meson-baryon molecule model~\cite{Oset:2001cn} describes it as a 
$K\Xi$ bound state.
Since the $\Lambda(1670)$ peaks just above the $\eta \Lambda$ threshold (1663~MeV),
it may be a cusp which is generated due to a strong coupling between the $\bar{K}N$ and $\eta \Lambda$
channels.

Recently, the Belle collaboration observed a narrow peak structure near 1670~MeV
in the invariant mass
of $K^-p$ pairs which were produced
in the three-body decay of $\Lambda_c^+ \rightarrow K^- p \pi^+$
~\cite{PhysRevLett.117.011801,Tanida:2019cif}.
After this observation, they searched for the resonant substructure of the 
$\Lambda_{c}^{+} \rightarrow \eta \Lambda \pi^{+}$ decay~\cite{Lee:2020xoz}. 
By selecting $\Lambda \eta$ pairs, pure isospin-zero amplitudes were investigated, and indeed,
a prominent peak of the $\Lambda(1670)$ was observed. 
The mass and width parameters are determined precisely to be 
$1674.3 \pm 0.8 \pm 4.9$~MeV and $36.1 \pm 2.4 \pm 4.8$~MeV, respectively.
They also measured relative branching fractions of 
$\Lambda_{c}^{+} \rightarrow \eta \Sigma^{0} \pi^{+}$ and $\Lambda(1670) \pi^{+}$, and 
$\eta \Sigma(1385)^{+}$ as
\begin{align*}
\frac{\mathcal{B}\left(\Lambda_{c}^{+} \rightarrow \eta \Lambda \pi^{+}\right) }
{ \mathcal{B}\left(\Lambda_{c}^{+} \rightarrow p K^{-} \pi^{+}\right)} 
&= 0.293 \pm 0.003 \pm 0.014, \\
\frac{\mathcal{B}\left(\Lambda_{c}^{+} \rightarrow \eta \Sigma^{0} \pi^{+}\right) }{\mathcal{B}\left(\Lambda_{c}^{+} \rightarrow p K^{-} \pi^{+}\right)}
&=0.120 \pm 0.006 \pm 0.006, \\
\frac{\mathcal{B}\left(\Lambda_{c}^{+} \rightarrow \Lambda(1670) \pi^{+}\right) \times \mathcal{B}(\Lambda(1670) \rightarrow \eta \Lambda) }{ \mathcal{B}\left(\Lambda_{c}^{+} \rightarrow p K^{-} \pi^{+}\right)}
&=(5.54 \pm 0.29 \pm 0.73) \times 10^{-2} , \\
\frac{\mathcal{B}\left(\Lambda_{c}^{+} \rightarrow \eta \Sigma(1385)^{+}\right) }{ \mathcal{B}\left(\Lambda_{c}^{+} \rightarrow p K^{-} \pi^{+}\right)}
&=0.192 \pm 0.006 \pm 0.016 .
\end{align*}
The three-body decay of the $\Lambda_c^+ \rightarrow K^- p \pi^+$ and $\Lambda_{c}^{+} \rightarrow \eta \Lambda \pi^{+}$ processes was theoretically studied in Ref.~\cite{Miyahara:2015cja} using the final state interaction model of Ref.~\cite{Oset:2001cn}. A peak structure of the $\Lambda(1670)$ in the $K^{-}p$ and $\eta\Lambda$ spectra was predicted~\cite{Miyahara:2015cja}. In Ref.~\cite{Ahn:2019rdr}, the $\Lambda_c^+ \rightarrow K^- p \pi^+$ and $\Lambda_c^+ \rightarrow K^0_{s} p\pi^0$ decays were studied by the effective Lagrangian approach. By introducing the several resonance contributions in the final state interaction, the Dalitz plot density was calculated in comparison with that of Ref.~\cite{PhysRevLett.117.011801}. The interpretation of the narrow peak structure observed by the Belle collaboration was discussed.

Among the various PWA results, the authors of 
Refs.~\cite{Kamano:2014zba,Kamano:2015hxa} 
found a possible evidence of a narrow $J^P=3/2^+$ $\Lambda$ resonance around 1670~MeV in
their ``Model B'' in addition to the ordinary $J^P=1/2^-$ resonance.
The new resonance cannot be established only by considering the total 
cross sections; it is necessary to examine its effects on the angular distributions of
differential cross sections
of the $K^- p \rightarrow \eta \Lambda$ reaction. 
A new experiment, J-PARC E72, was proposed to provide high precision data
of the differential cross sections of
$K^- p \rightarrow \eta \Lambda$ reaction to elucidate the resonance contributions
around 1670~MeV~\cite{J-PARC_P72}. 
In the near future, updated PWA including new data will clarify hyperon spectroscopy
in this energy region.

\section{$S=-2$ baryons}\label{sec:S-2}

\begin{figure}[tbp]
\begin{center}
  \figureBB{\includegraphics[width=15cm,bb=0 0 502 464]{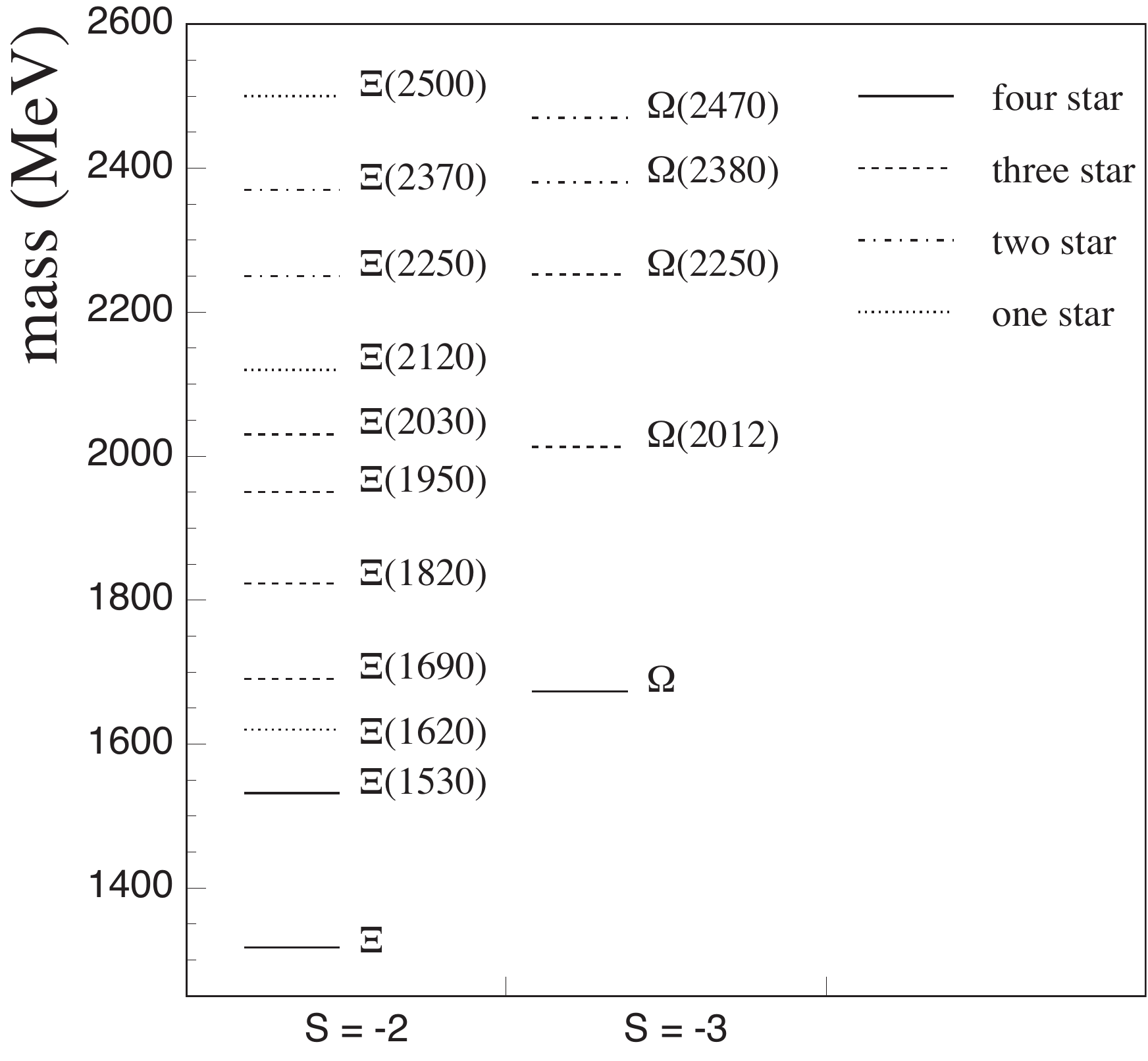}}
  {\includegraphics[width=15cm]{figs/mass_hyp2.pdf}}
\begin{minipage}[t]{16.5 cm}
\caption{\label{fig:mass_hype2}
The mass spectrum of 
$S=-2$ and $-3$ hyperons listed by the PDG~\cite{Zyla:2020zbs}.
}
\end{minipage}
\end{center}
\end{figure}

In contrast to the $S=-1$ baryons, less is known about $S=-2$ and $-3$
baryons. The spin and parity were assigned for only  three states, 
the $\Xi$, the $\Xi(1530)$, and the $\Omega^-$, and the PDG listed them as four-star 
states as shown in Fig.~\ref{fig:mass_hype2}~\cite{Zyla:2020zbs}.

More than ten $S=-2$ hyperons are listed in the PDG. However, the spin
and parity quantum numbers are assigned for the ground state $\Xi$ 
($J^P=1/2^+$) and the $\Xi(1530)$ ($J^P=3/2^+$).
For other states, the quantum numbers have not been determined yet.
The QCD dynamics is flavor-blind, and thus, we expect the $S=-2$ and $S=-3$ 
analogue states of the $S=-1$ baryons.
However, neither the first radial excitation of $J^P=1/2^+$ nor
the first orbital excitation with negative parity has been identified.
In this section, we review recent progress of spectroscopy of the $S=-2$ baryons.

\subsection{\it Spin and parity of the $\Xi$ and the $\Xi(1530)$}

The spin of the ground state $\Xi$ hyperon was determined from the decay angular correlation
of $\Xi^- \rightarrow \Lambda \pi^-$ and subsequent decay
$\Lambda \rightarrow p \pi^-$~\cite{Carmony:1964zza}. 
Byers and Fenster~\cite{Byers:1963zz} proposed a method to determine
spin of fermion, 
\[
(2 J+1)=\frac{\left[\langle\hat{p} \cdot \hat{n} \times \hat{\Lambda}\rangle^{2}+\langle\hat{p} \cdot \hat{\Lambda} \times(\hat{n} \times \hat{\Lambda})\rangle^{2}\right]^{1 / 2}}{\left(1-\alpha^{2}\right)^{1 / 2}|\langle \hat{p} \cdot \hat{\Lambda} \hat{n} \cdot \hat{\Lambda}\rangle|},
\] where
$\hat{\Lambda}$ is the $\Lambda$ direction in the $\Xi^-$ rest frame,
$\hat{p}$ is the proton direction in the $\Lambda$ rest frame,
$\alpha$ is the decay parameter of $\Xi\rightarrow \Lambda \pi^-$~\cite{Zyla:2020zbs}, 
$\hat{k}$ and $\hat{\Xi}$ are the directions of the incident $K^-$ beam and
$\Xi^-$, respectively, and
$\hat{n}=(\hat{k} \times \hat{\Xi}) /|\hat{k} \times \hat{\Xi}|$
is the normal to the production plane in the $\Xi^-$ rest frame,
and $\langle ~ \rangle$ represents an average of the enclosed experimental 
quantity.
The experimental result shows $(2 J+1)=1.53 \pm 0.88$, which is consistent 
with $J=1/2$. 
Its parity has not 
been measured directly, and is assigned 
as positive from the quark model expectation for the ground states.

\begin{figure}[tbp]
\begin{center}
  \figureBB{\includegraphics[width=10cm,bb=0 0 243 150]{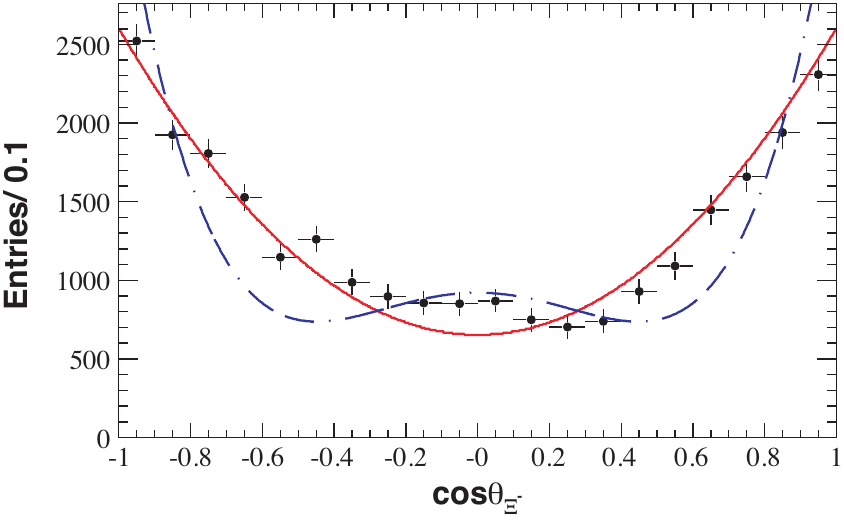}}
  {\includegraphics[width=10cm]{figs/BaBar_Xi1530_spin.pdf}}
\begin{minipage}[t]{16.cm}
\caption{\label{fig:xi_1530_spin_babar}
The $\cos \theta_{\Xi}$ distribution for the
$\Lambda_{c}^{+} \rightarrow \Xi^- \pi^+ K^{+}$ data in the
$\Xi(1530)^{0} \rightarrow \Xi^{-} \pi^{+}$ signal region
after efficiency correction.
The solid (dashed) curve corresponds to the parametrization of the
$\Xi(1530)$ angular distribution for the assumption of pure
spin 3/2 (5/2). Adapted from Ref.~\cite{Aubert:2008ty}.
}
\end{minipage}
\end{center}
\end{figure}

The spin and parity of the $\Xi(1530)$ was investigated using bubble chamber data
with the $K^-$ beam~\cite{Schlein:1963zza,ButtonShafer:1966zz}. Following
the method proposed by Ref.~\cite{Byers:1963zz},
they analyzed the angular correlation between the direction of the $\Lambda$ 
and the polarization of the $\Lambda$ in the $\Xi(1530) \rightarrow \Xi \pi
\rightarrow \Lambda \pi \pi$ decay, similar to 
the spin-parity measurement of the $\Lambda(1405)$ as described in the previous section.
Their observations suggested $P_{3/2}$ or $D_{5/2}$ amplitude for the $\Xi(1530)$.
A clear determination of the spin of the $\Xi(1530)$ is presented by 
the BaBar collaboration.
They measured the $\Lambda_{c}^{+} \rightarrow \Xi^- \pi^+ K^{+}$ decay 
and observed the $\Xi^{0}(1530) $ in the $ \Xi^-\pi^+$ invariant 
mass distribution~\cite{Aubert:2008ty}.
They assumed that the spin of the $\Lambda_c^+$ is $1/2$ and
applied the helicity formalism, where the spin quantization axis 
is chosen along the direction of the $\Xi^{0}(1530)$ in the
$\Lambda_c^+$ rest-frame, and the helicity angle $\theta_{\Xi^-}$
is defined as the angle between the direction of the $\Xi^-$ in the
rest-frame of the $\Xi^{0}(1530)$ and the quantization axis.
Figure~\ref{fig:xi_1530_spin_babar} shows the $\cos \theta_{\Xi}$
distribution and the calculation with the assumption of pure
spin 3/2 (5/2).
The measurement prefers spin 3/2, however significant deviations are
observed, indicating the interference with other amplitudes.
They also applied lineshape analysis to the $\Xi(1530)$ mass spectra, 
and concluded spin 3/2 for the $\Xi(1530)$.
In conjunction with previous analyses~\cite{Schlein:1963zza,ButtonShafer:1966zz},
the spin and parity is established as $J^P=3/2^+$.

\subsection{\it Theoretical studies for the $\Xi(1620)$ and the $\Xi(1690)$}

Above the $\Xi(1530)$, $S=-2$ baryon spectroscopy has not been well established until recently. In a lattice QCD study~\cite{Engel:2013ig}, $\Xi$ states with $J^{P}=1/2^{\pm}$ and $3/2^{\pm}$ were systematically calculated with the lightest pion mass of 255 MeV. It is shown that clean signals can be obtained for the ground state and the $\Xi(1530)$, while other excited states are found with sizable uncertainties. For instance, the lowest $1/2^{-}$ state was found in the range of 1.7-1.9 GeV. Certainly, there is no experimental data of the $S=-2$ meson-baryon scattering, and it is not possible to perform partial wave analysis to extract resonances.

Because there are no experimental inputs to constrain theoretical models, to study the excited $\Xi$ states, one need to extrapolate models constructed in other sectors ($S=0$ and $S=-1$) to $S=-2$ using flavor SU(3) symmetry. In conventional quark models~\cite{Isgur:1978xj,Capstick:1986bm}, the lowest negative parity $\Xi$ states are predicted around 1800 MeV, which are much heavier than the possible candidates $\Xi(1620)$ and $\Xi(1690)$ (see however Ref.~\cite{Pervin:2007wa} which predicted 1725 MeV for the lowest $1/2^{-}$ state). 

For the $\Xi$ resonances, dynamical scattering models have been constructed with chiral SU(3) symmetry~\cite{Ramos:2002xh,GarciaRecio:2003ks,Gamermann:2011mq,Sekihara:2015qqa}. A virtue in this approach is that the leading interaction of the $s$-wave meson-baryon scattering is model-independently given by the Weinberg-Tomozawa term, thanks to chiral symmetry (see Sec.~\ref{subsec:chiralSU3}). Nevertheless, there are cutoff degrees of freedom (subtraction constants), which should in principle be determined by the experimental input. In Refs.~\cite{Ramos:2002xh,GarciaRecio:2003ks,Gamermann:2011mq}, the cutoffs were determined from theoretical considerations. In Ref.~\cite{Ramos:2002xh}, the subtraction constants were chosen to be $a=-2$ at the regularization scale $\mu=630$ MeV, based on the natural size argument~\cite{Oller:2000fj}. Allowing some variation of the subtraction constants and the meson decay constants, a broad resonance was found around 1600 MeV. The residues of the pole indicate that the resonance couples strongly to $\pi\Xi$ and $\bar{K}\Lambda$. This behavior, together with the obtained mass, indicates that the resonance should be identified as $\Xi(1620)$. Because the $s$-wave meson-baryon scattering is considered, this approach predicts the $J^{P}=1/2^{-}$ assignment for the $\Xi(1620)$. 

In Ref.~\cite{GarciaRecio:2003ks}, the regularization condition $T(\sqrt{s}=\mu)=V(\mu)$ (which means $G(\mu)=0$) was adopted to determine the finite part of the loop function, where the scale $\mu$ was chosen to be the mass of the ground state baryon in each sector. This is the energy scale where the unitarized amplitude reduces to that of chiral perturbation theory, which reproduces the expected behavior of the amplitude from the crossing symmetry~\cite{Lutz:2001yb}. With this condition, two resonance poles were found in the $\Xi$ sector at $1565-123i$ MeV and $1663-2i$ MeV,  identified as the $\Xi(1620)$ and the $\Xi(1690)$, respectively. The coupling behavior of the lower energy resonance is similar to that found in Ref.~\cite{Ramos:2002xh}, while the higher energy state shows a small coupling to the $\pi\Xi$ channel, in accordance with the branching fraction of the $\Xi(1620)$ (see next section). 

SU(6) extension of the Weinberg-Tomozawa term was used to study the negative parity baryon resonances in Ref.~\cite{Gamermann:2011mq}. In this approach, by combining chiral SU(3) symmetry with spin SU(2), the scattering of the 35-plet meson ($0^{-}$ meson octet and $1^{-}$ meson nonet) and the 56-plet baryon ($1/2^{+}$ baryon octet and $3/2^{+}$ baryon decuplet) was analyzed. The subtraction constants were determined by the renormalization condition $T(\sqrt{s}=\mu)=V(\mu)$. Two $1/2^{-}$ $\Xi$ states, corresponding to the $\Xi(1620)$ and the $\Xi(1690)$, were found in a similar way with Ref.~\cite{GarciaRecio:2003ks}. In this approach, higher energy $\Xi$ states with $J^{P}=3/2^{-}$ and $5/2^{-}$ were also predicted. 

A detailed analysis focusing on the $\Xi(1690)$ was performed in Ref.~\cite{Sekihara:2015qqa}. The calculation was done in the particle basis with physical hadron masses without assuming isospin symmetry. The determination of the subtraction constants was carried out with the natural renormalization scheme $G(\mu)=0$ with $\mu=M_{\Lambda}, M_{\Sigma}$, or $M_{\Xi}$~\cite{Hyodo:2008xr}. Another set of subtraction constants was determined by fitting to the meson-baryon mass spectra of $\Lambda_{c}^{+}\to K^{+}(\bar{K}^{0}\Lambda)$ and $\Lambda_{c}^{+}\to K^{+}(K^{-}\Sigma^{+})$ decays observed by Belle~\cite{Abe:2001mb} (see next section in more detail). The results of the fitting turns out to be basically similar to those with $G(M_{\Lambda})=0$ case, finding a narrow resonance near the $\bar{K}\Sigma$ threshold. On the other hand, this pole disappears when the scale $\mu$ is chosen to be $\mu=M_{\Sigma}$ or $M_{\Xi}$. This indicates that the property of the $\Xi(1690)$ can be much affected by the proximity of the $\bar{K}\Sigma$ threshold.

The weak decay of $\Xi_{c}\to\pi MB$ was studied in Ref.~\cite{Miyahara:2016yyh}, based on the techniques developed in Refs.~\cite{Roca:2015tea,Miyahara:2015cja,Oset:2016lyh}. Adopting the $S=-2$ scattering amplitudes of models in Refs~\cite{Ramos:2002xh,GarciaRecio:2003ks,Sekihara:2015qqa}, the mass spectra of $MB=\pi\Xi, \bar{K}\Lambda$, and $\bar{K}\Sigma$ were predicted. It is shown that the ratio of the decay fractions into $\bar{K}^{0}\Lambda$ and $K^{-}\Sigma^{+}$ is useful to distinguish the $\Xi(1690)$ resonance peak from a $\bar{K}\Sigma$ threshold effect.

Before closing this section, we note that the theoretical works presented here have all been done before the new data shown in the next section. It is now desired to update the theoretical models in conjunction with the new experimental information.

\subsection{\it Experimental studies for the $\Xi(1620)$ and the $\Xi(1690)$ \label{sec:Xi1620_Xi1690} }

The $\Xi(1620)$ and $\Xi(1690)$ baryons are candidates of 
low-lying excitation states next to the $\Xi(1530)$. 
The latest version of the PDG has assigned to the $\Xi(1690)$ 
a three-star rating on a four-star scale, and a one-star rating to 
the $\Xi(1620)$.

Experimental evidence for the $\Xi(1620)$ was reported in
bubble chamber experiments using the $K^-$
beam~\cite{Apsell:1969sa,Ross:1972bf,Bellefon:1900zz,PhysRevD.16.2706}. 
They observed a peak structure around 1620~MeV, and the mass and
width are consistent within their large statistical uncertainties.
The authors of Refs.~\cite{Borenstein:1972sb} and~\cite{Hassall:1981fs}
searched for this resonance via the $K^-p$ reaction at 2.18~GeV and
6.5~GeV, respectively.
They did not observe a peak structure in the $\Xi\pi$ invariant 
mass spectra for their data, and thus, the $\Xi(1620)$ has not been regarded
as a firmly established state.

\begin{figure}[tbp]
\begin{center}
  \figureBB{\includegraphics[width=10cm,bb=0 0 245 163]{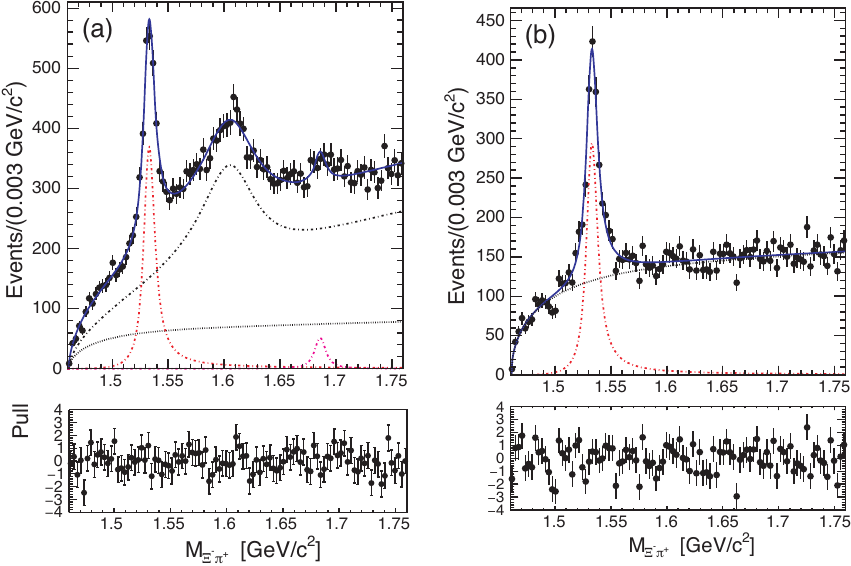}}
  {\includegraphics[width=10cm]{figs/sumihama_Xi_1620.pdf}}
\begin{minipage}[t]{16.cm}
\caption{\label{fig:xi_1620_belle}
The $\Xi^-\pi^+$ invariant mass spectra in 
$\Xi_{c}^{+} \rightarrow \Xi^{-} \pi_{H}^{+} \pi_{L}^{+}$ reaction, 
where the pion with the lower (higher) momentum is labeled 
$\pi_{L}^{+}\left(\pi_{H}^{+}\right)$.
(a) The $\Xi^-\pi^+_L$ invariant mass spectrum in the signal region.
(b)  The $\Xi^-\pi^+_L$ invariant mass spectrum in the sideband region.
Adapted from Ref.~\cite{Sumihama_2019}.
}
\end{minipage}
\end{center}
\end{figure}

Recently, the Belle collaboration measured the $\Xi_{c}^{+} \rightarrow \Xi^{-} \pi^{+} \pi^{+}$
decay, and studied resonance components in the $\Xi^-\pi^+$ system~\cite{Sumihama_2019}.
They observed a clear peak near 1610~MeV with 25$\sigma$ significance as shown
in Fig.~\ref{fig:xi_1620_belle} (a).
With the Breit-Wigner function, the mass and width were measured as 
$1610.4 \pm 6.0(\operatorname{stat})_{-4.2}^{+6.1}(\text{syst})$ and
$59.9 \pm 4.8(\text{stat})_{-7.1}^{+2.8}($syst$)$, respectively.
Figure~\ref{fig:xi_1620_belle} (b) shows
the $\Xi^-\pi^+$ invariant mass in the sideband region of the $\Xi_c^+$ in 
$M(\Xi^-\pi^+\pi^{+})$, where $\Xi^-\pi^+\pi^{+}$ were produced various reactions.
The peak of the $\Xi(1620)$ was not observed, and thus
the $\Xi(1620)$ was produced only from $\Xi_c^+$ decay.
The dependence of the $\Xi(1620)$ on the production mechanisms may be a hint to understand
the internal structure of $\Xi(1620)$.

The first experimental evidence of the $\Xi(1690)$ was reported 
from a bubble chamber experiment using a $K^-$ beam of
4.2~GeV~\cite{Dionisi:1978tg}.
They observed a strong enhancement in the $\Sigma\bar{K}$ invariant
mass spectra in both the neutral and negative-charge states near
the threshold, with weaker evidence in the $\Lambda\bar{K}$ channel.
Although the interpretation of this structure as the threshold enhancement
cannot be excluded, they concluded that the results of 
$\Sigma \bar{K}$-$\Lambda \bar{K}$ coupled channel analysis
is compatible with its interpretation as a resonance state.
The first direct observation of the $\Xi(1690)$ as
a resonance came from a hyperon beam experiment at
CERN~\cite{Biagi:1981cu,Biagi:1986zj}. 
$\Lambda K^-$ pairs were produced diffractively with a 
$\Xi^-$ beam of 116~GeV, 
a peak at 1690 MeV was observed in the
$\Lambda K^-$ invariant mass spectra.
The WA89 collaboration reported the first observation of the
$\Xi^-\pi^+$ decay mode of the $\Xi^0(1690)$ using the $\Sigma^-$ 
beam with 345~GeV~\cite{Adamovich_et_al__1998}.
They observed a clear peak in the $\Xi^-\pi^+$ invariant mass
distribution 
and
obtained the mass and width as
$M=1686 \pm 4~ \mathrm{MeV} $ and 
$\Gamma=10 \pm 6~ \mathrm{MeV}$, which are in
reasonable agreement with the previous measurements.

\begin{figure}[tbp]
\begin{center}
  \figureBB{\includegraphics[width=8cm,bb=309 545 529 691]{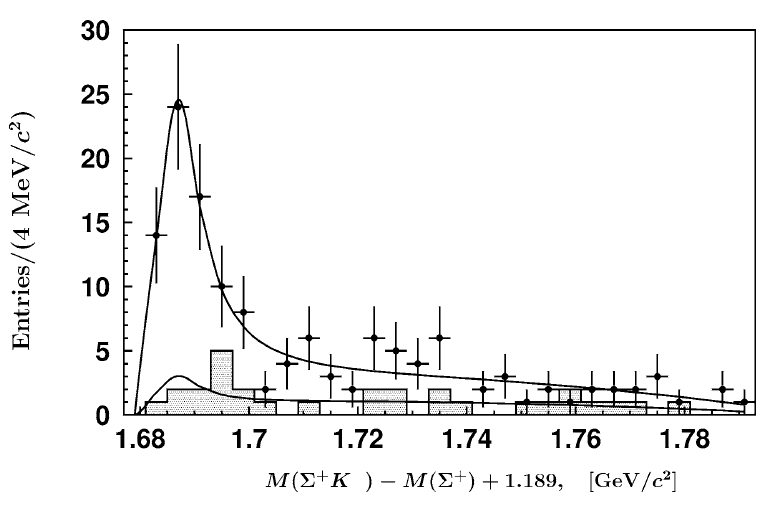}
  \includegraphics[width=8cm,bb=310 549 531 688]{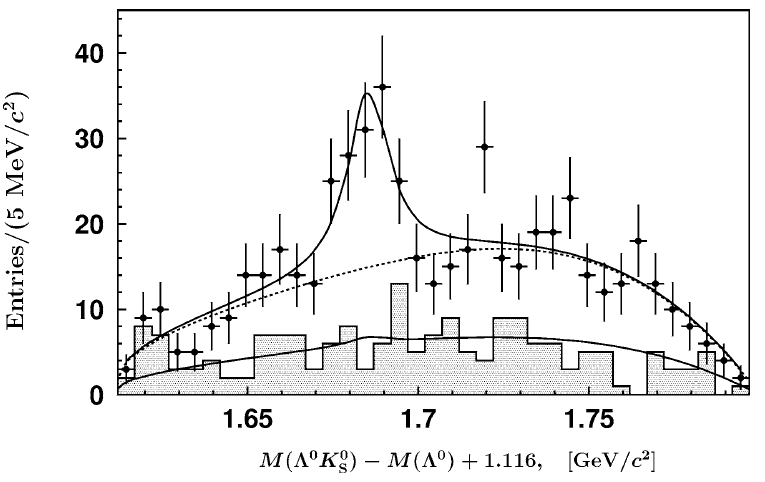}}
  {\includegraphics[width=8cm]{figs/Xi_1690_Belle_Lambda_c1.pdf}
  \includegraphics[width=8cm]{figs/Xi_1690_Belle_Lambda_c2.pdf}}
\caption{Top: Fitting for the $\Lambda_{c}^{+} \rightarrow \Xi(1690)^{0} K^{+}$
component: the invariant mass spectrum of $\Sigma^{+} K^{-}$ combinations from the
$\Lambda_{c}^{+} \rightarrow \Sigma^{+} K^{+} K^{-}$ signal area (points with 
error bars) and $\Lambda_c^+$ sidebands (shaded histogram) are shown.
Bottom: Fitting for the $\Lambda_{c}^{+} \rightarrow \Xi(1690)^{0} K^{+}$
component: the invariant mass spectrum of $\Lambda K_\mathrm{S}^0$
combinations from the 
$\Lambda_{c}^{+} \rightarrow \Lambda^{0} K_{\mathrm{S}}^{0} K^{+}$
signal area (points with error bars) and $\Lambda_c^+$ sidebands
(shaded histogram) are shown.
Adapted from Ref.~\cite{Abe:2001mb}.
~\label{fig:xi_1690_Belle_Lambda_c}
}
\end{center}
\end{figure}

The $\Xi$ resonances were also studied in the decay products of
$\Lambda_c^+$ by the Belle and BaBar collaborations.
The Belle collaboration measured the $\Lambda_c^+ \rightarrow \Sigma^+K^+K^-$
and $\Lambda_{c}^{+} \rightarrow \Lambda^{0} K_{\mathrm{S}}^{0} K^{+}$
decays, and studied resonant substructures in these 
decays~\cite{Abe:2001mb}.
Peaks corresponding to the $\Xi^{0}(1690)$ are seen in the
invariant mass spectra of both 
$\Sigma^+K^-$ and $\Lambda K_{\mathrm{S}}^{0}$ pairs 
(Fig.~\ref{fig:xi_1690_Belle_Lambda_c}).
The measured mass and width are consistent with previous
measurements. 
In addition, they obtained branching fractions of the 
$\Xi^{0}(1690)$ intermediate decay modes as
\[
\frac{\mathcal{B}\left(\Lambda_{c}^{+} \rightarrow \Xi(1690)^{0} K^{+}\right)}{\mathcal{B}\left(\Lambda_{c}^{+} \rightarrow \Sigma^{+} \pi^{+} \pi^{-}\right)} \times \mathcal{B}\left(\Xi(1690)^{0} \rightarrow \Sigma^{+} K^{-}\right)
= 0.023 \pm 0.005 (\text{stat}) \pm 0.005 (\text{syst}) ,~\mathrm{and}
\]
\[
\frac{\mathcal{B}\left(\Lambda_{c}^{+} \rightarrow \Xi(1690)^{0} K^{+}\right)}{\mathcal{B}\left(\Lambda_{c}^{+} \rightarrow \Lambda^{0} \bar{K}^{0} K^{+}\right)} \times \mathcal{B}\left(\Xi(1690)^{0} \rightarrow \Lambda^{0} \bar{K}^{0}\right)
= 0.26 \pm 0.08 (\text{stat}) \pm 0.03 (\text{syst}).
\]
Using known branching fractions of $\mathcal{B}\left(\Lambda_{c}^{+} \rightarrow \Sigma^{+} \pi^{+} \pi^{-}\right)$
and 
$\mathcal{B}\left(\Lambda_{c}^{+} \rightarrow \Lambda^{0} \bar{K}^{0} K^{+}\right)$, the ratio of the branching fractions was obtained as
\[
 \frac{\mathcal{B}\left(\Xi(1690)^{0} \rightarrow \Sigma^{+} K^{-}\right)}{\mathcal{B}\left(\Xi(1690)^{0} \rightarrow \Lambda^{0} \bar{K}^{0}\right)}=0.50 \pm 0.26.
\]
They also searched for the 
$\Lambda_{c}^{+} \rightarrow \Xi^{0}(1690) K^{+}$
decay in the 
$\Lambda_{c}^{+} \rightarrow\left(\Xi^{-} \pi^{+}\right) K^{+}$
decay mode, but did not find the $\Xi^{0}(1690)$ signal
in the $\Xi^-\pi^+$ invariant mass spectrum, which agree with the 
$\mathcal{B}\left(\Xi(1690)^{0} \rightarrow \Xi^- \pi^{+}\right)$
upper limit value from Ref.~\cite{Dionisi:1978tg}.

\begin{figure}[tbp]
\begin{center}
  \figureBB{\includegraphics[width=8cm,bb=0 0 246 156]{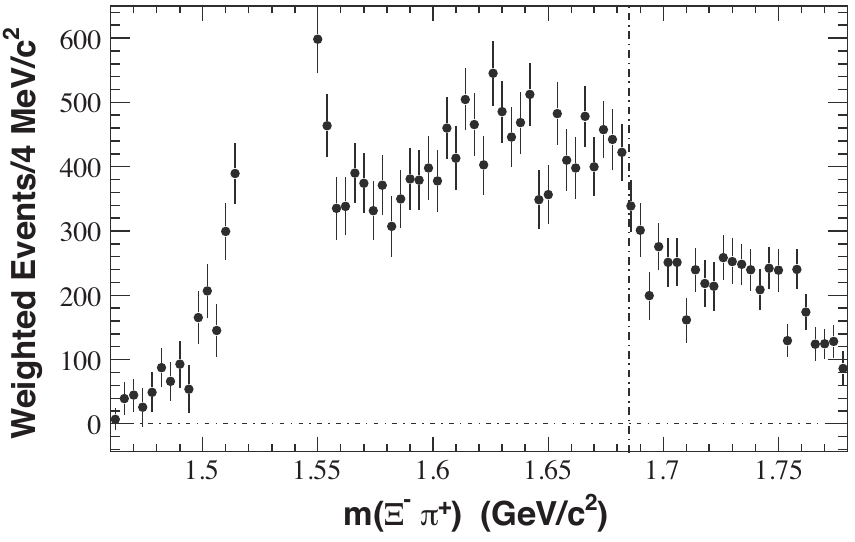}}
  {\includegraphics[width=8cm]{figs/Xi_1690_BaBar.pdf}}
\end{center}
\caption{The efficiency-corrected $\Lambda_c^+$ mass-sideband-subtracted
$P_{0}\left(\cos \theta_{\Xi^{-}}\right)$ moment of the $\Xi^-\pi^+$
system invariant mass distribution for the $\Lambda_c^+$ signal region.
The vertical dot-dashed line indicates the $\Xi(1690)^0$ mass value.
Adapted from Ref.~\cite{Aubert:2008ty}.
\label{fig:xi_1690_babar}}

\end{figure}

The BaBar collaboration examined the angular distribution of 
$\Lambda_c^+ \rightarrow K^+ \Xi^- \pi^+$ decay, and found an
indication that the $\Xi(1690)$ has $J^{P}=1/2^{-}$~\cite{Aubert:2008ty}.
Due to the small branching fraction of the $\Xi(1690)\rightarrow \Xi\pi$ decay,
the signal peak of the $\Xi(1690)$ was not seen. Instead,
they observed a dip structure in the 
$P_{0}\left(\cos \theta_{\Xi^{-}}\right)$ moment of the $\Xi^-\pi^+$
system invariant mass distribution around the $\Xi(1690)$ mass region
(Fig.~\ref{fig:xi_1690_babar}),
where $\theta_{\Xi^{-}}$ is the angle of $\Xi^-$ in the $\Xi^*$ rest-frame
with respect to the direction of $\Xi^*$ in the $\Lambda_c^+$ rest-frame.
This dip could occur as the result of the coherent addition of a small, resonant
$\Xi(1690)$ amplitude to the slowly increasing nonresonant $L^J = S^\frac{1}{2}$ 
amplitude.
The resultant amplitude will then yield a dip in overall intensity in the
$\Xi(1690)$ region. Since the dip is not visible in the $P$-wave amplitude,
it is indicated that the $\Xi(1690)$ decays strongly to the $\Xi^-\pi^+$ system in an
$S$-wave, and hence the $\Xi(1690)$ has spin-parity $1/2^{-}$.

\section{$S=-3$ baryons}\label{sec:S-3}


\subsection{\it Spin of the $\Omega^-$}

\begin{figure}[tbp]
\begin{center}
  \figureBB{\includegraphics[width=10cm,bb=0 0 236 157]{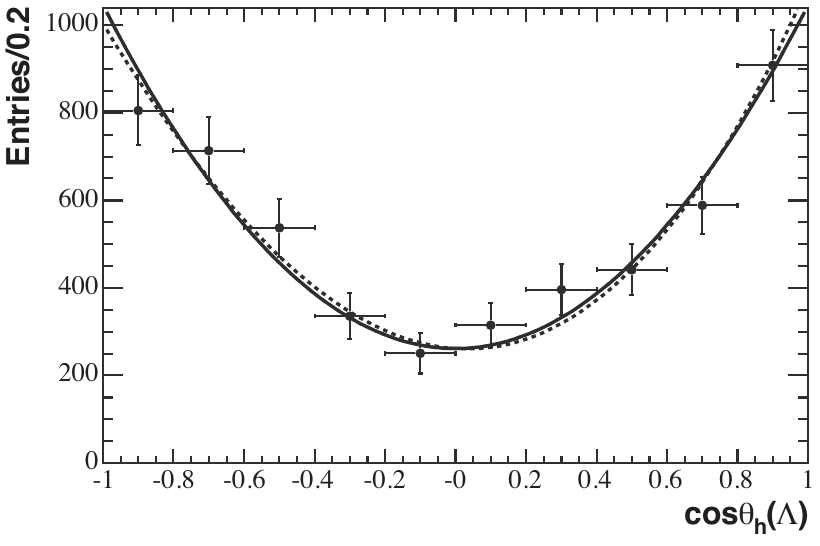}}
  {\includegraphics[width=10cm]{figs/Omega_spin_BaBar.pdf}}
\begin{minipage}[t]{16.cm}
\caption{\label{fig:omega_spin_babar}
The efficiency-corrected $\cos\theta_h(\Lambda)$ distribution for
$\Xi_c^0 \rightarrow \Omega^-K^+$ data. The dashed curve shows the 
 $J_\Omega = 3/2$  fit, in which $\beta$ allows for possible asymmetry
due to the parity violation in $\Xi_c^0$ and $\Omega^-$ weak decay.
The solid curve represents the corresponding fit with $\beta=0$.
Adapted from Ref.~\cite{Aubert:2006dc}.
}
\end{minipage}
\end{center}
\end{figure}

The $\Omega^-$ hyperon is composed of three strange quarks, and 
its spin had been also investigated using the $K^-p$ reaction data taken
with hydrogen bubble chamber~\cite{Deutschmann:1977mq,Baubillier:1978ip,Hemingway:1978hx}. However, due to the limited statistics, the spin was not determined but estimated to be
greater than $1/2$.
The BaBar collaboration presented an unambiguous measurement of the spin of the $\Omega^-$
hyperon production through exclusive process 
$\Xi_c^- \rightarrow \Omega^- K^+$ and 
$\Omega_c^0 \rightarrow \Omega^-\pi^+$~\cite{Aubert:2006dc}.
In the same manner with the spin measurement of $\Xi(1530)$, the helicity formalism
was applied to examine the implications of various $\Omega^-$ spin hypotheses for
angular distribution of the $\Lambda$ from the $\Omega^-$ decay.
The quantization axis was chosen along the direction of the $\Omega^-$ in the
charm baryon rest frame, and 
the helicity angle $\theta_h$ is defined as the angle between the direction of 
the $\Lambda$ in the rest frame of the $\Omega^-$ and the quantization axis.
An asymmetry parameter, $\beta$, was introduced 
in order to take into account of possible asymmetry due to parity violation in
the decay of charm baryon and the $\Omega^-$.
Figure~\ref{fig:omega_spin_babar} shows the $\cos\theta_h(\Lambda)$ distribution 
of the $\Xi_c^- \rightarrow \Omega^- K^+$ decay
with the fit results assuming $J_\Omega=3/2$.
The same analysis was also carried out using the $\Omega_c^0 \rightarrow \Omega^-\pi^+$ 
samples, supporting spin 3/2 results.
Thus, the spin of the $\Omega^-$ is firmly established as 3/2.

\subsection{\it Theoretical studies for the excited $\Omega$ states}

From the viewpoint of flavor SU(3) symmetry, the investigation of the excited $\Omega$ states is of theoretical importance. Unless we consider exotic configurations beyond the three-quark states, an $\Omega$ baryon belongs purely to the decuplet representation. The same is true for a $\Delta$, while a $\Sigma$ and a $\Xi$ can in principle be realized as a mixture of octet and decuplet states. Thus, by determining the masses of the $\Delta$ and $\Omega$ states (hopefully with $J^{P}$ quantum numbers), one can identify the decuplet members in the excited baryons. 

Excitation of the $\Omega$ baryons was studied by lattice QCD in Ref.~\cite{Engel:2013ig} for $J^{P}=1/2^{\pm},3/2^{\pm}$ quantum numbers. The first excited state is found around 2 GeV in the $1/2^{-}$ channel and $3/2^{-}$ channel. The signal in the $1/2^{-}$ channel is obtained with some noise, while in the $3/2^{-}$ channel a fairly good signal is obtained slightly above 2 GeV. The lowest state in the $1/2^{+}$ channel is predicted in the region 2.3-2.6 GeV.

In quark models, the first excited baryons belong to 70-plet of SU(6), which contains one SU(3) decuplet with quark spins combined into $S=1/2$. Together with the orbital angular momentum $\ell=1$, one obtains $J^{P}=1/2^{-}$ and $3/2^{-}$ states as a spin-orbit partner. In actual calculations, the negative parity excited states were found around 2 GeV. For instance, Ref.~\cite{Capstick:1986bm} predicted the $1/2^{-}$ state at 1950 MeV and $3/2^{-}$ state at 2000 MeV. In the model of Ref.~\cite{Pervin:2007wa}, the mass of the $1/2^{-}$ state is obtained at 1923 MeV and the $3/2^{-}$ state at 1953 MeV. 

In Ref.~\cite{Gamermann:2011mq}, a dynamical scattering model with SU(6) extension of the Weinberg-Tomozawa term was used to study negative parity excited baryons. The $1/2^{-}$ state was found at 1798 MeV, and the $3/2^{-}$ state at 1928 MeV, while the other states appear at much higher energies. In contrast to the quark models, the mass difference of the $1/2^{-}$ and $3/2^{-}$ states is not caused by the spin-orbit splitting, because these states belong to different SU(6) multiplets: the $1/2^{-}$ state is in 70-plet and the $3/2^{-}$ state in 56-plet.

\subsection{\it Experimental studies for the excited $\Omega$ states \label{sec:Omega_2012}}

The excited states of the $\Omega^-$
are poorly known; whereas the PDG lists ten excited $\Xi$ states,
only four excited $\Omega^-$ states are listed~\cite{Zyla:2020zbs} 
(Fig.~\ref{fig:mass_hype2}).
Since the isospin of the $\Omega$ is zero, the $\Omega^{*}\rightarrow \Omega\pi$
decays are strongly suppressed, and the possible decays of low-lying excited states are
expected as $\Xi K$  or $\Omega\pi\pi$ channels.
In the following, we review spectroscopy of $S=-3$ hyperons, especially,
recent observation of a low-lying excitation state of $\Omega^-$ by the Belle collaboration.

The experimental search for the excited $\Omega^-$ hyperons via the $K^-p$ reactions
were statistically limited 
due to the small cross sections for $S=-3$ production.
Biagi {\it et al.} reported the first observation of the $\Omega^*$ resonances
using $\Xi^-$ hyperon beam of 116~GeV at the CERN SPS~\cite{Biagi:1985rn}.
They measured inclusive production of $\Xi^-\pi^+K^-$ from a beryllium target,
and observed two peak structures, the $\Omega^-(2250)$ and the $\Omega^{-}(2380)$,
in the invariant mass spectra of $\Xi^-\pi^+K^-$.
The masses and widths of these two resonances are obtained as
$M_{1}=2251 \pm 12$~MeV and $\Gamma_{1}=48 \pm 20 $~MeV, and
$M_{2}=2384 \pm 12$~MeV and $\Gamma_{1}=26 \pm 23 $~MeV, respectively.
They also reported that the $\Omega^{-}(2250)$ decays predominantly into $\Xi^{0}(1530)K^-$,
and obtained 
the ratio of branching fractions
$\mathcal{B}(\Omega^-(2250) \rightarrow \Xi^{0}(1530)K^-)$/
$\mathcal{B}(\Omega^{-}(2250)\rightarrow \Xi^-\pi^+K^-)$
as $0.70\pm 0.20$, while that of the $\Omega^-(2380)$ was consistent with zero.
According to their observation, 
about a half of the
$\Omega^-(2380)\rightarrow \Xi^-\pi^+K^-$ decay occurs
through the quasi-two body $\Omega^-(2380)\rightarrow K^0(890) \Xi^-$ decay.
In the following year, Aston {\it et al.} confirmed the $\Omega^{-}(2250)$ with
the consistent mass and width parameters using the 11~GeV/$c$ $K^-$ beam 
and LASS spectrometer
at SLAC~\cite{Aston:1987bb}. However, the $\Omega^-(2380)$ was not reported. 
The excited $\Omega^-$ states decaying into $\Omega^-\pi^+\pi^-$ were searched for
by the same authors~\cite{Aston:1988yn}. They claimed observation of a resonance $\Omega(2470)$
with the mass and width of
$M =2474 \pm 12$~MeV and $\Gamma=72 \pm 33$~MeV.

\begin{figure}[tbp]
\begin{center}
  \figureBB{\includegraphics[width=10cm,bb=0 0 238 207]{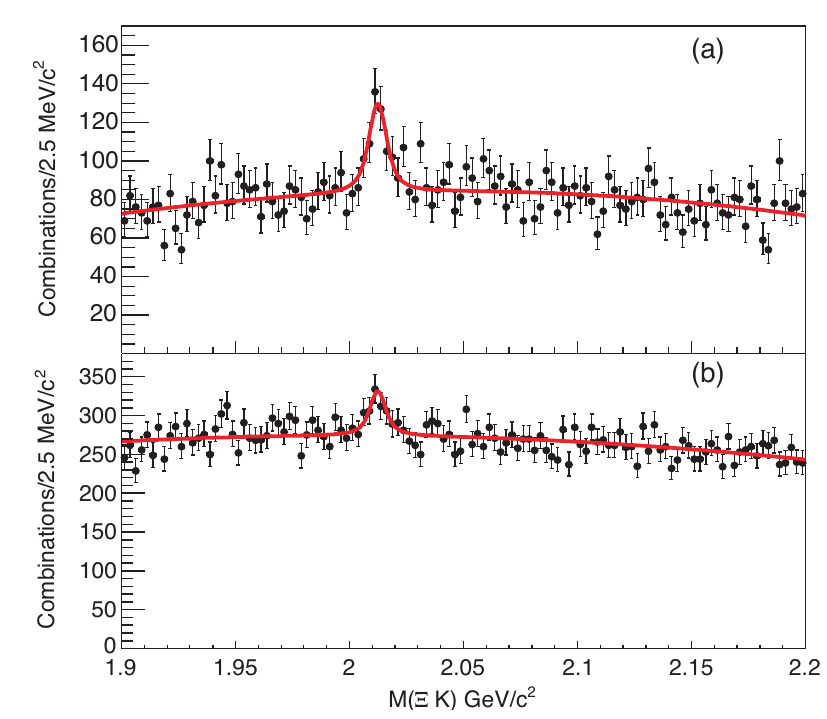}}
  {\includegraphics[width=10cm]{figs/omega_reso_belle.pdf}}
\begin{minipage}[t]{16.cm}
\caption{\label{fig:omega_2012_belle} The (a) $\Xi^0K^-$ and (b) $\Xi^-K^0_S$ invariant mass distributions
in data taken at the $\Upsilon(1 S)$, $\Upsilon(2 S)$, and $\Upsilon(3 S)$ resonance energies. 
The curves show a simultaneous fit to the two distributions with a common mass and width.
Adapted from Ref.~\cite{Yelton:2018mag}}
\end{minipage}
\end{center}
\end{figure}

Recently, the Belle collaboration searched for the excited $\Omega^-$ hyperons
in the $e^+e^-$ collision data taken at the $\Upsilon(1S)$, $\Upsilon(2S)$, 
$\Upsilon(3S)$ energies~\cite{Yelton:2018mag}. 
At these energies, below the $B\overline{B}$ threshold, 
the decays into $B\overline{B}$ pairs are forbidden, hence the decay into
three gluons is dominant. Owing to the flavor blind quark-gluon coupling, a 
substantial number of strange quarks can be produced from gluonic initial states, 
and hence the productions of the
multi-strangeness hadrons like the $\Omega^-$ are expected.
The authors of Ref.~\cite{Yelton:2018mag} investigated
the invariant mass spectra of $\Xi^0 K^-$ and $\Xi^- K^0_{S}$ pairs in these
event samples.
Figure~\ref{fig:omega_2012_belle} shows observed invariant mass distributions
of $\Xi^0K^-$ and $\Xi^- K^0_S$ pairs. One can see clear peaks around 2.1~GeV.
The $\Xi^0K^-$ combinations have strangeness $S=-3$, on the other hand,
the $\Xi^-K^0_S$ combinations may have $S=-1$ and thus have a large combinatorial 
background. 
From the simultaneous fit, the statistical significance is found as
8.3$\sigma$.
The mass and width are obtained 
$2012.4 \pm 0.7(\text{stat}) \pm 0.6(\text{syst})$~MeV and 
$\Gamma=6.4_{-2.0}^{+2.5}(\text{stat}) \pm 1.6(\text{syst})$~MeV, respectively.
In the event sample taken mostly at the $\Upsilon(4S)$ energy, they also find
a peak structure but with less significance, thus, $\Omega^{-}(2012)$ is found
primarily in the decay of the $\Upsilon(1S)$, $\Upsilon(2S)$ and  $\Upsilon(3S)$.
Using the same data sample, the Belle collaboration searched for the
$\Omega(2012) \rightarrow K\Xi(1530)\rightarrow K\pi\Xi$ decay~\cite{Jia:2019eav} 
which is predicted as dominant decay mode by models describing the $\Omega(2012)$ as
a $K\Xi(1530)$ molecule~\cite{Lin:2018nqd,Valderrama:2018bmv,Polyakov:2018mow,Huang:2018wth}.
No significant signals are observed in these channels, and 90\% 
credibility level upper limits on the ratios of the branching fractions
relative to $K\Xi$ decay modes are obtained.
These measurements are important input for understanding 
the structure of the $\Omega(2012)$.
At the same time, theoretical discussion is ongoing on the molecular picture for the $\Omega(2012)$ in conjunction with the data of Ref.~\cite{Jia:2019eav}. For instance, in Ref.~\cite{Lu:2020ste}, it is shown that the molecular picture is compatible with the upper limit of the $\Omega(2012) \rightarrow K\Xi(1530)\rightarrow K\pi\Xi$ decay in Ref.~\cite{Jia:2019eav}. For more details on the $\Omega(2012)$, 
see e.g. Ref.~\cite{Ikeno:2020vqv} and references therein.

\section{$S=+1$ baryons \label{sec:S+1}}


\subsection{\it Status of the $\Theta^{+}$ \label{sec:Theta+}}

Baryon resonances with strangeness $S=+1$ should contain an $\bar{s}$ quark. Baryons that consist of $uudd\bar{s}$ quarks
have an exotic flavor quantum number, and hence regarded as genuine pentaquark states.
In 1997, Diakonov {\it et al.} predicted such a pentaquark state in the chiral soliton model~\cite{Diakonov_1997},
which has a rather light mass around 1530~MeV, spin $1/2$, isospin 0 and strangeness $+1$. A peculiar feature is the narrow width of the predicted state.
The light and narrow exotic baryon resonance in this mass region may escape from the identification in the old $K^+n$ scattering experiments with low resolution.
In 2003, the LEPS collaboration claimed an evidence of a narrow resonance peak, now referred to as the $\Theta^+$,
consistent with the above prediction, in the $\gamma n \rightarrow K^+ K^- n$ reaction 
from a neutron in a carbon nuclei~\cite{Nakano:2003qx}.
Soon after the report by LEPS, some other experimental groups observed the peak at the same energy, while many other groups with higher statistics did not. 
Thus, the existence of the $\Theta^+$ has not been established. In the PDG, the $\Theta^{+}$ first appeared in the 2004 edition with three-star, changed to two-star in 2005, one-star in 2006, and omitted from the summary table in 2007.
More details can be found in Refs.~\cite{MartinezTorres:2010zzb,Torres:2010jh,Hicks:2012zz,Liu_2014,Liu_2019}. 
In the following section, we review a few recent results.

The J-PARC E19 collaboration reported the results of search for 
the $\Theta^+ $ via the $\pi^- p \rightarrow K^- X$ reaction with beam
momenta of 2.01 GeV~\cite{Moritsu:2014bht},
which is an update of the results
with pion beam of 1.92 GeV~\cite{Shirotori:2012ka}.
In both data, they did not observe any signals of the $\Theta^+$ in the missing mass distribution of
the $\pi^- p \rightarrow K^- X$ reaction.
These results are used to constrain the possible $J^{P}$ and the upper limit of the decay width in Ref.~\cite{Hyodo:2012hn}.

The ZEUS collaboration searched for a narrow structure in the $p K_S^0$ and $\bar{p} K_S^0$ systems produced
via $e p $ collision at the center-of-mass energy of 318~GeV for exchanged photon virtuality, $Q^2$, 
between 20 and 100 GeV${}^2$. 
No resonance peak was found in the $p(\bar{p})K_S^0$ invariant-mass distribution in the range $1.45-1.7$~GeV. 
Upper limits on the production cross section were set~\cite{HERA2016446}.

The CLAS collaboration has searched for the $\Theta^+$ using photon beam as the LEPS collaboration did in the $\gamma d \rightarrow K^+ K^- pn$ reaction~\cite{PhysRevC.79.025210}. The energy ranges of photon beams of these experiments overlap with each other, while the angular coverage of 
detectors are different. 
In the former analysis of the CLAS collaboration~\cite{McKinnon:2006zv}, they measured the 
momenta of a $K^+$, a $K^-$, and a proton, and identified a neutron using the missing mass technique.
On the other hand, the LEPS collaboration detected only a $K^+K^-$ pair, and assumed that the proton is
a spectator~\cite{PhysRevC.79.025210}, thus the kinematic conditions also differed.
The authors of Ref.~\cite{Camp:2017jca} performed an analysis following the method of
the LEPS collaboration using CLAS data, where only a $K^+K^-$ pair was detected.
In their preliminary results, no signals of the $\Theta^+$ were found in the $K^+n$ invariant 
mass distribution.

The LEPS collaboration has continued their search for the $\Theta^+$ in the $\gamma d \rightarrow K^+ K^- p n$ 
reaction.
They took data sample using almost identical experimental setup with their second result~\cite{PhysRevC.79.025210}
but about 2.6 times higher statistics.
In their preliminary results of inclusive analysis, where the proton and neutron 
contributions were not separated, a strong peak of the $\Theta^+$ was not observed. 
However, by applying exclusive analysis rejecting protons which were produced through background reactions, the
signal-to-noise ratio was improved and a signal enhancement was observed~\cite{Kato:2013hua}. 
They took further data with improved
proton detection efficiency~\cite{Yosoi:2019mno}, and their results will be presented in the near future.

Theoretical study of the $K^{+}d\to K^{0}pp$ reaction was carried out in Ref.~\cite{Sekihara:2019cot} using the two-step approach (see also the review of this reaction in Ref.~\cite{Sibirtsev:2006yw}). In this process, the $\Theta^{+}$ can be produced by direct formation $KN\to KN$ in the second step. Examining several momenta of the incident kaon, feasibility of possible experiments to search for the $\Theta^{+}$ was discussed. This analysis will be useful to design new experiments at $K^{+}$ beam facilities such as J-PARC.

\section{Summary and future prospects}


Because of the peculiar role of the strange quark in QCD, the strange baryons form a rich and complicated spectrum. In this review, we have discussed the physics of strange baryon spectrum from theoretical and experimental viewpoints. We have first overviewed the basic properties of QCD, emphasizing the role of symmetries and the developments of the scattering calculation in lattice QCD. Because the baryon excited states with strangeness are unstable against the strong decay, they should be treated as resonances in hadron-hadron scatterings. As an appropriate tool to study resonances, we have introduced the scattering theory and explained how the resonance pole of the scattering amplitude is related to the generalized eigenstate of the Hamiltonian. The quest for the internal structure of exotic hadrons is an important subject in hadron physics, and is also relevant for the strange baryons. We have summarized the classification of exotic hadrons, and explained the importance of defining a proper measure of the internal structure of hadron resonances. 

Next, we have reviewed the current status of selected baryon resonances. In the $S=-1$ sector, the most intensively studied resonance is the ``$\Lambda(1405)$'' in the energy region below the $\bar{K}N$ threshold. Thanks to the experimental developments, new and high-statistics data have been obtained for the low-energy $\bar{K}N$ scattering as well as the $\pi\Sigma$ mass spectrum. The theoretical framework to construct the low-energy meson-baryon scattering has been systematically developed from the unitarity of the scattering amplitude and chiral symmetry of QCD. The combination of these developments has revealed an interesting two-pole structure, leading to the new entry of the two-star resonance $\Lambda(1380)$ in the PDG Particle Listings~\cite{Zyla:2020zbs}. It is shown that the pole position of the $\Lambda(1405)$ is converging to the energy region around 1420 MeV, while the position of the $\Lambda(1380)$ pole is not yet settled quantitatively. In the energy region above the $\bar{K}N$ threshold, the $\Lambda^{*}$ resonances have been identified by the partial wave analysis of the $K^{-}p$ scattering data. At the same time, the three-body decay spectra of $\Lambda_{c}$ provide additional information of the baryon resonances.

In contrast to the $S=-1$ sector, theoretical calculation of the excited $\Xi$ baryons in the $S=-2$ sector had not been well established, because of the lack of the direct scattering experiment. In particular, predictions of negative parity resonances had not been very converging. This situation is now being changed by the new data of three-body decays of charmed baryons. Accurate $\pi\Xi$ mass spectra have been obtained in the $\Xi_{c}\to \pi\pi\Xi$ decay, leading to the identification of the $\Xi(1620)$ and the $\Xi(1690)$, whose properties are constrained by the decay branching ratios. The $S=-3$ sector shows a similar trend; theoretical investigation is now stimulated by the new data from the $\Upsilon$ decays, and intensive discussion on the nature of the newly found $\Omega(2012)$ is ongoing. The $S=+1$ sector is a unique channel to study the flavor exotics. Although the existence of the $\Theta^{+}$ was not established, the investigation of this sector should be continued for the understanding of the formation mechanism of hadrons in QCD.

Based on the current status described above, let us indicate several directions of the future prospects:
\begin{itemize}

\item The $\Lambda(1405)$/$\Lambda(1380)$ pole positions \\
Quantitative determination of the pole positions is an important step to clarify the nature of resonances. 
In the PDG, evidence of existence of the $\Lambda(1380)$ is only {\it fair}, and 
we need more data and analysis to confirm the two-pole structure of 
the $\Lambda$ resonances in this energy region. New results from 
the E31 collaboration at J-PARC~\cite{J-PARC_P31},
the BGO-OD collaboration at ELSA~\cite{jude2020strangeness}, 
the LEPS2 collaboration at SPring-8~\cite{Yosoi:2019mno}, and the AMADEUS collaboration
at DA$\Phi$NE~\cite{Piscicchia:2018umq} are expected in the near future.
Together with a detailed understanding of the reaction mechanism, the $\pi\Sigma$ mass spectra in these reactions will provide clues to investigate the pole structure in this energy region. To determine the position of the $\Lambda(1380)$ pole, it is desirable to have experimental constraints in the energy region far below the $\bar{K}N$ threshold. A possible strategy for this purpose is to determine the scattering lengths of $\pi\Sigma$ and $\pi\Lambda$, which constrains the meson-baryon scattering amplitude directly. A method to determine the $\pi\Sigma$ scattering lengths from the $\Lambda_{c}$ decay was proposed in Ref.~\cite{Hyodo:2011js}. To further sharpen the precision of the position of the $\Lambda(1405)$ pole near the $\bar{K}N$ threshold, accurate measurements of the kaonic hydrogen and kaonic deuterium will provide threshold constraints for the $\bar{K}N$ system~\cite{J-PARC_P57,Scordo:2018ufs,Meissner:2004jr,Meissner:2006gx,Hoshino:2017mty}. The near-threshold $\bar{K}N$ interaction will also be constrained by the high-precision data of the correlation function measurements in the high-energy collisions.

\item Determination of resonance parameters \\
Traditionally, the resonance parameters (mass and width) have been extracted by fitting the peak of the spectrum with the Breit-Wigner parametrization. As we have discussed in Section~\ref{sec:resonances}, the use of the Breit-Wigner function is valid only for an isolated narrow resonance. In particular, the effect of nearby thresholds should be considered seriously, because there are several thresholds of meson-baryon channels in the energy region of strange baryon resonances. Several methods have been proposed to extract the resonance parameters with including the threshold effect~\cite{Flatte:1976xu,Hanhart:2015cua}, and theoretical effort should be continued further to develop general tools for the analysis of near-threshold resonances. In view of the accumulation of new and precise experimental data of heavy hadron decays at LHCb~\cite{Palano:2018fcr}, Belle~II~\cite{Snowmass}, and BES~III~\cite{Ablikim:2019hff}, developing a theoretical framework to properly treat the final state interactions in the whole region of the Dalitz plot is also an important issue to be explored in the future.
Another approach to study hyperon resonances is to produce them
using the kaon beam.
Recently, new experiments have beam proposed at JLab~\cite{Amaryan:2020xhw} and J-PARC~\cite{J-PARC_LOI_Xi},
utilizing the neutral and charged kaon beams, respectively.
The pole positions of 
excited $\Lambda$, $\Sigma$, $\Xi$, and $\Omega$ hyperons
will be presented from PWA by these experiments in the near future.

\item Internal structure of hadron resonances \\
The identification of strange baryons with exotic internal structure is an important subject to understand the nonperturbative dynamics of QCD. We emphasize that this is not a simple task. First of all, we have to clearly define what is meant by the ``exotic structure''. Although there are many studies to clarify the internal structure of hadrons, they are in general based on various different definitions of the structure. This sometimes causes controversial discussion. To end the controversy, therefore, it is needed to establish a theoretically well-defined measure to characterize the structure of hadron resonances. Unless a good measure is established, we should keep in mind that there is no ``smoking-gun'' experimental observable which directly determines the internal structure of hadrons. Rather, we need to accumulate several clues from various experiments to finally elucidate the exotic nature. In the following, let us give a couple of promising attempts to relate the internal structure with the experimental observables. The compositeness of near-threshold resonances can be related to the pole position and the two-body scattering length~\cite{Kamiya:2015aea,Kamiya:2016oao}. To obtain an intuitive picture of the structure, it is useful to extract the spatial distribution of the resonance wave function, through the form factors and radiative decay~\cite{Darewych:1985dc,Burkhardt:1991ms,Sekihara:2008qk,Sekihara:2010uz,Sekihara:2013sma}. At high-energy exclusive productions, the constituent-counting rule in perturbative QCD can be used to determine the internal quark-gluon configurations~\cite{Kawamura:2013iia}. It is shown that the production yields in the high-energy collision experiments reflect the internal structure of hadrons due to the different coalescence mechanisms~\cite{ExHIC:2010pzl,Cho:2011ew,Cho:2017dcy}. By examining various criteria, we can approach the nature of strange baryon resonances. 

\item Lattice QCD \\
First principle calculations of QCD are very welcome for the study of strange baryon spectrum. It is worth noting that the lattice calculations and real experiments can be complementary with each other. For instance, while the $\pi\Sigma$ scattering is impossible in experiments, the elastic scattering of the lowest-energy channel would be the first target in lattice QCD, before going to  more involved coupled-channel calculations. Although the meson-baryon scattering calculation on the lattice is not yet completely matured, the developments in other sectors is encouraging. 

\end{itemize}
We hope that the contents of this review stimulate the further developments in this field, which eventually shed new light on the physics of baryon spectrum with strangeness in the near future.

\section*{Acknowledgments}
This work is supported in part by the Grants-in-Aid for Scientific Research
from JSPS (Nos.
19H05150, 
16K17694, 
and 16H06007) 



\end{document}